\documentclass[10pt,journal,compsoc]{IEEEtran}
\usepackage[table]{xcolor}
\usepackage[textsize=scriptsize,backgroundcolor=yellow!40]{todonotes}
\usepackage[english]{babel}
\usepackage{subcaption}
\usepackage{url}
\usepackage{colortbl}
\usepackage{amssymb}
\usepackage{stmaryrd}
\usepackage{amsmath}
\usepackage{mathrsfs}
\usepackage{amssymb}
\usepackage[left]{lineno}
\usepackage[nonumberlist,acronym,sanitize=none]{glossaries}
\glsdisablehyper
\usepackage{comment}
\usepackage[pdftex, colorlinks=true, hyperfootnotes=true, hyperindex=true,
            plainpages=false, pagebackref=false, pdfpagelabels=true, pdfstartview=FitH,
            linkcolor=blue, citecolor=blue, urlcolor=blue,
            bookmarks, bookmarksopen, bookmarksdepth=3]{hyperref}
\usepackage[capitalise,nameinlink]{cleveref}
\Crefname{algocf}{Algorithm}{Algorithms}
\usepackage{tkz-base}
\usetikzlibrary{decorations.pathmorphing,trees,snakes,arrows,shapes,automata}
\usepackage{paralist}
\usepackage{multicol}
\usepackage{booktabs}
\usepackage[shortlabels]{enumitem}
\usepackage{multirow}
\usepackage{rotating}
\usepackage{etaremune}
\usepackage{marginnote}
\usepackage{mathtools}
\usepackage{mathabx}
\usepackage{ifthen}
\usepackage[normalem]{ulem}
\usepackage{lineno}
\usepackage{soul}
\usepackage{xfrac}
\usepackage{lipsum}
\usepackage{makecell}
\usepackage{siunitx}
\usepackage{adjustbox}
\usepackage{microtype}
\usepackage{wrapfig}
\usetikzlibrary{DECLARE} 
\usepackage{float}
\renewcommand{\arraystretch}{1.5}
\renewcommand\tabcolsep{5pt}
\newcolumntype{d}{>{\columncolor{gray!10}}c}
\newcolumntype{m}{>{\columncolor{gray!10}}l}
\newenvironment{iiilist}%
{\begin{inparaenum}[\itshape(i)\upshape]}%
{\end{inparaenum}}

\def\GoodExampleMark{\checkmark}
\def\BadExampleMark{$\times$}

\RequirePackage{xparse}
\NewDocumentEnvironment{AuthNote}{+o+o}{%
	\IfValueT{#2}{\marginnote{\scriptsize{#2}}}%
	\begin{scriptsize}
		\colorbox{gray}%
		{\color{white} Note\IfValueT{#1}{ (#1)}:}%
		\quad%
		\color{brown}
}{%
	\normalcolor
	\end{scriptsize}
}

\RequirePackage{pifont}

\RequirePackage{lipsum}
\newcommand{\LipsumGray}[1][]{{\color{gray}\ifthenelse{\equal{#1}{}}{\lipsum}{\lipsum[#1]}}}

\RequirePackage{siunitx}
\newcolumntype{D}[1]{S[
	table-omit-exponent,
	round-mode=places,
	round-integer-to-decimal,
	round-precision={#1}]} 
\input{addon/macros-for-requirements-list}
\input{addon/macros-DECLARE} 
\input{addon/macros-DECLARE-drawing} 
\graphicspath{{figures/}}
\newacronym[\glslongpluralkey={Business Processes}]{bp}{BP}{Business Process}
\newacronym{wf}{WF}{workflow}
\newacronym{bpi}{BPI}{Business Process Intelligence}
\newacronym{bpm}{BPM}{Business Process Management}
\newacronym{bpms}{BPMS}{Business Process Management System}
\newacronym{bpmn}{BPMN}{Business Process Model and Notation}
\newacronym{cpn}{CPN}{colored Petri net}
\newacronym{dfg}{DFG}{Directly-Follows Graph}
\newacronym{soa}{SOA}{Service-Oriented Architecture}
\newacronym{kpi}{KPI}{Key Performance Indicator}
\newacronym{wfms}{WfMS}{Workflow Management System}
\newacronym{pn}{PN}{Petri net}
\newacronym{xes}{XES}{eXtensible Event Stream}
\newacronym{yawl}{YAWL}{Yet Another Workflow Language}
\newglossaryentry{task}{%
	name={task},description={the non-divisible, elementary activity}}
\def\paramx {\ensuremath{x}}
\def\paramy {\ensuremath{y}}

\def\letterx {\ensuremath{a}}
\def\lettery {\ensuremath{b}}
\def\letterz {\ensuremath{c}}

\newcommand{\Task}[1] {\ensuremath{\scalebox{0.85}{\normalfont\textsf{#1}}}}
\def\taska {\Task{a}}
\def\taskb {\Task{b}}
\def\taskc {\Task{c}}

\newglossaryentry{promod}{%
	name={process model},description={the model of a process}
}
\def\LogAlph {\ensuremath{\Sigma}}
\newglossaryentry{logalph}{
	name={log alphabet},description={the process alphabet, as reflected in a log},%
	symbol={\LogAlph}}
\def\Evt {\ensuremath{e}}
\newglossaryentry{evt}{
	name={event},description={a record of an instantaneous fact during the process enactment},%
	symbol={\Evt}}
\def\Trc { \ensuremath{\tau} }
\def\StrTrc { \ensuremath{\sigma} }
\newglossaryentry{trace}{
	name={trace},description={a sequence of \glsplural{evt}},%
	symbol={\Trc}}
\def\EvtLog {\ensuremath{L}}

\newglossaryentry{evtlog}{
	name={event log},description={a collection of \glstext{evttrace}s},%
	symbol={\EvtLog}} 
\newcommand{\DclrSty}[1] {\textsc{#1}}
\def\Declare {\DclrSty{Declare}}
\newglossaryentry{declare}{%
	name={\Declare},description={a declarative process modelling language and notation}}
\def\DeclaModel {\ensuremath{\mathcal{M}}}
\newglossaryentry{declamodel}{%
	name={declarative \glsentrytext{promod}},description={\glsentrydesc{promod}, expressed by means of constraints},
	symbol={\DeclaModel}
}

\newglossaryentry{mindeclamodel}{%
	name={discovered \glsentrytext{declamodel}},description={\glsentrydesc{declamodel}, discovered from an \glsentrytext{evtlog}},
	symbol={\DeclaModel}
}
\newglossaryentry{minerful}{%
	name={\DclrSty{miner}ful},description={A declarative process discovery algorithm}}
\newacronym{mf}{Mf}{\gls{minerful}}
\newglossaryentry{minerfulVac}{%
	name={MINERful Vacuity Checker},description={\glsentrytext{minerful}} algorithm with semantical vacuity detection}
\newacronym{mfv}{Mf-Vchk}{\gls{minerfulVac}}
\newacronym{dmm}{DMM}{Declare Maps Miner}
\newglossaryentry{decmapmin}{%
	name={Declare Maps Miner},description={the declarative process discovery algorithm \glsentrytext{decmapmin}}}
\newacronym{dmm2}{DM2}{Declare Miner 2}
\newglossaryentry{decmapmin2}{%
	name={Declare Miner 2},description={improvement of \glsentrytext{decmapmin} algorithm}}
\newglossaryentry{janus}{%
	name={Janus},description={the declarative process discovery algorithm \glsentrytext{janus}}}
\def\Subsum {\ensuremath{\sqsubseteq}}

\newglossaryentry{subsum}{%
	name={subsumption},description={is subsumed by},%
	symbol={\Subsum}}

\newglossaryentry{relaxop}{%
	name={relaxation},description={relaxation operator, climbing the \glsentrytext{subsum} hierarchy}}

\newglossaryentry{actv}{%
	name={activation},description={the activation of a constraint}}
\newglossaryentry{activator}{name={activator},description={the event that signals the occurrence of the activation in the trace}}

\newglossaryentry{target}{%
	name={target},description={target}}

\def\Cns {\ensuremath{C}}
\newglossaryentry{con}{%
	name={constraint},description={a temporal business rule},
	symbol={\Cns}
}

\newglossaryentry{welldef}{%
	name={well-defined},description={of \glsentrytext{con}s for which a finite non-empty trace exists that complies with them}
}
\newglossaryentry{cnspar}{%
	name={parameter},description={a parameter of a \glsentrytext{con}},
}
\newglossaryentry{cnsarity}{%
	name={arity},description={number of parameters of a \glsentrytext{con}},
}
\newglossaryentry{exi}{
	name={existence},
	description={constrains single tasks}
}
\newglossaryentry{exicon}{
	name={\glsentrytext{exi} \glsentrytext{con}},
	description={constrains single tasks}
}
\newglossaryentry{posicon}{
	name={position \glsentrytext{con}},
	description={constrains the position of tasks}
}
\newglossaryentry{cardicon}{
	name={cardinality \glsentrytext{con}},
	description={limits the number of tasks}
}
\newglossaryentry{rela}{
	name={relation},
	description={constraint on pairs of tasks}
}
\newglossaryentry{relacon}{
	name={\glsentrytext{rela} \glsentrytext{con}},
	description={constraint on pairs of tasks}
}
\newglossaryentry{unirelacon}{
	name={unidirectional \glsentrytext{relacon}},
	description={constraint on pairs of tasks, out of which one is the activation, as the other is the target}
}
\newglossaryentry{unifwrelacon}{
	name={\glsentrytext{fw}-\glsentrytext{unirelacon}},
	description={constraint on pairs of tasks, having the first parameter as the activation, and the second one as the target}
}
\def\FwCns {\ensuremath{\mathit{fw}}}
\newglossaryentry{fw}{
	name={forward},
	description={forward constraint},
	symbol={\FwCns}
}

\newglossaryentry{unibwrelacon}{
	name={\glsentrytext{bw}-\glsentrytext{unirelacon}},
	description={constraint on pairs of tasks, having the second parameter as the activation, and the first one as the target}
}
\def\BwCns {\ensuremath{\mathit{bw}}}
\newglossaryentry{bw}{
	name={backward},
	description={backward constraint},
	symbol={\BwCns}
}

\newglossaryentry{corelacon}{
	name={coupling \glsentrytext{con}},
	description={constraint based on pairs of relation constraints}
}
\newglossaryentry{nega}{
	name={negative},
	description={of a constraint, that negates a coupling relation constraint}
}
\newglossaryentry{negacon}{
	name={\glsentrytext{nega} \glsentrytext{con}},
	description={constraint negating a coupling relation constraint}
}
\def\CnsTmp {\ensuremath{\mathcal{C}}}
\newglossaryentry{cnstemp}{%
	name={template},description={the template of a \glsentrydesc{con}},
	symbol={\CnsTmp}}
\def\CnsTmpPrm {\ensuremath{\CnsTemp'}}
\def\CnsTmpSec {\ensuremath{\CnsTemp''}}
\newcommand{\CnsTmpFunc}[2] {\ensuremath{\CnsTmp(#1\ifthenelse{\equal{#2}{}}{}{,#2})}}
\newcommand{\CnsTmpFuncPrm}[2] {\ensuremath{\CnsTmpPrm(#1\ifthenelse{\equal{#2}{}}{}{,#2})}}
\newcommand{\CnsTmpFuncSec}[2] {\ensuremath{\CnsTmpSec(#1\ifthenelse{\equal{#2}{}}{}{,#2})}}

\newglossaryentry{cnstype}{%
	name={type},description={the type of a \glsentrydesc{cnstemp}}}
\def\CnsRep {\ensuremath{\mathfrak{C}}}
\newglossaryentry{cnsrep}{name={repertoire},description={the repertoire of \glsentrytext{declare} \glsentrytext{temp}s},
	symbol={\CnsRep}}
\newglossaryentry{cnsuniv}{name={\glsentrytext{con}s universe},description={the set of \glsentrytext{declare} \glsentrytext{temp}s over the process alphabet reflected in the log}}
\def\CnsInstRelation {\ensuremath{\Gamma}}

\newglossaryentry{cnsinst}{%
	name={\glsentrytext{cnstemp} instantiation relation},description={the assignment relation instantiating \glsentrytext{cnstemp}s into \glsentrytext{con}s, namely assigning \glsentrytext{task}s to \glsentrytext{cnspar}s.},
	symbol={\CnsInstRelation}}

\newcommand{\CnsInterpFun} {\ensuremath{\mathscr{I}}}
\newglossaryentry{cnsinterp}{
	name={interpretation function},description={function interpreting a \glsentrytext{declamodel}},
	symbol={\CnsInterpFun}}

\def\RelaConTemp {\ensuremath{\mathcal{R}}}
\newglossaryentry{relacontemp}{%
	name={relation template},description={the template of a relation \glsentrydesc{con}},
	symbol={\RelaConTemp}}

\def\ExiConTemp {\ensuremath{\mathcal{E}}}
\newglossaryentry{exicontemp}{%
	name={existence template},description={the template of an existence \glsentrydesc{con}},
		symbol={\ExiConTemp}}

\def\Supp {\ensuremath{\sigma}}

\newglossaryentry{support}{%
	name={support},description={the support of a \glsentrydesc{con}},
	symbol={\Supp}}

\def\Conf {\ensuremath{\kappa}}
\newglossaryentry{conf}{%
	name={confidence},description={the confidence level of a \glsentrydesc{con}},
	symbol={\Conf}}

\def\IntF {\ensuremath{\iota}}
\newglossaryentry{intf}{%
	name={interest factor},description={the interest factor of a \glsentrydesc{con}},
	symbol={\IntF}}

\def\CnsEvalFunctor {\ensuremath{\eta}}
\newglossaryentry{cnseval}{
	name={evaluation},description={evaluation of a \glsentrytext{con} or a \glsentrytext{declamodel} over a \glsentrytext{evttrace} or an \glsentrytext{evtlog}},
	symbol={\CnsEvalFunctor}}

\def\UniqTxt {AtMostOne}

\def\RespTxt {Response}
\def\AltRespTxt {AlternateResponse}

\def\ChaRespTxt {ChainResponse}
\def\PrecTxt {Precedence}
\def\AltPrecTxt {AlternatePrecedence}

\def\ChaPrecTxt {ChainPrecedence}

\def\SuccTxt {Succession}

\def\NotSuccTxt {NotSuccession}

\def\UniqTmp {\ensuremath{\DclrSty{\UniqTxt}}}

\def\RespTmp {\ensuremath{\DclrSty{\RespTxt}}}
\def\AltRespTmp {\ensuremath{\DclrSty{\AltRespTxt}}}
\def\ChaRespTmp {\ensuremath{\DclrSty{\ChaRespTxt}}}
\def\PrecTmp {\ensuremath{\DclrSty{\PrecTxt}}}
\def\AltPrecTmp {\ensuremath{\DclrSty{\AltPrecTxt}}}
\def\ChaPrecTmp {\ensuremath{\DclrSty{\ChaPrecTxt}}}

\def\NotSuccTmp {\ensuremath{\DclrSty{\NotSuccTxt}}}

\newcommand{\Uniq}[1] {\ensuremath{\DclrSty{\UniqTxt}(#1)}}

\newcommand{\Resp}[2] {\ensuremath{\DclrSty{\RespTxt}(#1,#2)}}

\newcommand{\AltResp}[2] {\ensuremath{\DclrSty{\AltRespTxt}(#1,#2)}}

\newcommand{\ChaResp}[2] {\ensuremath{\DclrSty{\ChaRespTxt}(#1,#2)}}

\newcommand{\Prec}[2] {\ensuremath{{\DclrSty{\PrecTxt}}(#1,#2)}}
\newcommand{\AltPrec}[2] {\ensuremath{\DclrSty{\AltPrecTxt}(#1,#2)}}

\newcommand{\ChaPrec}[2] {\ensuremath{\DclrSty{\ChaPrecTxt}(#1,#2)}}

\newcommand{\Succ}[2] {\ensuremath{\DclrSty{\SuccTxt}(#1,#2)}}

\newcommand{\NotSucc}[2] {\ensuremath{\DclrSty{\NotSuccTxt}(#1,#2)}}

\newglossaryentry{fulfilment}{name={fulfilment},description={satisfaction of a constraint on a trace in which the activation occurs}} 
\newglossaryentry{condridet}{
	name={process drift detection},description={automated identification of changes in the process execution},%
}
\newacronym{dvd}{VDD}{Visual Drift Detection}
\def\DriftMap {Drift Map}
\def\DriftChart {Drift Chart}

\def\Windo{\ensuremath{\mathrm{win}}}
\def\WinSize{\ensuremath{\Windo_{\mathrm{size}}}} 
\def\WinStep{\ensuremath{\Windo_{\mathrm{step}}}} 
\def\WinNum{\ensuremath{\#_{\Windo}}} 
\def\CnsNum{\ensuremath{\#_{\mathrm{cns}}}} 
\def\Errtcsm{\ensuremath{\mathrm{Erratic}}}
\def\Sprconfm{\ensuremath{\mathrm{Spread}}}
\newacronym{edfg}{eDFG}{extended Directly-Follows Graph}

\def\SeqNum{\ensuremath{\#_{\mathrm{seq}}}} 

\def\ActNum{\ensuremath{\#_{\mathrm{act}}}} 

\ifCLASSOPTIONcompsoc
  \usepackage[nocompress]{cite}
\else
  \usepackage{cite}
\fi
\ifCLASSINFOpdf
\else
\fi
\makeatletter
\def\endthebibliography{%
	\def\@noitemerr{\@latex@warning{Empty `thebibliography' environment}}%
	\endlist
}
\makeatother

\begin{document}
%

\title{Visual Drift Detection for Event Sequence Data of Business Processes}

%
%
%
%

\author{Anton~Yeshchenko, 
        Claudio~Di~Ciccio, 
        Jan~Mendling,
        and~Artem~Polyvyanyy
\IEEEcompsocitemizethanks{\IEEEcompsocthanksitem Anton Yeshchenko and Jan Mendling are with the Vienna Universtiy of Economics and Business, Vienna, Austria
\protect\\
E-mails: {anton.yeshchenko@wu.ac.at}, {jan.mendling@wu.ac.at}
\IEEEcompsocthanksitem Claudio Di Ciccio is with the Sapienza University of Rome.
\protect\\
E-mail: {claudio.diciccio@uniroma1.it}
\IEEEcompsocthanksitem Artem Polyvyanyy is with the University of Melbourne.
\protect\\
E-mail: {artem.polyvyanyy@unimelb.edu.au}}
%
%
}

%
%

\markboth{IEEE Trans. Vis. Comput. Graph.,~Vol.~X, No.~X, X-20XX}%
{Yeshchenko \MakeLowercase{\textit{et al.}}: A Comprehensive Approach for the Visual Analysis of Process Drifts}
%



\IEEEtitleabstractindextext{%
\begin{abstract}
Event sequence data is increasingly available in various application domains, such as business process management, software engineering, or medical pathways. Processes in these domains are typically represented as process diagrams or flow charts. So far, various techniques have been developed for automatically generating such diagrams from event sequence data. An open challenge is the visual analysis of drift phenomena when processes change over time.
In this paper, we address this research gap. 
Our contribution is a system for fine-granular process drift detection and corresponding visualizations for event logs of executed business processes.
We evaluated our system both on synthetic and real-world data. 
On synthetic logs, we achieved an average F-score of \num{0.96} and outperformed all the state-of-the-art methods. 
On real-world logs, we identified all types of process drifts in a comprehensive manner. 
Finally, we conducted a user study highlighting that our visualizations are easy to use and useful as perceived by process mining experts.
In this way, our work contributes to research on process mining, event sequence analysis, and visualization of temporal data.

\end{abstract}

}

\maketitle

\IEEEdisplaynontitleabstractindextext

%
\IEEEpeerreviewmaketitle
%
%
\IEEEraisesectionheading{\section{Introduction}\label{sec:intro}}
\begin{figure*}
	\includegraphics[width=1.0\linewidth]{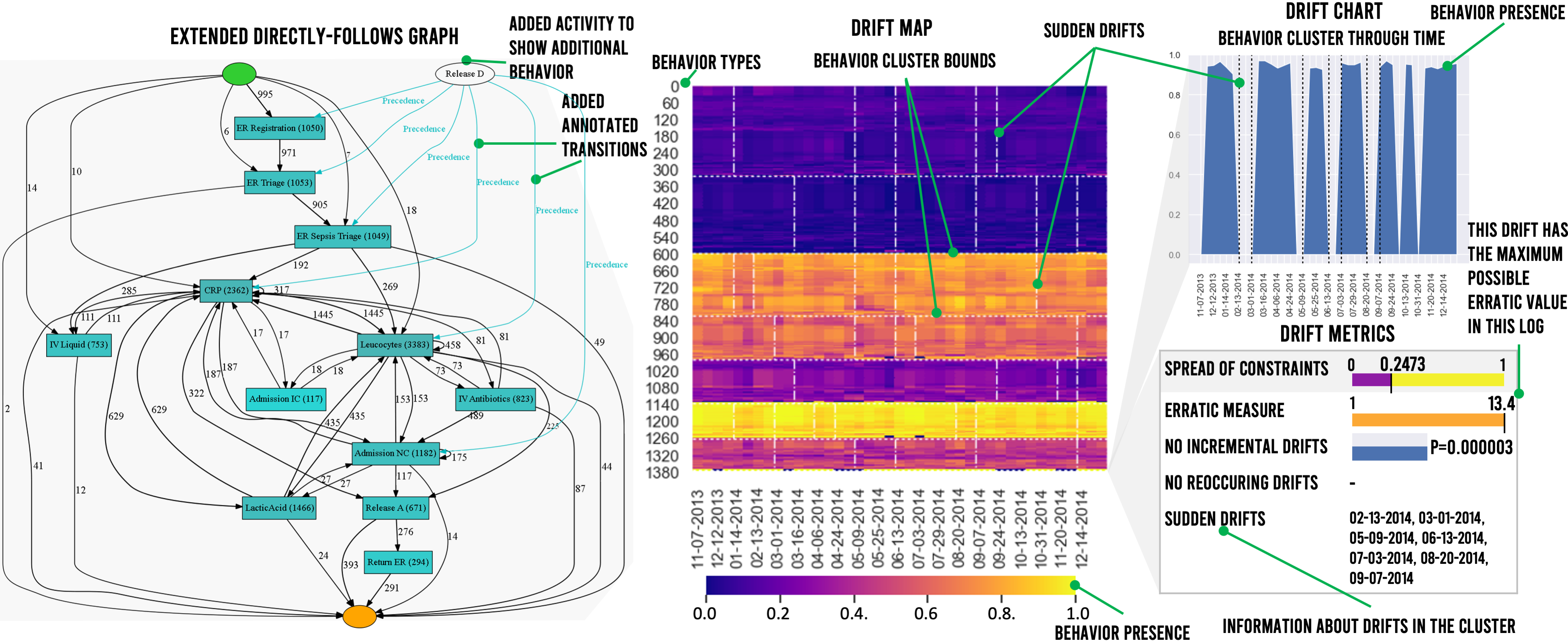}%
	\captionof{figure}{The \acrfull{dvd} approach visualization (here using as input the Sepsis event log~\cite{DBLP:conf/emisa/MannhardtB17}). In the center, a Drift Map shows the degree to which clusters of behaviour change over time (on the x axis). The intensity of the color indicates the confidence associated to the behavioral constraints (on the y axis).  Vertical dashed lines signal drift points. On the top-right corner, a Drift Chart depicts the oscillations of the confidence values that determine the drift points of a cluster. On the bottom-right corner, Drift Metrics document the detected erratic behavior. On the left-hand side, the extended Directly Follows Graph illustrates the behavior of the cluster as a workflow diagram.}%
	\label{fig:approach-example}%
\end{figure*}%
%
%
%
%
Event sequence data is increasingly available in various application domains and the design of suitable analysis techniques is an ongoing research challenge. Research by Aigner et al.~\cite{aigner2007visual,aigner2011visualization}, provides an excellent overview of time-oriented visualizations concluding that most available techniques plot temporal data in a \emph{continuous} way. Examples of this visualization type are the Time Line Browser~\cite{cousins1991visual}, History Flow~\cite{viegas2004studying}, ThemeRiver~\cite{havre2002themeriver}, and TimeNets~\cite{kim2010tracing}. Various domains such as business process management, software engineering, and medical pathways use process diagrams, flow charts, and similar models to describe temporal relations between \emph{discrete} activities and events~\cite{mendling2008metrics}. Techniques from process mining are concerned with generating such visual models from event sequence data~\cite{DBLP:books/sp/Aalst16}.

Business process management is a discipline concerned with organizing activities and events in an efficient and effective way~\cite{DBLP:books/sp/DumasRMR18}. To this end, business processes are analyzed, designed, implemented, and monitored. A business process in this context can be a travel request or an online order of a textbook. \emph{Event sequence data} plays an important role in process analysis. An individual case of a textbook order by the first author on the 4th of April is also referred to as a \emph{trace}, and a multiset of such traces is called an \emph{event log}. 
In process mining, process discovery algorithms have proven to be highly effective in generating process models from event logs of stable behavior~\cite{DBLP:books/sp/Aalst16}. However, many processes are not stable but change over time. In data mining, such change over time is called \emph{drift}. Furthermore, to the detriment of \emph{process analysts}, drift is a concept that has been addressed only to a limited extent in BPM.


Recent works have focused on integrating ideas from research on concept drift into process mining~\cite{Denisov/BPM2018:MiningConceptDriftinPerformanceSpectraofProcesses,DBLP:conf/simpda/HompesBADB15,DBLP:conf/s-bpm-one/SeeligerNM17,DBLP:conf/otm/ZhengW017,DBLP:conf/er/OstovarMRHD16}. The arguably most advanced technique is proposed in~\cite{DBLP:journals/tkde/MaaradjiDRO17}, where Maaradji et al. present a framework for detecting process drifts based on tracking behavioral relations over time using statistical tests. A strength of this approach is its statistical soundness and ability to identify a rich set of drifts, making it a suitable tool for validating if an intervention at a known point in time has resulted in an assumed change of behavior. 
However, a key challenge remains. In practice, the existence of different types of drifts in a business process is not known beforehand, and analysts are interested in distinguishing what has and what has not changed over time. This need calls for a more fine-granular analysis as compared to what recent techniques have offered.

In this paper, we present a \emph{design study}~\cite{sedlmair2012design} on how to support process analysts with visualizations to better understand drift phenomena~\cite{DBLP:journals/corr/abs-2007-15272} associated with business processes. Specifically, we develop 
a novel system for \gls{condridet}, named \gls{dvd}, which addresses the identified research gap. Our system aims to support process analysts by facilitating the \textit{visual analysis}~\cite{ware2012information} of process drifts. \Cref{fig:approach-example} schematically illustrates the main visual cues it offers to the users to this end.  We integrate various formal concepts grounded in the rigor of temporal logic, {\Declare} constraints~\cite{Aalst.etal/CSRD09:DeclarativeWFsBalancing,DBLP:journals/tmis/CiccioM15} and time series analysis~\cite{CharlesTruonga/SParxiv:SelectiveReviewOfOfflineChangePointDetectionMethods}. Key strengths of our system are clustering of declarative behavioral constraints that exhibit similar trends of changes over time and automatic detection of changes in terms of drift points. For each of these analysis steps, we provide different visualizations, including the Extended Directly-Follows Graph, the Drift Map, Drift Charts, and various measures to indicate the type of drift. These features allow us to detect and explain drifts that would otherwise remain undetected by existing techniques. The paper presents an evaluation that demonstrates these capabilities.

The remainder of the paper is structured as follows. \Cref{sec:background} illustrates the problem of process drift detection and formulates five requirements for its analysis. Then, \Cref{sec:preliminaries} states the preliminaries. \Cref{sec:approach} presents the concepts of our drift detection system, while \Cref{sec:evaluation} evaluates the system using benchmark data and a user study. Finally, \Cref{sec:conclusion} summarizes the results and concludes with an outlook on future research.

\section{Process Drift Analysis}
\label{sec:background}
%
%

In this section, we discuss the analysis of drift phenomena for business processes. First,~\cref{sec:drifts} illustrates an example of drift in a business process. \cref{sec:analyst} then characterizes the specific analysis task of the analysts and identifies requirements for supporting process analysts for visually inspecting drift.

%
\subsection{Drift in Business Processes}
\label{sec:drifts}
%
%
Business processes are collections of inter-related events and activities involving a number of actors and objects~\cite{DBLP:books/sp/DumasRMR18}. They define the steps by which products and services are provided to customers. Arguably, any work or business operation can be understood as a business process, though more specific terms are used in different industries: manufacturing processes, clinical pathways, service provisioning, or supply chains~\cite{dos2019process}. Analyzing and improving these processes is difficult due to their complexity and their division of labour with separate agents being responsible for different activities. 


As an example of a business process, consider the log of a hospital on handling sepsis patients~\cite{DBLP:conf/emisa/MannhardtB17} displayed by our system in~\cref{fig:approach-example}. The diagram on the left-hand side is a Directly-Follows Graph showing potential sequences of the process. One individual patient is a case of this process, and his or her sequences through the process is a trace.
The process typically starts with the registration and admission of the patient with \Task{ER Registration}. A first diagnosis is performed with the \Task{ER Triage} activity followed by an \Task{ER Sepsis Triage}. The patients suspected of sepsis are treated with infusions of antibiotics and intravenous liquid (\Task{IV Antibiotics} and \Task{IV liquid}). The majority of the patients are admitted to the normal care ward (\Task{Admission NC}), while some are admitted to intensive care (\Task{Admission IC)}. In some cases, the admission type changes during the treatment process. At the end of the treatment, and due to different reasons, patients are dispatched (with \Task{Release A-D} activities). 

The hospital is now interested in this question: 
\emph{Has the process of treating sepsis patients changed over time, and which parts of it now work differently than in the past?}
The described problem is typical for many process domains. 
The objective is to explain the change of the process behavior in a dynamically changing non-stationary environment based on some \textit{hidden context}~\cite{DBLP:journals/csur/GamaZBPB14}. The data mining and machine learning community use the term \textit{concept drift} to refer to any change of the conditional distribution of the output given a specific input. Corresponding techniques for \emph{concept drift detection} identify drift in data collections, either in an \textit{online} or \textit{offline} manner, with applications in prediction and fraud detection~\cite{TsymbalAlexey/:TheProblemOfConceptDrift:DefinitionsAndRelatedWork}.

Recently, the availability of event logs of business processes has inspired various process mining techniques~\cite{DBLP:books/sp/Aalst16}. Those techniques mainly support process monitoring and analysis. 
Classical process mining techniques have implicitly assumed that logs are not sensitive to time in terms of systematic change~\cite{DBLP:books/sp/Aalst16}. 
For instance, sampling-based techniques explicitly build on this assumption for generating a process model with a subset of the event log data~\cite{DBLP:conf/caise/BauerSGGW18}. 
A significant challenge for adopting concept drift for process mining is to represent behavior in a time-dependent way. 
The approach reported in~\cite{DBLP:journals/tkde/MaaradjiDRO17} uses causal dependencies and tracks them over time windows. Support for more detailed analysis is limited so far. 
Specifically relevant is the question if a process show concept drift and which of its activities relate to it. 

\begin{figure}[t]
	\includegraphics[width=\columnwidth]{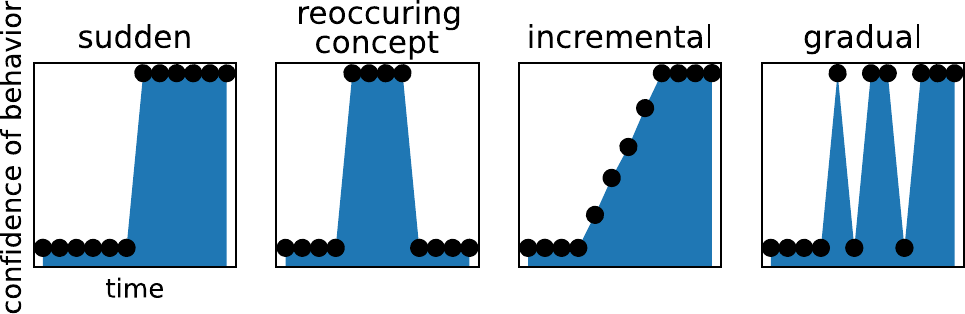}
	\vspace{-3mm}
	\caption{Different types of drifts, cf.\ Fig.\ 2 in~\cite{DBLP:journals/csur/GamaZBPB14}.
	}
	\label{fig:driftypes2}
	\vspace{-3mm}
\end{figure}

Prior research on data mining has described different archetypes of drift (see \cref{fig:driftypes2}). 
%
%
We use the example of the sepsis process to illustrate the potential causes of drifts. 
A \textit{sudden drift} is typically caused by an intervention. 
A new guideline could eliminate the need to conduct triage in two steps, as it is currently done.
As a result, we would not see second triage events in our log in the future. 
An \textit{incremental drift} might result from a stepwise introduction of a new type of infusion. 
A \textit{gradual drift} may stem from a new guideline to consider intensive care already for patients with less critical symptoms.
Finally, a \textit{reoccurring drift} might result from specific measures taken in the holiday season from July to August when inflammations are more likely due to warm weather.
Existing process mining techniques support these types of drifts partially.

\begin{sloppypar}
The following are four illustrative cases from the sepsis process: 
\begin{compactdesc}
	\item[10 Jan.\ 2014:] $\langle \Task{ER Registration},\; \Task{ER Triage},\; \Task{ER Sepsis Triage},\;
	\\ \Task{IV Antibiotics},\; \Task{Release A} \rangle$
	\item[15 Jan.\ 2014:] $\langle \Task{ER Registration},\; \Task{ER Triage},\; \Task{ER Sepsis Triage},\;
	\\ \Task{IV Antibiotics},\; \Task{Release A} \rangle$
	\item[04 Feb.\ 2014:] $\langle \Task{ER Registration},\; \Task{ER Triage},\; \Task{IV Antibiotics},\;
	\\ \Task{Release A} \rangle$
	\item[06 Feb.\ 2014:] $\langle \Task{ER Registration},\; \Task{ER Triage},\; \Task{IV Antibiotics},\;
	\\ \Task{Release A} \rangle$
\end{compactdesc}

We observe a sudden drift here due to the introduction of a new guideline. After 04 Feb. 2014, the sepsis triage is integrated with the general triage step. Therefore, in formal terms, from case 3 onwards, the behavioral rule that the sepsis triage occurs after the general triage abruptly decreases in the number and share of observations. Several rule languages with a rich spectrum of behavioral constraints have been proposed~\cite{DBLP:journals/jcss/DeutchM12,DBLP:conf/apn/PolyvyanyyWCRH14,DBLP:conf/simpda/PrescherCM14}. In rule languages based on linear temporal logic such as {\Declare}, 
we can formally state that the rule  $\AltResp{\Task{ER Triage}}{\Task{ER Sepsis Triage}}$ drops in confidence. We will make use of such rules in our technique.

\end{sloppypar}

\subsection{Analysis Tasks of Process Analysts}
\label{sec:analyst}
%
%
We frame our design study in the tasks of process analysts. \emph{Process analysts} are typically in charge of steering process improvement projects, gathering 
information about current process performance, modeling the process as-is, analyzing weaknesses and changes over time, developing redesign options, and bringing them into implementation~\cite{DBLP:books/sp/DumasRMR18}. 
%
The analysis of changes based on the evidence brought by event logs entails the challenge of detecting and understanding process drifts.
Such a complex task with data requires interactive support to explore and investigate various aspects of the information source at hand~\cite{DBLP:books/sp/Aalst16}.
Based on the experience gained in projects with industry partners, we identified five requirements for process drift analysis~\cite{DBLP:conf/er/YeshchenkoCMP19a}:



%
\begin{requidescr} 
	\item[Identify drifts:\namedlabel{req:identify}] The points at which a business process undergoes drifts should be identified based on precise criteria; 
	\item[Categorize drifts:\namedlabel{req:categorize}] Process drifts should be categorized according to their types; 
	\item[Drill down and roll up analysis:\namedlabel{req:drill}] Process drifts should be characterized at different levels of granularity, e.g., drifts that concern the entire process or only its parts; 
	\item[Quantitative analysis:\namedlabel{req:quanti}] Process drifts should be associated with a degree of change, a measure that quantifies to which extent the drift entails a change in the process;
	\item[Qualitative analysis:\namedlabel{req:quali}] Process drifts should convey changes in a business process to process analysts effectively.
\end{requidescr}
\setlength\intextsep{0pt}
\begin{table}[tb]
	\centering
	\caption{Process drift detection in process mining.}
	\label{table-approaches}
	\resizebox{0.8\columnwidth}{!}{%
\begin{tabular}{l r r r r r}
\toprule
\textbf{Approach} & \textbf{R1} & \textbf{R2}& \textbf{R3} & \textbf{R4} & \textbf{R5} \\

\midrule
ProDrift~\cite{DBLP:journals/tkde/MaaradjiDRO17,DBLP:conf/er/OstovarMRHD16} & + & +/- & - & - & - \\
TPCDD~\cite{DBLP:conf/otm/ZhengW017} & + & - & - & - & - \\
Process Trees~\cite{quteprints121158} & + & - & - & - & + \\
Performance Spectra~\cite{Denisov/BPM2018:MiningConceptDriftinPerformanceSpectraofProcesses} & - & - & +/- & - & + \\
Comparative Trc.\ Clustering~\cite{DBLP:conf/simpda/HompesBADB15} & - & - & - & + & + \\
Graph Metrics On Proc.Graphs~\cite{DBLP:conf/s-bpm-one/SeeligerNM17} & + & - & - & + & + \\ \midrule
Eventpad~\cite{DBLP:conf/vizsec/CappersMEW18} & + & - & - & - & + \\ 
ViDX~\cite{DBLP:journals/tvcg/XuMR017} & + & - & +/-  & +/-  & + \\ 
Eventthread3~\cite{DBLP:conf/bigdataconf/GuoJCGZC19} & - & - & + & + & + \\ \midrule
\textbf{\Gls{dvd} (this paper)}  & \textbf{+} & \textbf{+} & \textbf{+} & \textbf{+} & \textbf{+} \\
\bottomrule
\end{tabular}
%
	}
\end{table}

\noindent

\Cref{table-approaches} provides an overview of the state-of-the-art methods for process drift analysis with reference to the five listed requirements. Notice that collectively these methods address (at least partially) all the requirements, whereas
each method addresses only a subset. In particular, \ref{req:categorize} and \ref{req:drill} remain mostly open challenges.

Approaches like ProDrift~\cite{DBLP:journals/tkde/MaaradjiDRO17} and Graph Metrics on Process Graphs~\cite{DBLP:conf/s-bpm-one/SeeligerNM17} put an emphasis on requirement \ref{req:identify}. 
The evaluation of ProDrift in~\cite{DBLP:journals/tkde/MaaradjiDRO17} shows that sudden and gradual drifts are found accurately, thus partly addressing requirement~\ref{req:categorize}, although with a reported high sensitivity to the choice of method parameters. ProDrift relies on the automatic detection of changes in business process executions, which are analyzed based on causal dependency relations studied in process mining~\cite{1316839}. The Tsinghua Process Concept Drift Detection approach (TPCDD)~\cite{DBLP:conf/otm/ZhengW017} uses two kinds of behavioral relationships: direct succession and weak order. The approach computes those relations on every trace, so as to later identify the change points with the help of clustering. The only type of drift that TPCDD detects is sudden drift.

The other approaches emphasize requirement~\ref{req:quali}.
The approach based on Process Trees~\cite{quteprints121158} uses ProDrift for drift detection, and aims at explaining how sudden drifts influence process behavior. To this end, process trees are built for pre-drift and post-drift sections of the log and used to explain the change. The Performance Spectra approach~\cite{Denisov/BPM2018:MiningConceptDriftinPerformanceSpectraofProcesses}
focuses on drifts that show seasonality. The technique filters the control-flow and visualizes identified flow patterns. It is evaluated against a real-world log, in which recorded business processes show year-to-year seasonality. 
A strength of the Comparative Trace Clustering approach~\cite{DBLP:conf/simpda/HompesBADB15} is its ability to include non-control-flow characteristics in the analysis. 
Based on these characteristics, it partitions and clusters the log. Then, the differences between the clusters indicate the quantitative change in the business processes, which addresses requirement~\ref{req:quanti}.
The Graph Metrics on Process Graphs approach~\cite{DBLP:conf/s-bpm-one/SeeligerNM17} discovers a first model, called a reference, using the Heuristic Miner on a section of the log~\cite{DBLP:books/sp/Aalst16}. Then, it discovers models for other sections of the log and uses graph metrics to compare them with the reference model. 
The technique interprets significant differences in the metrics as drifts.
The reference model and detection windows get updated once a drift is detected.

Works that emphasize the visualization analysis of drifts for event sequence data mainly approach change as a type of anomaly. Eventpad~\cite{DBLP:conf/vizsec/CappersMEW18} allows the users to import event sequences for interactive exploration by filtering the visual representation using constraints and regular expressions. The overview provided by the system helps to uncover change patterns. Eventpad supports the requirements \ref{req:identify} and \ref{req:quali}. The ViDX system~\cite{DBLP:journals/tvcg/XuMR017} offers an interactive visualization system to discover seasonal changes. Note that Performance Spectra~\cite{Denisov/BPM2018:MiningConceptDriftinPerformanceSpectraofProcesses} build on similar design ideas. The user of the ViDX system can select the sequences that are considered normal and the system highlights the sequences that deviate from this norm. The system also supports a calendar view, which helps to identify where drifts happen in the timeline. The system supports requirements \ref{req:identify},  \ref{req:quali}, and partially \ref{req:drill} and \ref{req:quanti}. The EventThread3 system~\cite{DBLP:conf/bigdataconf/GuoJCGZC19} relies on an unsupervised anomaly detection algorithm and the interactive visualization system to uncover changes in event sequence data. Seven connected views allow the analyst to inspect the flow-based overview of the event sequence data with additional information on anomalous sequences. The system supports the thorough analysis of anomalous behavior (requirements \ref{req:drill}, \ref{req:quanti} , and \ref{req:quali}) but neither identifies the exact point in time in which the change of behavior happened, nor classifies the changes.

Beyond these specific works on process drift, there are numerous related works on the visualization of event sequence data~\cite{DBLP:series/hci/AignerMST11,DBLP:journals/corr/abs-2006-14291}. 
The \emph{summarization of event sequence data} can be supported by visual representations of different types. Chen et al.~\cite{DBLP:journals/tvcg/ChenXR18} use several connected views including a raw sequence representation and an abstraction based on the minimum description length principle. The work by Gou et al.~\cite{DBLP:journals/tvcg/GuoXZGZC18} splits the event data into threads and stages. In this way, they summarize complex and long event sequences. Zhang et al.~\cite{DBLP:journals/tvcg/ZhangCD19} combine the raw event sequence visual representation with annotated line plots together with custom infographics emphasizing use-case related characteristics. Wongsuphasawat et al.~\cite{DBLP:conf/chi/WongsuphasawatGPWTS11} introduce an interactive event sequence overview system called LifeFlow, which builds upon the Icicle plot and represents temporal spacing within event sequences. Monroe et al.~\cite{DBLP:journals/tvcg/MonroeLLPS13} present the event sequence analysis system EventFlow, which offers different types of aggregation and simplification. Law et al.~\cite{DBLP:journals/tvcg/LawLMB19} introduce an interactive system that supports flexible analysis of event sequences by combining querying and mining.  Wongsuphasawat and Gotz~\cite{DBLP:journals/tvcg/WongsuphasawatG12} extend the directed graph event sequence representation with colored vertical rectangles used as the transitions between events. Tanahashi and Ma~\cite{DBLP:journals/tvcg/TanahashiM12} describe design considerations for visualizing event sequence data. This work concerns the usage of color and layout when designing visualizations. Other papers explain how to effectively visualize the \emph{alignment of sequences}. Albers et al.~\cite{DBLP:journals/tvcg/AlbersDG11} present a hierarchically structured visual representation for genome alignments. Cappers et al.~\cite{DBLP:journals/tvcg/CappersW18} visualize event sequences aligned by user-defined temporal rules. Malik et al.~\cite{DBLP:conf/iui/MalikDMOPS15} introduce the cohort comparison system CoCo, which uses automated statistics together with a user interface for exploring differences between datasets. Zhao et al.~\cite{DBLP:conf/chi/ZhaoLDHW15} introduce a novel visualization system based on the matrices arranged in a zig-zagging pattern that allows for less overlapping edges than common Sankey based visualizations. Xu et al.~\cite{DBLP:journals/tvcg/XuMR017} achieve visualization of changes and \emph{drifts} in event sequence data through compound views consisting of Marey's graph, line plots, bar charts, calendar views, and custom infographics.

This discussion, summarized in \cref{table-approaches}, witnesses that none of the state-of-the-art methods covers the full set of the five requirements of visualizing process drifts. The approach described in the following addresses this research gap.

%
%
\section{Preliminaries}
\label{sec:preliminaries}
%
%
This section defines the formal preliminaries of our approach. \Cref{sec:notioneventlog}  gives an overview of the event log, the main input data type used in process mining. \Cref{sec:notiondfg,sec:declare} describe process representation languages: the former introduces the directly-follows graphs for procedural models, and the latter illustrates the representation of the process. \Cref{sec:declare,sec:declaresubsumption} discuss the {\Declare} specification and the techniques to discover and simplify those models from event logs, respectively. 
\Cref{sec:clustering} describes time series clustering, and \cref{sec:changepoint} illustrates change point detection methods, which are the main instruments of our approach.

\subsection{Event log}
\label{sec:notioneventlog}
Event logs capture actual execution sequences of business processes. They represent the input for process mining techniques.
An event log $L$ (\emph{log} for short) is a collection of recorded traces that correspond to process enactments.
In this paper, we abstract the set of activities of a process as a finite non-empty alphabet $\Sigma = \{ \letterx, \lettery, \letterz, \ldots \}$. Events record the execution of activities, typically indicating the activity and the completion timestamp. A trace $\sigma$ is a finite sequence of events. For the sake of simplicity, we shall denote traces by the sequence of activities they relate to, $a_i \in \sigma, 1 \leq i \leq n$, sorted by their timestamp.  
In the following examples, we also resort to the string-representation of traces (i.e., $\sigma=a_1 a_2 \cdots a_n$) defined over $\Sigma$.
Case 1 of the sepsis process from~\cref{sec:drifts} is an example of a trace. 
An event log $L$ is a multiset of traces, as the same trace can be repeated multiple times in the same log: denoting the multiplicity $m \geqslant 0$ as the
exponent of the trace, we have that 
%
$L=\{ \sigma_1^{m_1}, \sigma_2^{m_2}, \ldots, \sigma_N^{m_N} \}$ (if $m_i=0$ for some $1 \leqslant i \leqslant N$ we shall simply omit $\sigma_i$).
The size of the log is defined as $|L| = \sum_{i=1}^{N}{m_i}$ (i.e., the multiplicity of the multiset).
Cases 1-4 of the sepsis process in~\cref{sec:drifts} constitute an example of event log of size \num{4}.
The size of the sepsis log is \num{1050}~\cite{DBLP:conf/emisa/MannhardtB17}.
A sub-log $L' \subseteq L$ is a log $L'= \lbrace \sigma_1^{m'_1}, \sigma_2^{m'_2}, \ldots, \sigma_N^{m'_N} \rbrace$ such that $m'_i \leqslant m_i$ for all $1 \leqslant i \leqslant N$.
A log consisting of cases 1-3 from the example log $L$ in \cref{sec:drifts} is a sub-log of $L$.
\begin{table*}[tbp]
	\caption{Example {\Declare} constraints.} 
	\label{tab:declare:verbose:gfx}
	\centering
	\resizebox{0.8\textwidth}{!}{%
		%
%
\renewcommand{\arraystretch}{1.6}
\rowcolors{2}{white}{gray!12.5}
\begin{tabular}{ l p{8cm} l l l l }
	\toprule
	\textbf{Constraint} & 
	\textbf{Explanation} & 
	\multicolumn{4}{c}{\textbf{Examples}}
	\\
	\midrule
%
%
%
	
	$\Uniq{\taska}$ & 
	If $\taska$ occurs, then it occurs at most \emph{once} &
	\GoodExampleMark \Task{bcc} & \GoodExampleMark \Task{bcac} &
	\BadExampleMark \Task{bcaac} & \BadExampleMark \Task{bcacaa}
	\\
	
%
%
	$\Resp{\taska}{\taskb}$ &
	If {\taska} occurs, then {\taskb} occurs eventually after {\taska} &
	\GoodExampleMark \Task{baabc} & \GoodExampleMark \Task{bcc} &
	\BadExampleMark \Task{bcba} & \BadExampleMark \Task{caac}
	\\
	$\AltResp{\taska}{\taskb}$ &
	If {\taska} occurs, then {\taskb} occurs eventually afterwards, and no other {\taska} recurs in between &
	\GoodExampleMark \Task{cacb} & \GoodExampleMark \Task{abcacb} &
	\BadExampleMark \Task{caacb} & \BadExampleMark \Task{bacacb}
	\\
	$\ChaResp{\taska}{\taskb}$ &
	If {\taska} occurs, then {\taskb} occurs immediately afterwards &
	\GoodExampleMark \Task{cabb} & \GoodExampleMark \Task{abcab} &
	\BadExampleMark \Task{cacb} & \BadExampleMark \Task{bca}
	\\
	$\Prec{\taska}{\taskb}$ &
	If {\taskb} occurs, then {\taska} must have occurred before &
	\GoodExampleMark \Task{cacbb} & \GoodExampleMark \Task{acc} &
	\BadExampleMark \Task{ccbb} & \BadExampleMark \Task{bacc}
	\\
	$\AltPrec{\taska}{\taskb}$ &
	If {\taskb} occurs, then {\taska} must have occurred before and no other {\taskb} recurs in between &
	\GoodExampleMark \Task{cacba} & \GoodExampleMark \Task{abcaacb} &
	\BadExampleMark \Task{cacbba} & \BadExampleMark \Task{abbabcb}
	\\
	$\ChaPrec{\taska}{\taskb}$ &
	If {\taskb} occurs, then {\taska} occurs immediately beforehand &
	\GoodExampleMark \Task{abca} & \GoodExampleMark \Task{abaabc} &
	\BadExampleMark \Task{bca} & \BadExampleMark \Task{baacb}
	\\
	$\NotSucc{\taska}{\taskb}$ &
	{\taska} occurs if and only if {\taskb} does not occur afterwards &
	\GoodExampleMark \Task{bbcaa} & \GoodExampleMark \Task{cbbca} &
	\BadExampleMark \Task{aacbb} & \BadExampleMark \Task{abb}
	\\
	\bottomrule
\end{tabular}%
	}
\end{table*}
\subsection{Directly-Follows Graph}
\label{sec:notiondfg}
%
%
The first output that process mining tools generate for providing an overview of a business process is the Directly-Follows Graph (DFG, also referred to as \textit{process map}).
Given an event log $L$, a DFG is a tuple $G(L) = ({A}_{L}, {\mapsto}_{L}, {A}_{L}^{start}, {A}_{L}^{end})$~\cite{DBLP:books/sp/Aalst16, DBLP:books/sp/DumasRMR18}. In a DFG, each node in set ${A}_{L}$ represents an activity class, and each arc denotes a tuple in the directly-follows relation ${\mapsto}_{L}$ \emph{discovered} from the event log.
\Cref{fig:approach-example} shows a DFG of the sepsis log on the left-hand side.
For instance, for a specific patient we observe that the \Task{ER Triage} activity is followed by \Task{ER Sepsis Triage}, resulting into a corresponding tuple in the directly-follows relation.
Each arc is annotated with a number representing frequency of occurrence in the event log to indicate the importance of that transition between tasks in the process. $G(L)$ explicitly encodes start and end of the discovered process with sets of activities ${A}_{L}^{start}, {A}_{L}^{end}$, respectively.
DFGs are known to be simple and comprehensive~\cite{DBLP:conf/icpm/LeemansPW19, DBLP:conf/caise/LeemansFA15}. Indeed, they are used as a visual overview for processes both in open-source and commercial process mining tools like Fluxicon Disco\footnote{\url{https://fluxicon.com/disco/}} and Celonis\footnote{\url{https://www.celonis.com/}}, and pm4py~\cite{DBLP:journals/corr/abs-1905-06169}. They are also used as an intermediate data structure by several process discovery algorithms~\cite{DBLP:conf/icpm/LeemansPW19, DBLP:journals/tkde/AalstWM04}. 

As shown in~\cite{weijters2006process}, the complexity of DFG mining is linear in the number of traces ($O(|L|)$) and quadratic in the number of activities ($O(|\Sigma|^2)$). 

%
\subsection{{\Declare} modeling and mining}
\label{sec:declare}
%
%
Fine-granular behavior of a process can be represented in a declarative way.
A declarative process specification represents this behavior by means of \emph{constraints}, i.e., temporal rules that specify the conditions under which activities may, must, or cannot be executed.
In this paper, we focus on {\Declare}, a well-known standard for declarative process modeling \cite{Aalst.etal/CSRD09:DeclarativeWFsBalancing} based on linear temporal logic.
%
{\Declare} provides a \emph{repertoire} of template constraints \cite{Polyvyanyy.etal/FAOC2016:ExpressivePowerBehavioralProfiles,DiCiccio.etal/IS2017:ResolvingInconsistenciesRedundanciesDeclare}). 
Examples of {\Declare} constraints are
\Resp{\taska}{\taskb}
and
\ChaPrec{\taskb}{\taskc}.
The former constraint applies the {\RespTmp} template on tasks {\taska} and {\taskb}, 
and states that if {\taska} occurs then {\taskb} must occur later on within the same trace.
In this case, {\taska} is named \emph{activation}, because it is mentioned in the if-clause, thus triggering the constraint, whereas {\taskb} is named \emph{target}, as it is in the consequence-clause~\cite{DiCiccio.etal/IS2017:ResolvingInconsistenciesRedundanciesDeclare}.
\ChaPrec{\taskb}{\taskc} asserts that if {\taskc} (the activation) occurs, then {\taskb} (the target) must have occurred immediately before.
Given an alphabet of activities $\Sigma$, we denote the number of all possible constraints that derive from the application of {\Declare} templates to all activities in $\Sigma$ as $\CnsNum \subseteq O(\Sigma^2)$~\cite{DiCiccio.etal/IS2017:ResolvingInconsistenciesRedundanciesDeclare}. For the sepsis log, $\CnsNum = \num{3424}$.
\Cref{tab:declare:verbose:gfx} shows some of the templates of the {\Declare} repertoire, together with the examples of traces that satisfy (\GoodExampleMark) or violate (\BadExampleMark) them.

Declarative process mining tools can measure to what degree constraints hold true in a given event log~\cite{DBLP:conf/cidm/MaggiMA11}.
To that end, diverse measures have been introduced~\cite{DBLP:conf/icpm/CecconiGCMM20}.
Among them, we consider here \emph{support} and \emph{confidence}~\cite{DBLP:journals/tmis/CiccioM15}.
Their values range from \num{0} to \num{1}.
In \cite{DBLP:journals/tmis/CiccioM15}, the support of a constraint is measured as the ratio of times that the event is triggered and satisfied over the number of activations.
Let us consider the following example event log:
$\EvtLog = \lbrace {\StrTrc_{1}^{4}}, {\StrTrc_{2}^{1}}, {\StrTrc_{3}^{2}} \rbrace$,
having
${\StrTrc_{1}} =
\Task{baabc}  $,
${\StrTrc_{2}} =
\Task{bcc} $, and
${\StrTrc_{3}} =
\Task{bcba} $.
The size of the log is $4+1+2=7$.
The activations of \Resp{\taska}{\taskb} that satisfy the constraint amount to \num{8} because two {\taska}'s occur in ${\StrTrc_{1}}$ that are eventually followed by an occurrence of {\taskb}, and ${\StrTrc_{1}}$ has multiplicity \num{4} in the event log. The total amount of the constraint activations in $L$ is \num{10} (see the violating occurrence of {\taska} in ${\StrTrc_{3}}$). The support thus is \num{0.8}.
By the same line of reasoning, the support of \ChaPrec{\taskb}{\taskc} is $\frac{7}{8}=\num{0.875}$ (notice that in ${\StrTrc_{2}}$ only one of the two occurrences of {\taskc} satisfies the constraint).
To take into account the frequency with which constraints are triggered, confidence scales support by the ratio of traces in which the activation occurs at least once.
Therefore, the confidence of \Resp{\taska}{\taskb} is $\num{0.8} \times \frac{6}{7} \approx \num{0.69}$ because {\taska} does not occur in ${\StrTrc_{2}}$. As $\taskb$ occurs in all traces, the confidence of \ChaPrec{\taskb}{\taskc} is \num{0.875}.

As shown in \cite{DBLP:journals/tmis/CiccioM15,DBLP:journals/is/MaggiCFK18}, the computation of constraint measures on an event log $\EvtLog$ is performed efficiently as the mining algorithms have a complexity that is
\begin{iiilist}
\item linear with respect to the number of traces, $O(|L|)$, 
\item quadratic to the total number of events, $O(\sum_{\StrTrc \in \EvtLog}|\StrTrc|^2)$, and 
\item linear to the number of constraints, $O(\CnsNum)$, hence quadratic with respect to the number of activities in the event log as $\CnsNum \subseteq O(\Sigma^2)$.
\end{iiilist} 
This complexity corresponds to that of mining \gls{dfg}, as previously discussed in \cref{sec:notiondfg}.

\subsection{Subsumption of {\Declare} rules}
\label{sec:declaresubsumption}
\begin{figure}[btp]
	\centering%
	\begin{footnotesize}
		\begin{tikzpicture}[scale = 0.5,x = 6em]

	\tikzstyle{block} = [draw,text width=14em,align=center]
	\tikzstyle{conlabel} = [text width=13em,align=center]
	
	\node[block] (precab)  at ( 0, -2) {$\Prec{\paramx}{\paramy}$};
	\node[block] (aprecab) at ( 0, -4) {$\AltPrec{\paramx}{\paramy}$};
	\node[block] (cprecab) at ( 0, -6) {$\ChaPrec{\paramx}{\paramy}$};
	
	\draw[-open triangle 45] (aprecab) -- (precab);
	\draw[-open triangle 45] (cprecab) -- (aprecab);

\end{tikzpicture}
	\end{footnotesize}
	\caption{The subsumption relation restricted to \ChaPrec{\paramx}{\paramy}, \AltPrec{\paramx}{\paramy}, and \Prec{\paramx}{\paramy}}
	\label{fig:subsum:hierarchy:example}
\end{figure}
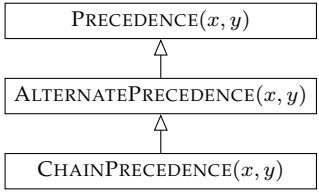

%
%
For one event log, there are typically a large amount of {\Declare} constraints. Efficient abstraction can be achieved by pruning out constraints that are subsumed by others. To this end, 
we outline here the concept of \emph{subsumption} for declarative constraints, and its impact on the support and confidence measures. 
For technical details, see~\cite{DiCiccio.etal/IS2017:ResolvingInconsistenciesRedundanciesDeclare}.
As it can be noticed in \cref{tab:declare:verbose:gfx}, {\ChaPrecTmp} imposes a stricter rule on the process execution than {\AltPrecTmp}, which in turn exerts a stricter rule than {\PrecTmp}: for example,
$\Cns = \Prec{\taskb}{\taskc}$ requires that every occurrence of {\taskc} is preceded at some point before by {\taskb};
$\Cns' = \AltPrec{\taskb}{\taskc}$ adds to the statement of {\Cns} that no other {\taskc} can recur between {\taskc} and the preceding occurrence of {\taskb}; on top of that,
$\Cns'' = \ChaPrec{\taskb}{\taskc}$ excludes that \emph{any} other task between {\taskc} and the preceding {\taskb} occurs (not just \taskc). As a consequence, every trace that satisfies $\Cns''$ is also compliant with $\Cns'$, and every trace that satisfies the latter, in turn, complies with $\Cns$.
For example, let
$\EvtLog' = \lbrace {\StrTrc_{4}^{2}}, {\StrTrc_{5}^{1}}, {\StrTrc_{6}^{3}} \rbrace$ be an event log in which
${\StrTrc_{4}} =
\Task{bccabc}  $,
${\StrTrc_{5}} =
\Task{bacabc} $, and
${\StrTrc_{6}} =
\Task{bcaabc} $.
$\StrTrc_{4}$ satisfies $\Cns$ but does not comply with either of $\Cns'$ and $\Cns''$.
$\StrTrc_{5}$ satisfies $\Cns$ and $\Cns'$ but not $\Cns''$.
Finally,
$\StrTrc_{6}$ satisfies $\Cns$, $\Cns'$ and $\Cns''$.
Notice that it is not possible to find an example of trace satisfying, e.g., $\Cns$ and $\Cns''$ but not $\Cns'$.
We say that $\Cns''$ is subsumed by $\Cns'$ and $\Cns'$ is subsumed by $\Cns$. Subsumption enjoys the properties of transitivity, reflexivity, and anti-symmetry, thus being a partial order.
The repertoire of {\Declare} constitutes a \emph{subsumption hierarchy}.
\Cref{fig:subsum:hierarchy:example} depicts the fragment of subsumption hierarchy related to the aforementioned constraints as an is-a relation.
Interestingly, the subsumption hierarchy among constraints induces a partial order also on the sub-multisets of traces in an event log, the homomorphism being the relation with respect to constraints: considering the example above,
$\lbrace {\StrTrc_{4}^{2}}, {\StrTrc_{5}^{1}}, {\StrTrc_{6}^{3}} \rbrace$ satisfies $\Cns$,
$\lbrace {\StrTrc_{5}^{1}}, {\StrTrc_{6}^{3}} \rbrace$ 
satisfies $\Cns'$,
and
$\lbrace {\StrTrc_{6}^{3}} \rbrace$ 
satisfies $\Cns''$.
Therefore, by definition, support and confidence are monotonically non-decreasing along the subsumption hierarchy~\cite{DiCiccio.etal/IS2017:ResolvingInconsistenciesRedundanciesDeclare}.
On $\EvtLog'$, e.g., we have that the support of $\Cns''$, $\Cns'$, and $\Cns$ is \num{0.71}, \num{0.85}, and \num{1.0}, respectively.
Their confidence coincides with support as $\taskc$ (the activation) occurs in all traces, for simplicity.
We shall take advantage of this property to reduce the number of constraints to represent the behavior of identified clusters.

An array of algorithms have been introduced to automatically detect and remove redundant constraints. The techniques described in \cite{DBLP:journals/tmis/CiccioM15,DBLP:journals/is/SmedtWSV18} resort to auxiliary data structures that are heuristically optimized for the repertoire of {\Declare}, and require linear time with respect to the number of constraints, $O(\CnsNum)$. In \cite{DiCiccio.etal/IS2017:ResolvingInconsistenciesRedundanciesDeclare} a general and more effective approach for declarative languages has been proposed. It first creates a priority list for the elimination of possibly redundant constraints ($O(\CnsNum \cdot \mathrm{log}_2(\CnsNum))$) and then linearly scans that list for redundancy checking. The check is based on the incremental comparison of the finite-state automata underlying the process model and the constraints. We resort to techniques optimized for {\Declare} as a pre-processing phase pruning the vast majority of redundancies and operate with the small-sized automata of {\Declare} constraints for the final removal of redundancies.
\subsection{Time series clustering}
\label{sec:clustering}
%
%

Plotting the confidence and support of different {\Declare} constraints over time produces a time series. 
A \emph{time series} is a sequence of ordered data points $\left\langle t_1, t_2, \ldots, t_d\right\rangle = T \in \mathbb{R}^d$ consisting of
$d \in \mathbb{N^{+}}$ real values.
The illustrations of drift types in \Cref{fig:driftypes2} are in essence time series.
A \emph{multivariate time series} is a set of $n \in \mathbb{N^{+}}$ time series $D=\{T_{1}, T_{2}, \ldots, T_{n}\}$.
We assume a multivariate time series to be piece-wise stationary except for its \emph{change points}.

In our approach, we take advantage of the time series clustering algorithms.
Time series clustering is an unsupervised data mining technique for organizing data points into groups based on their similarity~\cite{Aghabozorgi:2015:TCD:2799194.2799230}.
The objective is to maximize data similarity within clusters and minimize it across clusters.
More specifically, the \emph{time-series clustering} is the process of partitioning $D$ into non-overlapping clusters of multivariate time series, $C=\{C_{1}, C_{2}, \ldots,C_{m}\} \subseteq 2^{D}$, with $C_{i} \subseteq D$ and $1\leq m \leq n$, for each $i$ such that $1 \leq i \leq m$, such that homogeneous time series are grouped together based on a \emph{similarity measure}.
A \emph{similarity measure} $\mathrm{sim}(T,T')$ represents the distance between two time series $T$ and $T'$ as a non-negative number. Time-series clustering is often used as a subroutine of other more complex algorithms and is employed as a standard tool in data science for anomaly detection, character recognition, pattern discovery, visualization of time series~\cite{Aghabozorgi:2015:TCD:2799194.2799230}. As discussed in~\cite{Aghabozorgi:2015:TCD:2799194.2799230} the hierarchical clustering computation is polynomial in the number of time series (which, in turn, is proportional to the number of constraints), hence $O(|D|^3) = O(\CnsNum^3)$.
%
\subsection{Change point detection}
\label{sec:changepoint}
%
%

Change point detection is a technique for identifying the points in which multivariate time series exhibit changes in their values~\cite{CharlesTruonga/SParxiv:SelectiveReviewOfOfflineChangePointDetectionMethods}. 
Let $D^{j}$ denote all elements of $D$ at position $j$, i.e., $D^{j}=\{T_{1}^{j}, T_{2}^{j}, ..., T_{n}^{j}\}$, where $T^{j}$ is a $j$-th element of time series $T$.
The objective of change point detection algorithms is to find $k \in \mathbb{N^{+}}$ changes in $D$, where $k$ is previously unknown. Every element $D^{j}$ for $0 < j \leqslant k$ is a point at which the values of the time series undergo significant changes.
Change points are often represented as vertical lines in time series charts.

To detect change points, the search algorithms require a \emph{cost function} and a \emph{penalty} parameter as inputs.
The former describes how homogeneous the time series is. It is chosen in a way that its value is high if the time series contains many change points and low otherwise. The latter is needed to constrain the search depth. 
The supplied penalty should strike a good balance between finding too many change points and not finding any significant ones.
Change point detection is a technique commonly used in signal processing and, more in general, for the analysis of dynamic systems that are subject to changes~\cite{CharlesTruonga/SParxiv:SelectiveReviewOfOfflineChangePointDetectionMethods}. In the worst case, the change point detection algorithm has a quadratic performance~\cite{killick2012optimal} in the number of time series in the cluster $O(|D|^2) = (\CnsNum^2)$.
%
%

%
%
\section{A System for Visual Drift Detection}
\label{sec:approach}
%
%
\begin{figure*}[bt]%
	\includegraphics[width=1.0\textwidth]{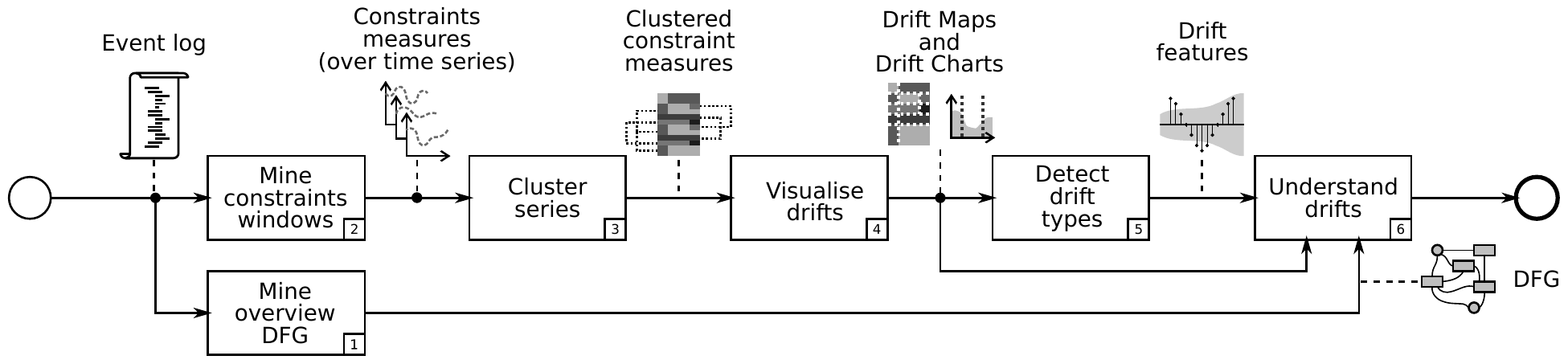}%
	\caption{The Visual Drift Detection approach.}
	\label{fig:approach}
\end{figure*}

%
In this section, we introduce the \gls{dvd} system. Its overall design idea is to cut the log into time windows and compute the confidence of behavioral constraints on the traces within those windows, so that the results can be visualized over time.
%
\Cref{fig:approach} illustrates the steps of \gls{dvd} for generating the visualizations.


\noindent
\textbf{Step 1: Mining Directly-Follows Graph as an overview.} 
In the first step, we mine a DFG from an input event log to get an overview of the behavior captured in the log.
\\
\textbf{Step 2: Mining constraints windows.} Concurrently with the first step, we split the log into sub-logs. From each sub-log, we mine the set of {\Declare} constraints and compute their confidence. As a result, we obtain several time series.
\\
\textbf{Step 3: Clustering Time Series.} In this step, we 
cluster those time series into groups of constraints that exhibit similar confidence trends (henceforth, \emph{behavior clusters}). 
\\
\textbf{Step 4: Visualizing Drifts.} In this step, we detect drift points for the whole log and each cluster separately. We plot drift points in {\DriftMap}s and {\DriftChart}s to effectively communicate the drifts to the user.
\\
\textbf{Step 5: Detecting Drift Types.} In this step, 
we use an array of methods to further analyze drift types. 
We employ multi-variate time series change point detection algorithms to spot \emph{sudden drifts} in both the entire set of constraints and in each cluster. 
We use stationarity analysis to determine if clusters exhibit \emph{gradual drifts} and autocorrelation plots to check if \emph{reoccurring drifts} are present. While Step~4 is concerned with estimating \emph{the extent} of drift presence, Step~5 is intended to show and \emph{explain} those drifts.
\\
\textbf{Step 6: Understanding drift behavior.} In the final step, we present semantic information on the identified drifts. Step~6 produces a
minimized list of constraints and a projection of these constraints onto the Directly-Follows Graph to explain the behavior in the drift cluster.

\noindent
In the following, we detail these steps.


\subsection{Mining Directly-Follows Graph as an Overview} 
\label{sec:mining:dfg}
The first step takes as input a log $L$ and produces the \acrfull{dfg}. 
The DFG includes an arc
$ a \stackrel{n}{\longrightarrow} a' $
if a sub-sequence $\langle a, a' \rangle$ is observed in any traces of the log ($n$ indicates the total number of such observations).
The process analyst typically starts the analysis by exploring the paths of the DFG.

\subsection{Mining constraints windows}
\label{sec:declare:shift}

Performed in parallel with the mining of the DFG, this step takes as input a log $L$ and two parameters ({\WinSize} and {\WinStep}). It returns a multivariate time series $D$ based on the confidence of mined {\Declare} constraints. 

In this step, we first sort the traces in the event log $L$ by the timestamp
of their start events. Then, we extract a sub-log
from $L$ as a window of size $\WinSize \in \mathbb{N^{+}}$, with $1 \leqslant \WinSize \leqslant |L|$. Next, we shift the sub-log window by a given step ($\WinStep \in \mathbb{N^{+}}$, with $1 \leqslant \WinStep \leqslant \WinSize$). Notice that we have sliding windows if $\WinStep < \WinSize$ and tumbling windows if $\WinStep = \WinSize$. Thus, the number of produced sub-logs is equal to:
$\WinNum = \left\lfloor \frac{|L| - \WinSize - \WinStep}{\WinStep} \right\rfloor$. Having {\WinSize} set to \num{50} and {\WinStep} set to \num{25}, {\WinNum} is \num{39} for the sepsis log. 

For every sub-log $L_j \subseteq L$ thus formed ($1 \leqslant j \leqslant \WinNum$), we check all possible {\Declare} constraints that stem from the activities alphabet of the log, amounting to {\CnsNum} (see \cref{sec:declare}). 
For each constraint $i \in 1..\CnsNum$, we compute its confidence over the sub-log $L_j$, namely $\mathrm{Conf}_{i,j} \in [0,1]$.
This generates a time series $T_i =  ( \mathrm{Conf}_{i,1}, \ldots, \mathrm{Conf}_{i,\WinNum} ) \in [0,1]^{\WinNum}$ for every constraint $i$. In other words, every time series $T_i$ describes the confidence of all the {\Declare} constraints discovered in the $i$-th window of the event log. The multivariate time series $D=\{T_{1}, T_{2}, \ldots, T_{\CnsNum}\}$ encompasses the full spectrum of all constraints.
Next, we detail the steps of slicing the {\Declare} constraints and explaining the drifts.
%

\subsection{Clustering Time Series}
\label{sec:cluster-changepoint}
The third step processes the previously generated multivariate time series of {\Declare} constraints $D$ to derive  a set $C$ of clusters exhibiting similar confidence trends.
For instance, if we observe confidence values over five time windows for \Resp{\taska}{\taskb} as $(0.2, 0.8, 0.9, 0.8, 0.9)$ and for \ChaPrec{\taskb}{\taskc} we have $(0.23, 0.8, 0.9, 0.9, 0.9)$, it is likely that the two time series for these constraints might end up in the same cluster due to their small difference.
The aim of this step is to identify drift points at a fine-granular level. 
%
To this end, we use time-series clustering techniques~\cite{Aghabozorgi:2015:TCD:2799194.2799230} for grouping together similarly changing pockets of behavior of the process. Each time series describes how one constraint changes its confidence over time. By clustering, we find all the time series that share similar trends of values, hence, we find all similarly changing constraints. 
We use \emph{hierarchical clustering}, as it is reportedly one of the most suitable algorithms when the number of clusters is unknown~\cite{Aghabozorgi:2015:TCD:2799194.2799230}.
As a result, we obtain a partition of the multivariate time series of {\Declare} constraint confidence values into behavior clusters.

\subsection{Visualizing Drifts}
\label{sec:visualdrift}
The fourth step generates visual representations of drifts. To this end, we construct a graphical representation called {\DriftMap}. {\DriftMap}s depict clusters and their constraints' confidence measure evolution along with the time series and their drift points.
We allow the user to drill down into every single cluster and its drifts using dedicated diagrams that we call {\DriftChart}s.

{\DriftMap}s (see \cref{fig:approach-example}, in the center) 
plot all drifts on a two-dimensional canvas. The visual representation we adopt is inspired by~\cite{ware2012information}. The x-axis is the time axis, while every constraint corresponds to a point on the y-axis. We add vertical lines to mark the identified change points, i.e., drift points, and horizontal lines to separate clusters. Constraints are sorted by the similarity of the confidence trends. The values of the time series are represented through the plasma color-blind friendly color map~\cite{ware2012information} from blue (low peak) to yellow (high peak).
To analyze the time-dependent trend of specific clusters, we build {\DriftChart}s (see \cref{fig:approach-example}, on the right).
They have time on the x-axis and average confidence of the constraints in the cluster on the y-axis.
We add vertical lines as in {\DriftMap}s.

{\DriftMap}s offer users a global overview of the clusters and the process drifts. {\DriftChart}s allow for a visual categorization of the drifts according to the classification introduced in \cite{DBLP:journals/csur/GamaZBPB14}, as we explain next. These visualizations help the analyst determine if drifts exist at all, which kind of pattern they exhibit over 
time, 
and which kind of behavior is stable or drifting.



\begin{figure}[tb]
	\centering
	\begin{subfigure}[t]{0.34\columnwidth}
		\centering
		\includegraphics[width=1\linewidth]{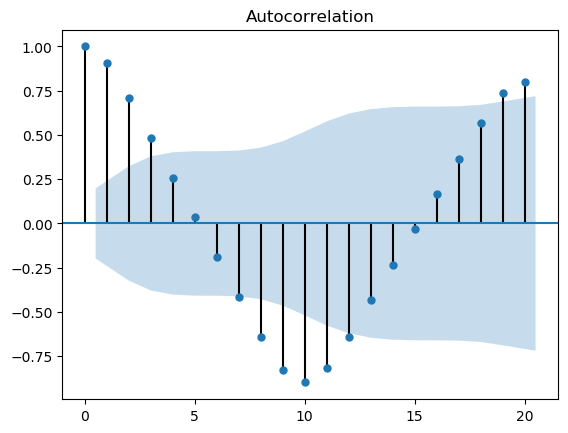}
		\caption{Example of autocorrelated time series}
		\label{fig:method:auto_cosine}
	\end{subfigure}%
	~~~
	\begin{subfigure}[t]{0.325\columnwidth}
		\centering
		\includegraphics[width=1\linewidth]{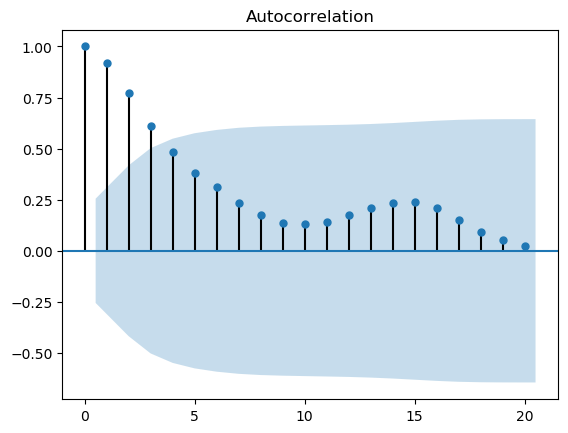}
		\caption{Example of low autocorrelation}
		\label{fig:method:autono}
	\end{subfigure}%
	
	\caption{Example of autocorrelation plots (sepsis log).}
	\label{fig:method:auto}
	\vspace{-2mm}
\end{figure}

We use autocorrelation plots to identify the process changes that follow a seasonal pattern, namely the \emph{reoccuring concept} drift. 
Autocorrelation is determined by comparing a time series with the copy of itself with a lag (delay) of some size~\cite{box2015time}. Autocorrelation plots are useful to discover seasonality in the data.  
The vertical axis in the plot shows the correlation coefficient between elements. The horizontal axis shows the size of the lag between the time series and its copy, refer to~\cref{fig:method:auto}.
The cosine-wave shaped graph in~\cref{fig:method:auto_cosine} shows a high seasonality as the peaks share the same value, while the x-axis indicates the steps needed for the season to reoccur. 
The plot exhibits a seasonal behavior that changes every \num{10} steps from positive to negative correlation.
This means that the values in the time series in step 0 are the opposite of those in step 10 and match those in step 20.
%
\Cref{fig:method:autono}, in contrast, shows the graph with an autocorrelation suggesting that the time series does not exhibit seasonality. We determine whether the step lags are significantly autocorrelated via statistical time series analysis~\cite{box2015time}. We classify only significant autocorrelations as an evidence of reoccurring drifts. 

%

\subsection{Detecting Drift Types}
\label{sec:detecting-drift-types}
After clustering the behavior of the log, we support the analyst with visual cues to understand and categorize the drifts within behavior clusters. To this end, we calculate several measures and display them in our visualization system (see Drift Metrics in~\cref{fig:approach-example}). These measures are introduced for guiding the analyst in the analysis of the drifts.
First, we aid visual analysis by providing a ranking of drifts to assist in focusing on the interesting clusters and filter out outliers. 
We do so by computing the erratic measure (\cref{sec:erratic-measure}). 
Then, we categorize drifts using time series coefficients to identify sudden drifts (\cref{sec:changepoint-detection}). The sudden drifts are highlighted on {\DriftChart}s and summarized as a list of timestamps indicating when they happened. 
We then report on statistics that aids in the identification of incremental and gradual drifts (\cref{sec:stationarity}).


\subsubsection{Finding erratic behavior clusters}
\label{sec:erratic-measure}
As we are interested in the extent to which the confidence levels of constraints change over time, we calculate the following measures.

First, to quantify the overall behavior change in the log we introduce a measure we name \emph{range of confidence}. This measure shows what the average change of the value of {\Declare} constraint is in the whole log. We compute this measure as follows. For all constraint time series  $T_i = (T_{i,1},\ldots,T_{i,\WinSize})$, where $1 \leq i \leq |D|$, we calculate the difference between maximum and minimum values. Then, we average the difference on the number of time series:
\begin{equation}
\label{eqn:spreadconst}
\Sprconfm(D) =  \frac{\sum_{i=1}^{|D|} \mathrm{max}(T_i)-\mathrm{min}(T_i)}{|D|}
\end{equation}
%
Second, to find the most interesting (\emph{erratic}) behavior clusters, we define a measure 
inspired by the idea of finding the length of a poly-line in a plot. The rationale is that straight lines denote a regular trend and have the shortest length, whilst more irregular wavy curves evidence more behavior changes, and their length is higher.
We are, therefore, mostly interested in long lines.

We compute our measure as follows.
We calculate for all constraints $i$ such that $1 \leq i \leq \CnsNum$ the Euclidean distance $\delta : [0,1]\times[0,1]\to\mathbb{R_+}$ between consecutive values in the time series $T_i = (T_{i,1},\ldots,T_{i,\WinSize})$, i.e.,
$\delta(T_{i,j},T_{i,j+1})$ for every $j$ s.t.\ $1 \leqslant j \leqslant \WinSize$.
For every time series $T_i$, we thus derive the overall measure $\Delta(T_i)=\sum_{j=1}^{\WinSize-1}{\delta(T_{i,j},T_{i,j+1})}$.
Thereupon, to measure how erratic a behavior cluster is, we devise the \textit{erratic measure} as follows:
\begin{equation}
\label{eqn:erratic}
\Errtcsm(C) =  \sum_{i=1}^{|C|} \sqrt{1 + ( \Delta(T_i) \times \WinNum)^{2}}
\end{equation}
The most erratic behavior cluster has the highest $\Errtcsm$ value.  

\subsubsection{Detect sudden drifts: Change point detection}
\label{sec:changepoint-detection}
For each cluster of constraints, we search for a set of \emph{sudden drifts}. This means that we look for a set of $k \in \mathbb{N^{+}}$ change points in the time series representing a drifting cluster.
To detect change points, we use the \emph{Pruned Exact Linear Time (PELT)} algorithm~\cite{killick2012optimal}. This algorithm performs an exact search, but requires the input dataset to be of limited size. Our setup is appropriate as, by design, the length of the multivariate time-series is limited by the choice of parameters {\WinSize} and {\WinStep}. Also, this algorithm is suitable for cases in which the number of change points is unknown a priori~\cite[p.~24]{CharlesTruonga/SParxiv:SelectiveReviewOfOfflineChangePointDetectionMethods}, as in our case. We use the \emph{Kernel cost function}, detailed in~\cite{CharlesTruonga/SParxiv:SelectiveReviewOfOfflineChangePointDetectionMethods}, which is optimal for our technique,
and adopt the procedures described in~\cite{killick2012optimal} to identify the optimal \emph{penalty} value. 

\subsubsection{Detect incremental and gradual drifts: Stationarity}
\label{sec:stationarity}

Stationarity is a statistical property of a time series indicating that there is no clear tendency of change over time. It is useful in the context of time series analysis to suggest the presence of a pronounced trend. 
%
Here, we rely on \emph{parametric tests} as a rigorous way to detect non-stationarity. One of the most used techniques are the Dickey-Fuller Test and the Augmented Dickey-Fuller Test~\cite{cheung1995lag}. It tests the null hypothesis of the presence of a unit root in the time series. If a time series has a unit root, it shows a systematic trend that is unpredictable and not stationary.

In particular, we use the Augmented Dickey-Fuller test to detect \emph{incremental} and \emph{gradual drifts}. Those drifts represent a slow change that goes undetected by change point detection algorithms. 
If a time series is non-stationary, this signifies that there is a trend in time series. 
Combined with the analysis of the {\DriftChart}s and the erratic measure, we can differentiate between the incremental and gradual drift. Non-stationary time series with a smoothly increasing {\DriftChart} represent an incremental drift. A {\DriftChart} that shows erratic behavior (or such that the erratic measure is large) indicate a gradual drift.
The highlighted cluster in \cref{fig:driftypes2} is stationary as suggested by the Augmented Dickey-Fuller test. This means there is no clear trend in the drift.

\subsection{Understanding drift behavior}
\label{sec:understand-drift}

\cref{sec:declare:shift,sec:cluster-changepoint,sec:visualdrift,sec:detecting-drift-types} describe techniques that provide various insights into 
drifts in the event log. However, knowing that a drift exists and that it is of a certain type is not sufficient for process analysis. Explanations are required to understand the association between the evidenced drift points and the change in the behavior that led to them. 
In this section, we describe the two visual aids that we employ to explain that association. 

\subsubsection{List of {\Declare} constraints} 
The first report that we generate is the list of {\Declare} constraints that are associated with drifts of a selected cluster. To this end, we use
the {\Declare} subsumption algorithm described in~\cref{sec:declaresubsumption}. 
Reporting these constraints together with the analysis and plots from previous sections help to understand \emph{what} part of the process behavior changes over time and \emph{how}.

Once a highly erratic drift with a seasonal behavior is found, we look up the constraints associated with that drift. For the sepsis case in \cref{fig:approach-example}, e.g., we detect the constraints summarized in~\cref{table:example:table:constraints}.
That drift relates to {\PrecTmp} constraints indicating that before \emph{Release D} can occur, \emph{Leucocytes}, \emph{CRP} and \emph{ER Triage} must occur.

\begin{table}[bt]
	\caption{Example of constraints present in the drift.}
	\label{table:example:table:constraints}
	\centering
	\resizebox{0.75\columnwidth}{!}{
\rowcolors{2}{white}{gray!12.5}
\begin{tabular}{r r r r D{1} D{1} D{1}}
	\toprule
	\textbf{Cluster}    & \textbf{Constraint} & \textbf{Activity 1}    & \textbf{Activity 2}         \\ \midrule
	\cellcolor{white}   & {\PrecTmp}       & \Task{Leucocytes}     & \Task{Release D}      \\ 
	\multirow{-1}{*}{1} & {\PrecTmp}       & \Task{CRP} & \Task{Release D}    \\ 
    \cellcolor{white}  & {\PrecTmp}       & \Task{ER Triage} & \Task{Release D}    \\ \bottomrule

\end{tabular} 
%
%

	}
\end{table}

\subsubsection{Extended Directly-Follows Graph}
\label{sec:edfg}
The process analyst also benefits from a graphical representation of the drifting constraints. To this end, we build upon the Directly-Follows Graphs (DFGs) as shown in~\cref{fig:driftypes2} on the left-hand side.
%
%
Our technique extends the DFG with additional arcs that convey the meaning of the {\Declare} constraints. We distinguish three general types of constraints: \emph{immediate} (e.g., {\ChaPrec{$a$}{$b$}}, imposing that $b$ can occur only if $a$ occurs immediately before), \emph{eventual} (e.g., \Succ{$a$}{$b$}, dictating that, if $a$ or $b$ occur in the same trace, $b$ has to eventually follow $a$), and {negated} (e.g., {\NotSucc{$a$}{$b$}}, imposing that $a$ cannot follow $b$). We annotate them with green, blue, and red colors, respectively. 
This way, the user is provided with an overview of the log and which parts of the business process are affected by drifts.

\subsection{Computational Complexity}

As discussed in \cref{sec:preliminaries}, \emph{Step 1} involves DFG mining algorithms that are linear in the number of traces ($O(|L|)$) and quadratic in the number of activities ($O(|\Sigma|^2)$). 
\emph{Step 2}, that is, mining constraint windows, is linear in the number of traces ($O(|L|)$) and quadratic in the number of activities ($O(|\Sigma|^2)$) too. The subsumption of {\Declare} constraints runs in $O(\CnsNum \cdot \mathrm{log}_2(\CnsNum))$ where $O(\CnsNum) \subseteq O(|\Sigma|^2)$.
\emph{Step 3}, clustering time series, is polynomial in the number of time series and, therefore, of constraints ($O(\CnsNum^3)$). 
\emph{Step 4}, sudden drift detection, runs in $O(\CnsNum^2)$ in the worst case.
The tasks of detecting gradual drifts and reoccurring drifts are constant operations, as they are performed on the averaged time series. 
Finally, \emph{Step 5}, understanding drift behavior, has the same asymptotic complexity as \emph{Step 1}.
We note that all the applied computations present at most polynomial complexity.

%
%
\section{Evaluation}
\label{sec:evaluation}
%
%


This section presents the evaluation of our visualization system. This evaluation represents the deploy step that completes the core phase of the design study methodology by~\cite{sedlmair2012design}. \Cref{setup} describes our implementation. Using this implementation, our evaluation focuses on the following aspects. 
\Cref{synthetic} evaluates our drift point detection technique for its capability to rediscover change points induced into synthetic logs. \Cref{real-world} presents insights that our system reveals on real-world cases. \Cref{performance} presents experimental results on computational complexity. \Cref{user} summarizes findings from a user study with process mining experts who evaluated the visualizations of our system on a real-world event log.
With this part of the evaluation, we focus on target users, their questions and their measurements~\cite{meyer2015nested}.
Finally, \cref{subsec:discussion,subsec:limitations} discuss how our system addresses the requirements for process drift detection and limitations of the approach, respectively.


\subsection{Implementation and user interaction}\label{setup}

\setlength\intextsep{0pt}
\begin{table}
	\centering
	\caption{Event logs used in the evaluation.}
	\label{table-event-logs}
\rowcolors{2}{white}{gray!12.5}
\begin{tabular}{l r r r}
	\toprule
	\textbf{Origin} & \textbf{Event log}  & \textbf{Related work}                           &  \\ \midrule
	Synthetic & ConditionalMove & ProDrift  2.0~\cite{DBLP:conf/er/OstovarMRHD16} & \\
	Synthetic & ConditionalRemoval & ProDrift  2.0~\cite{DBLP:conf/er/OstovarMRHD16} & \\
	Synthetic  & ConditionalToSequence  & ProDrift  2.0~\cite{DBLP:conf/er/OstovarMRHD16} & \\
	Synthetic  & Loop & ProDrift  2.0~\cite{DBLP:conf/er/OstovarMRHD16} & \\ 
	Real-world  & Italian help desk\footnotemark[1] & Process~Trees~\cite{quteprints121158}           &  \\
	Real-world  & BPI2011\footnotemark[3]       & ProDrift  2.0~\cite{DBLP:conf/er/OstovarMRHD16} &  \\
	Real-world  & Sepsis\footnotemark[7]       & - &  \\
	 \bottomrule
\end{tabular} 
\end{table}
%

For the implementation of our approach, we integrate several state-of-the-art techniques and tools.
To discover {\Declare} constraints, we use \gls{minerful}%
\footnote{\scriptsize \url{https://github.com/cdc08x/MINERful}}
because of its high performance~\cite{DBLP:journals/tmis/CiccioM15}.
For change point detection, we integrate the \textit{ruptures} python library%
\footnote{\scriptsize \url{https://github.com/deepcharles/ruptures}}.
For time series clustering, we resort to the \textit{scipy} library%
\footnote{\scriptsize \url{https://docs.scipy.org/doc/scipy/reference/generated/scipy.cluster.hierarchy.linkage.html}}. 

To attain the most effective outcome, we tune the clustering parameters such as the weighted method for linking clusters
(distance between clusters defined as the average between individual points) 
and the correlation metric (to find individual distances between two time-series). 
To enhance {\DriftMap} visualizations, we sort the time series of each cluster by the mean squared error distance metric. We implemented both the {\DriftMap} and {\DriftChart} using the python library \textit{matplotlib}.%
\footnote{\url{https://matplotlib.org/}}
For the Augmented Dickey-Fuller test and autocorrelation we use the \textit{statmodels} python library\footnote{\url{http://www.statsmodels.org/}}.
To discover the 
Directly-Follows Graph, we extended the pm4py
process mining python library%
\footnote{\url{http://pm4py.org}, \url{https://github.com/pm4py}}~\cite{DBLP:journals/corr/abs-1905-06169}.
Our overall system is implemented in Python 3. Its source code and the parameters used for our experiments are publicly available.%
\footnote{\scriptsize \url{https://github.com/yesanton/Process-Drift-Visualization-With-Declare}}  

We found that varying the window size affects the results only marginally. 
Experimenting with parameters, we observed that producing sub-logs out of \num{60} windows provided a good balance between detail and stability of the results. 
Therefore, we recommend the following set-up for the involved parameters: $\smash{\WinStep = \frac{|L|}{60 + 1}}$, and $\smash{\WinSize = 2\cdot\WinStep}$ for smooth visual representation.

We use \emph{hierarchical clustering} for time series clustering, as it is reportedly one of the most suitable algorithms when the number of clusters is unknown~\cite{Aghabozorgi:2015:TCD:2799194.2799230}. We found that the Ward linkage method and the Euclidean distance function produce the best results. To detect change points, we use the \emph{Pruned Exact Linear Time (PELT)} algorithm~\cite{killick2012optimal}. This algorithm performs an exact search but requires the input dataset to be of limited size. Our setup is appropriate as by design the length of the multivariate time-series is limited by the choice of parameters {\WinSize} and {\WinStep}. Also, this algorithm is suitable for cases in which the number of change points is unknown a priori~\cite[p.~24]{CharlesTruonga/SParxiv:SelectiveReviewOfOfflineChangePointDetectionMethods}, as in our case. We use the \emph{Kernel cost function}, detailed in~\cite{CharlesTruonga/SParxiv:SelectiveReviewOfOfflineChangePointDetectionMethods}, which is optimal for our technique,
and
adopt the procedures described in~\cite{killick2012optimal} to identify the optimal \emph{penalty} value. 

\begin{figure}
	\includegraphics[width=1.0\linewidth]{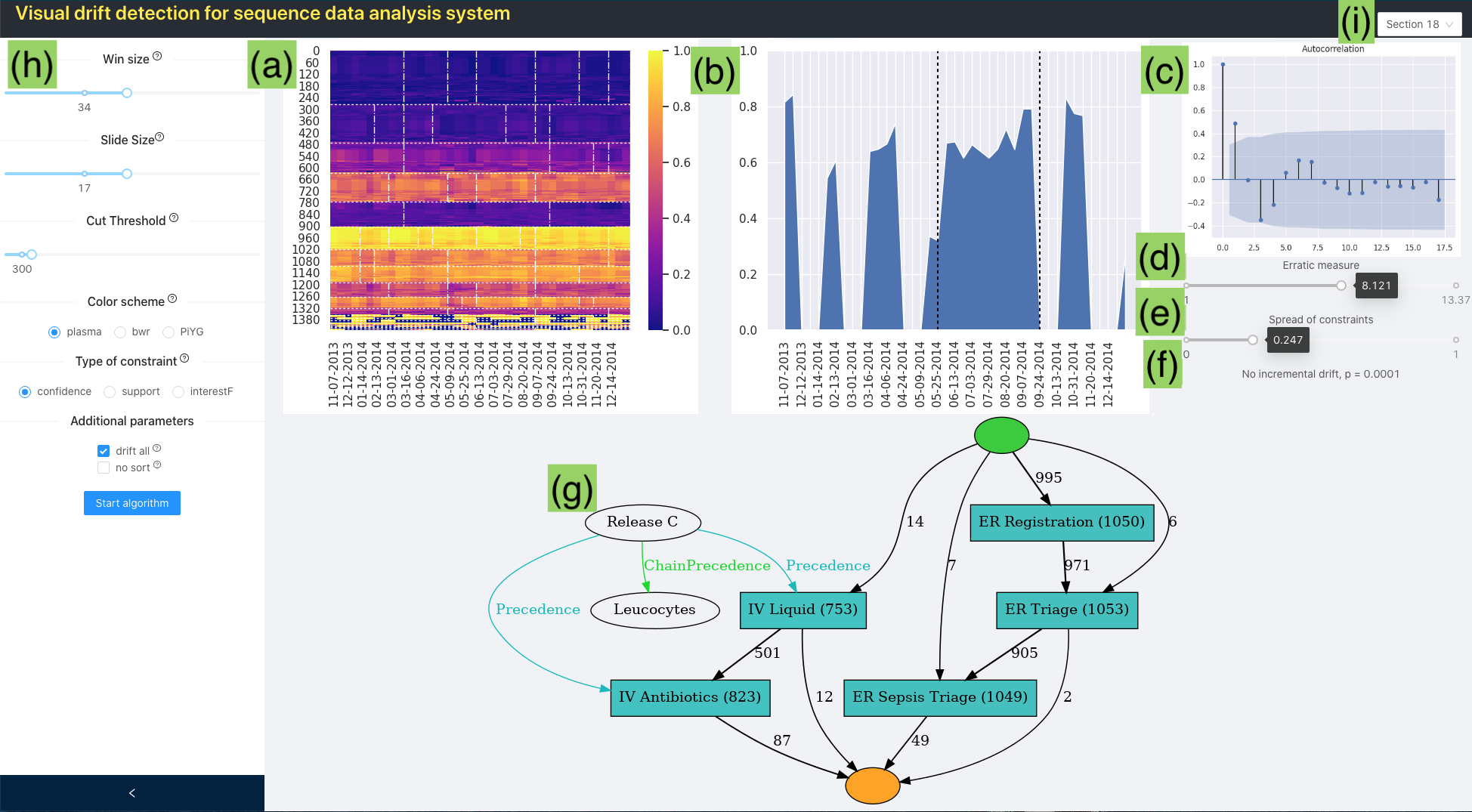}
	\caption[The VDD system UI]{The user interface of the \gls{dvd} system, running on the Sepsis event log~\cite{DBLP:conf/emisa/MannhardtB17}. (a) {\DriftMap}. (b) {\DriftChart}. (c) Autocorrelation plot. (d) Erratic measure. (e) Spread of constraints view. (f) Incremental drifts test. (g) Extended Directly-Follows Graph. (i) Behavior cluster selection menu.}
	\label{fig:tool-screenshot}
\end{figure}

The \gls{dvd} system web application is shown in~\cref{fig:tool-screenshot}. We describe the tool and user interaction in detail in the demo paper~\cite{DBLP:conf/icpm/YeshchenkoMCP20} and in the walk-through video.\footnote{\scriptsize \url{https://youtu.be/mHOgVBZ4Imc}}
The user starts with uploading an event log file. 
Then, she can tune analysis parameters including $\WinStep$, $\WinSize$, {\Declare} constraint type, cut threshold for hierarchical clustering, as well as look-and-feel parameters such as the color scheme, as shown in~\cref{fig:tool-screenshot}(h). Default values are suggested based on the characteristics of the input log. Multiple views are displayed and updated in the maun panel~\cref{fig:tool-screenshot}(a-g). The user can select the behavior cluster to focus on~\cref{fig:tool-screenshot}(i), thus triggering an update in the other views~\cref{fig:tool-screenshot}(b-g).

The application of our system with a multi-national company highlights the importance of such exploratory analysis strategies. Understanding changes over time is of key importance to process analysts to identify factors of change and effects of management interventions into the process. The user interaction of our system supports the visual identification of drifts and helps to drill down into the behavior that is associated with those drifts, thereby helping the analysts formulate and validate hypotheses about factors of change in the process.

\subsection{Evaluation on synthetic data}\label{synthetic}
For our evaluation, we make use of synthetic and real-world event logs.\footnote{\scriptsize \url{https://doi.org/10.4121/uuid:0c60edf1-6f83-4e75-9367-4c63b3e9d5bb}}\footnote{\scriptsize \url{https://doi.org/10.4121/uuid:a7ce5c55-03a7-4583-b855-98b86e1a2b07}}\footnote{\scriptsize \url{https://doi.org/10.4121/uuid:d9769f3d-0ab0-4fb8-803b-0d1120ffcf54} (preprocessed as in~\cite{DBLP:conf/er/OstovarMRHD16})}\footnote{\scriptsize \url{https://data.4tu.nl/repository/uuid:915d2bfb-7e84-49ad-a286-dc35f063a460}}
In this way, we can compare the effectiveness of our approach with earlier proposals.
\Cref{table-event-logs} summarizes the event logs used in the evaluation and indicates which prior papers used these logs.

To demonstrate the accuracy with which our technique detects drifts, we first test it on synthetic data in which drifts were manually inserted, thereby showing that we accurately detect drifts at the points in which they occur. 
We compare our results with the state-of-the-art algorithm ProDrift~\cite{DBLP:conf/er/OstovarMRHD16} 
on real-world event logs.

\begin{figure}[tbp]
	\centering
	\begin{subfigure}[t]{0.31\columnwidth}
		\centering
		\includegraphics[width=1\linewidth]{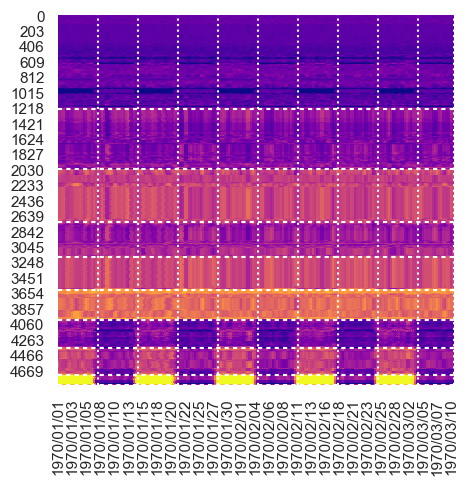}
		\caption{ConditionalMove}
		\label{fig:conditionalMove}
	\end{subfigure}%
	\begin{subfigure}[t]{0.31\columnwidth}
		\centering
		\includegraphics[width=1\linewidth]{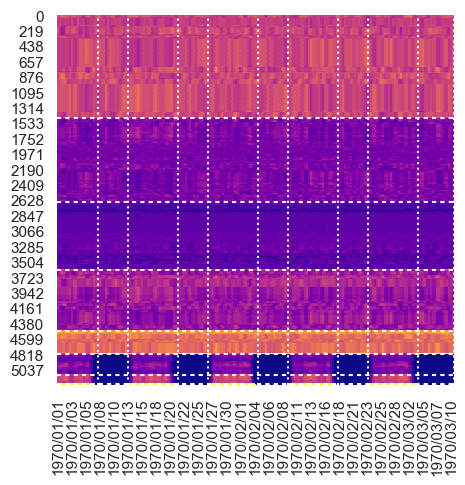}
		\caption{ConditionalRemoval}
		\label{fig:conditionalRemoval}
	\end{subfigure}
	\begin{subfigure}[t]{0.36\columnwidth}
		\centering
		\includegraphics[width=1\linewidth]{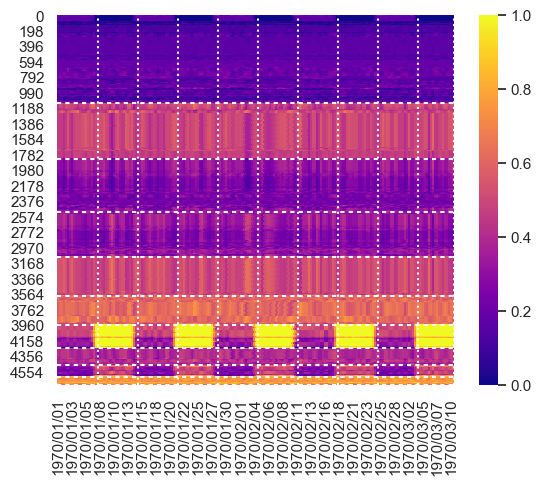}
		\caption{ConditionalToSequence}
		\label{fig:conditionalToSequence}
	\end{subfigure}
	
	
	\begin{subfigure}[t]{0.31\columnwidth}
		\centering
		\includegraphics[width=1\linewidth]{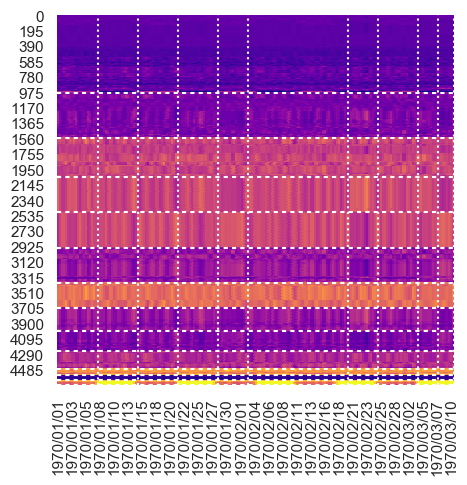}
		\caption{Loop}
		\label{fig:loop}
	\end{subfigure}%
	\begin{subfigure}[t]{0.36\columnwidth}
		\centering
		\includegraphics[width=1\linewidth]{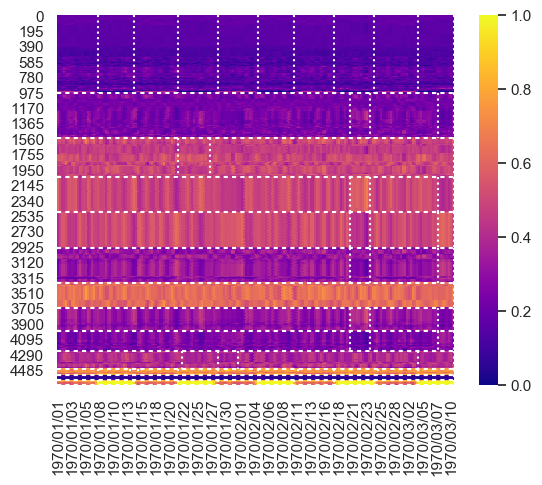}
		\caption{Loop, drifts by cluster}
		\label{fig:loop-in-cluster}
	\end{subfigure}
	\begin{subfigure}[t]{0.31\columnwidth}
		\centering
		\includegraphics[width=1\linewidth]{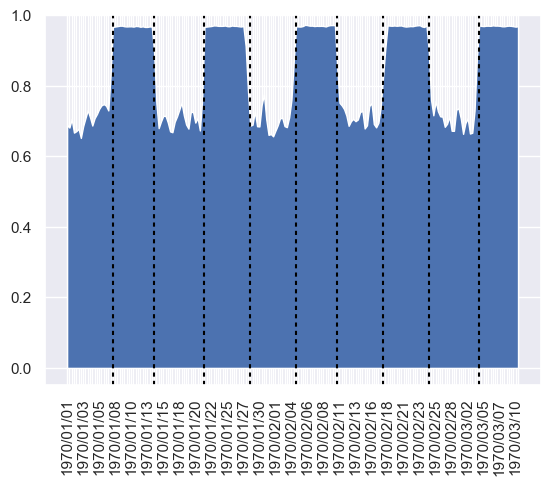}
		\caption{Loop, most erratic cluster}
		\label{fig:loop-detail}
	\end{subfigure}
	\vspace{-1mm}
	\caption{Evaluation results on synthetic logs.}
	\label{fig:manmade}
\end{figure}

Ostovar et al.~\cite{DBLP:conf/er/OstovarMRHD16} published a set of synthetic logs that they altered by artificially injecting drifting behavior: {ConditionalMove}, {ConditionalRemoval}, {ConditionalToSequence}, and {Loop}.%
\footnote{\scriptsize \url{http://apromore.org/platform/tools}}
\Cref{fig:manmade} illustrates the results of the application of the \gls{dvd} technique on these logs.
By measuring \emph{precision} as the fraction of correctly identified drifts over all the ones retrieved by \gls{dvd} and \emph{recall} as the fraction of correctly identified drifts over the actual ones, we computed the F-score (harmonic mean of precision and recall) of our results for each log.
Using the default settings and no constraint set clustering, we achieve the F-score of \num{1.0} for logs {ConditionalMove}, {ConditionalRemoval}, {ConditionalToSequence}, and \num{0.89} for the {Loop} log. 
When applying the cluster-based change detection for the {Loop} log, we achieve an F-score of \num{1.0}. 
he {\DriftChart} in~\cref{fig:loop-detail} illustrates the trend of confidence for the most erratic cluster for the \emph{Loop} log.
The {\DriftMap} for the \emph{Loop} log is depicted in~\cref{fig:loop-in-cluster}. In contrast to~\cite{DBLP:conf/er/OstovarMRHD16} we can see which behavior in which cluster contributes to the drift.

%

\subsection{Evaluation on real-world data}\label{real-world}
%

Next, we evaluate our system with three real-world event logs. In the next subsections we describe all processing steps for each of the logs. 

\subsubsection{Sepsis log}
\begin{figure}[tbp]
	\centering
	\begin{subfigure}[t]{0.34\columnwidth}
		\centering
		\includegraphics[width=1\linewidth]{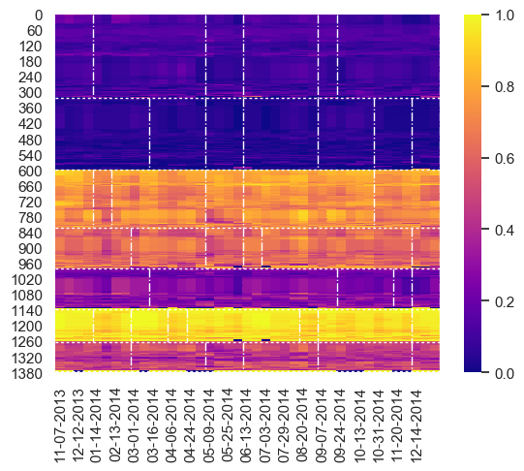}
		\caption{Drift Map}
		\label{fig:sepsis:overall}
	\end{subfigure}%
	\begin{subfigure}[t]{0.325\columnwidth}
		\centering
		\includegraphics[width=1\linewidth]{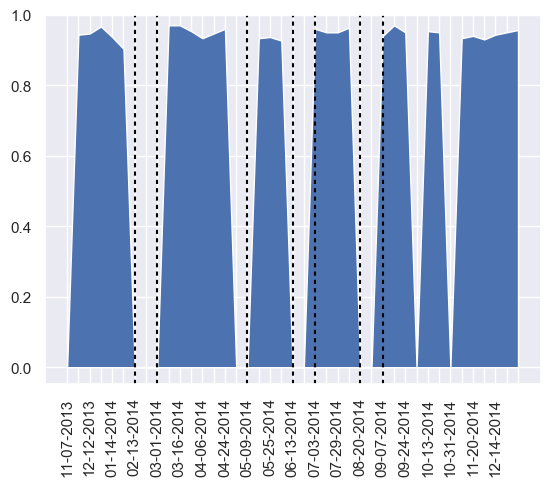}
		\caption{Drift in cluster 8}
		\label{fig:sepsis:8}
	\end{subfigure}%
	\begin{subfigure}[t]{0.325\columnwidth}
		\centering
		\includegraphics[width=1\linewidth]{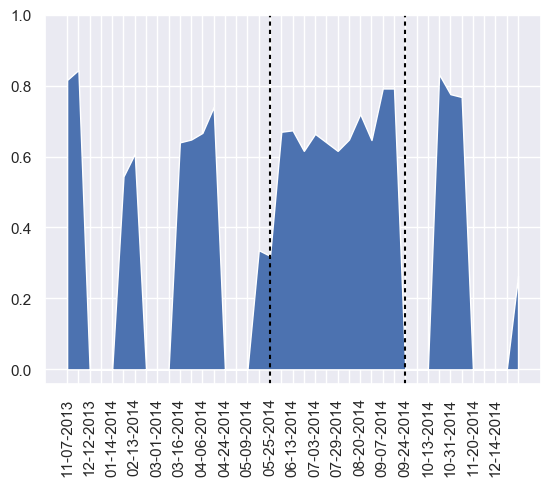}
		\caption{Drift in cluster 12}
		\label{fig:sepsis:12}
	\end{subfigure}
	
	\caption{Sepsis \gls{dvd} visualizations.}
	\label{fig:sepsis}
	\vspace{-2mm}
\end{figure}

%
%
The sepsis log describes the cases of patients affected by sepsis. This condition occurs to some patients as a response to infections. The process that generated this log captures the activities executed from the registration of the patients until they are discharged. 
Prior process mining techniques offer limited insights into this log~\cite{DBLP:conf/emisa/MannhardtB17}. We use the processing steps and the multiple outputs of our system to get an understanding of changes in this log over time.
\\
\textbf{Step 1: Mining Directly-Follows Graph as an overview.} 
\\
The directly-Follows Graph from this log shows \num{12} activities. The most frequent activity is \emph{Leucocytes} with \num{3386} instances, followed by the activity \emph{CRT} with \num{3262} occurrences. In contrast, the activity \Task{Admission IC} only occurred \num{117} times. The main path through the DFG is \Task{ER registration} $\,\to\,$ \Task{ER Triage} $\,\to\,$ \Task{ER Sepsis Triage} $\,\to\,$ \Task{CRT} $\,\to\,$ \Task{Leucocytes} $\,\to\,$ \Task{Admission NC}.
Next to this frequent path, various variants exist that correspond to the less frequent behavior -- including, e.g., the \Task{Addmission IC} and \Task{IV Liquid} activities.
\\
\textbf{Step 2: Mining {\Declare} constraints windows.} 
\\
For determining the {\WinSize} and {\WinNum} we consider the number of traces in the log.
The log contains \num{1050} cases spanning over the period of time of a year and four months. We chose a {\WinSize} of \num{50} and a {\WinStep} of \num{25}. 
\\
\textbf{Step 3-4: Finding Drifts and Visual Drift Overview.}
\Cref{fig:sepsis:overall} depicts the overall drift behavior. The {\DriftMap} hardly shows any strong patterns of change over time. Apparently, most of the major clusters of behavior do not evidence drifts. Drilling down into clusters with less constraints offers us insights into quite erratic behavior though (see~\cref{fig:sepsis:8} and~\cref{fig:sepsis:12}). 
\\
\textbf{Step 5: Detecting Drift Types.}
Using the Augmented Dickey-Fuller test, we test the hypothesis that there is a unit root present in the data. If so, the time-series is considered to be non-stationary. The analysis of cluster 8 and cluster 12 shows a $p$ value of \num[scientific-notation=true]{0.000003} and \num[scientific-notation=true]{0.000077}, respectively,
suggesting that the data does not have a unit root, i.e., it is \emph{stationary}. This means that the behavior does not have an upward or downward trend of change.

\begin{figure}[tb]
	\centering
	\begin{subfigure}{0.33\columnwidth}
		\centering
		\includegraphics[width=1\linewidth]{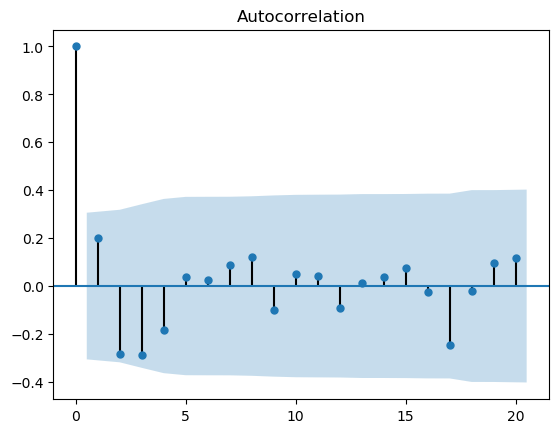}
		\caption{Cluster 8}
		\label{fig:sepsis:autocorrelation:8}
	\end{subfigure}
	\begin{subfigure}{0.33\columnwidth}
		\centering
		\includegraphics[width=1\linewidth]{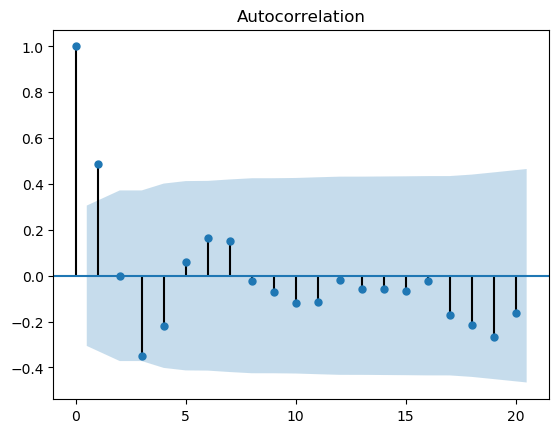}
		\caption{Cluster 10}
		\label{fig:sepsis:autocorrelation:12}
	\end{subfigure}
	\caption{Autocorrelation plots for the sepsis log.}
	\label{fig:sepsis:autocorrelation}
\end{figure}

The autocorrelation plots shown in~\cref{fig:sepsis:autocorrelation} display negative correlation in steps 2-3 and positive autocorrelation in steps 6-7 -- see \cref{fig:sepsis:autocorrelation:8} and \cref{fig:sepsis:autocorrelation:12}. That means that there is significant seasonality in the data. 
\\
\textbf{Step 6: Understanding drift behavior.} 
In order to understand the behavior behind some of the drifts we discovered in previous steps, we explore their list of constraints and the derived extended DFG. 
Based on the inspection of the {\DriftMap} in \cref{fig:sepsis:overall} and the erratic measures in \cref{table:sepsis-err-clusters}, we focus on the drifts in~\cref{fig:sepsis:8} and~\cref{fig:sepsis:12}.

\begin{table}[tb]
	\caption{Sepsis log erratic clusters.}
	\label{table:sepsis-err-clusters}
	\centering
\rowcolors{2}{white}{gray!12.5}
\begin{tabular}{r D{3}}
	\toprule
	\textbf{Drift number} & \textbf{{\Errtcsm} measure} \\ 
	
	\midrule
    without drift & 40\\
    \textit{11} & 245.3050142 \\
    \textit{12} & 324.8405101 \\
    \textit{10} & 415.0076954 \\
    
    \textit{7} & 417.1348637 \\
    \textit{9} & 495.7949553 \\
    \textit{8} & 534.8147645 \\
	\bottomrule
\end{tabular} 

\end{table}

\begin{figure}[tbp]
	\includegraphics[width=\columnwidth]{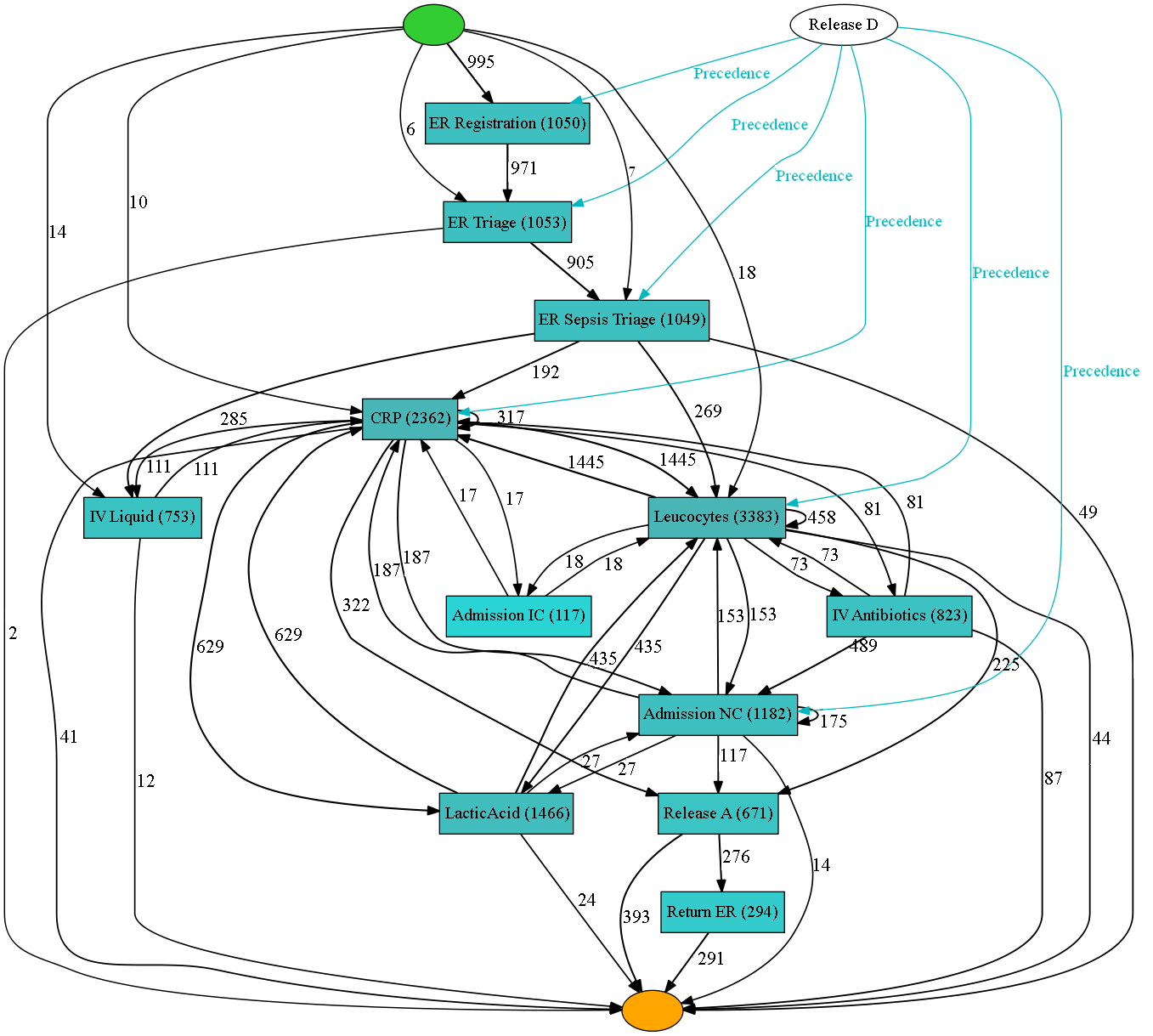}
	\vspace{-3mm}
	\caption{Extended Directy-Follows Graph for cluster 8, derived from the sepsis log.}
	\label{fig:sepsis-dfg-8}
	\vspace{-3mm}
\end{figure}

\begin{table}[bt]
	\caption{Sepsis log constraints.}
	\label{table:sepsis-drifts-constrains}
	\centering
	\resizebox{0.9\columnwidth}{!}{
\rowcolors{2}{white}{gray!12.5}
\begin{tabular}{r r r r D{1} D{1} D{1}}
	\toprule
	                         \textbf{Cluster} & \textbf{Constraint} & \textbf{Activity 1}         & \textbf{Activity 2}           \\ \midrule
	                        \cellcolor{white} & {\PrecTmp}       & \Task{IV Antibiotics}          & \Task{Release C}         \\
	                        \cellcolor{white} & {\AltPrecTmp}       & \Task{IV Antibiotics}       & \Task{Release C}     \\
	                        \cellcolor{white} & {\PrecTmp}       & \Task{IV Liquid}               & \Task{Release C} 		    \\
	                        \cellcolor{white} & {\AltPrecTmp}       & \Task{IV Liquid}            & \Task{Release C}                 		\\
	                        \multirow{-5}{*}{8} & {\AltPrecTmp}          & \Task{Leucocytes}      & \Task{Release C}         \\ \midrule
	                        \cellcolor{white} & {\ChaPrecTmp}       & \Task{CRP}				   & \Task{Release D}               \\
	                        \cellcolor{white} & {\PrecTmp}       & \Task{IV Antibiotics}        & \Task{Release D}               \\
	                        \cellcolor{white} & {\AltPrecTmp}       & \Task{IV Antibiotics}                  & \Task{Release D}     \\
	                        \cellcolor{white} & {\PrecTmp}       & \Task{IV Liquid}                  & \Task{Release D} 		    \\
	                        \cellcolor{white} & {\AltPrecTmp}       & \Task{IV Liquid}    & \Task{Release D}                 		\\
	                        \cellcolor{white} & {\PrecTmp}       & \Task{LacticAcid}                  & \Task{Release D}         \\
	                      \multirow{-7}{*}{12} & {\AltPrecTmp}          & \Task{LacticAcid}                  & \Task{Release D}         \\ \bottomrule
\end{tabular} 
%
%

	}
\end{table}
\noindent
\Cref{table:sepsis-drifts-constrains} shows the {\Declare} constraints of these clusters. We observe that the drifts are related to specific activities, namely \Task{Release C} for cluster 8 and \Task{Release D} for cluster 12. We conclude that there are reoccurring drift patterns indicating, thus there are seasonal factors affecting \Task{Release C} and a \Task{Release D}.
%
%
We highlight the process behavior that is subject to drifts via the extended Directly-Follows Graphs. \Cref{fig:sepsis-dfg-8} shows the extended DFG highlighting the activities involved in the drift behavior of cluster 8. 
For this case, we observe that activity \Task{Release D} was executed after several activities in certain parts of the timeline, as shown in~\cref{fig:sepsis:8}. 

%
\subsubsection{Italian help desk log}
%
%
\begin{figure*}[tbp]
	\includegraphics[width=1.0\textwidth]{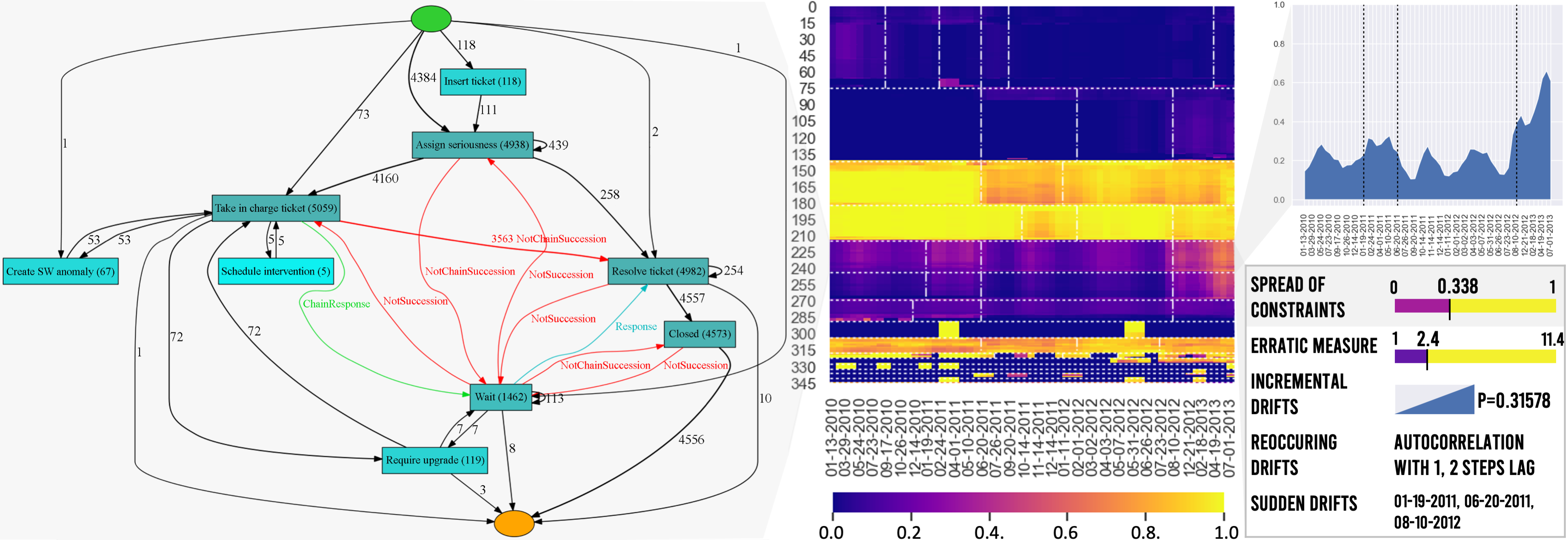}
	\caption{The \gls{dvd} system visualization of the Italian help desk event log.}
	\label{fig:approach-italian-help-desk-log}
\end{figure*}

\begin{figure}[tbp]
	\centering
	\begin{subfigure}[t]{0.297\columnwidth}
		\centering
		\includegraphics[width=1\linewidth]{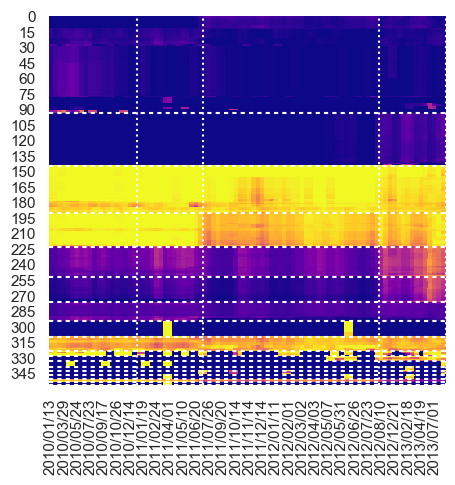}
		\caption{Overall change points}
		\label{fig:italianhelp:alldrifts}
	\end{subfigure}%
	\begin{subfigure}[t]{0.36\columnwidth}
		\centering
		\includegraphics[width=1\linewidth]{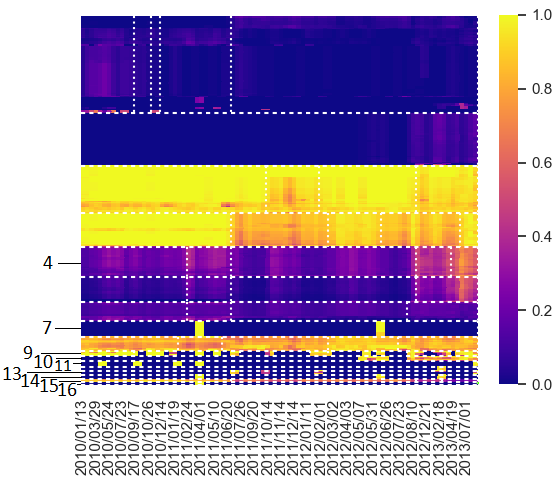}
		\caption{Drifts by cluster}
		\label{fig:italianhelp:separate}
	\end{subfigure}
	\begin{subfigure}[t]{0.33\columnwidth}
		\centering
		\includegraphics[width=1\linewidth]{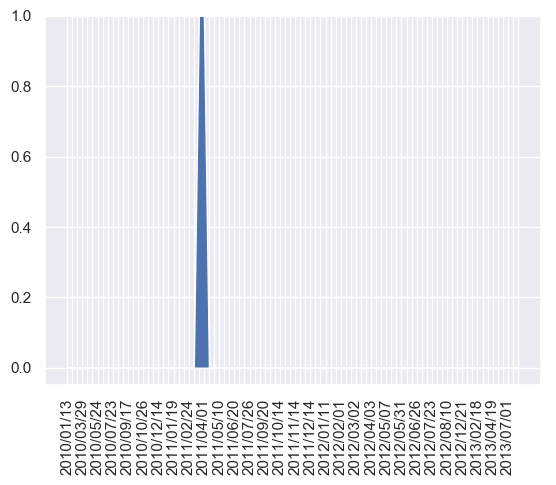}
		\caption{Most erratic cluster}
		\label{fig:italianhelp:outlier}
	\end{subfigure}
	
	\caption{Italian help desk log \gls{dvd} visualizations.}
	\label{fig:italianhelp}
	\vspace{-2mm}
\end{figure}

Next, we focus on the event log of the help-desk of an Italian software company. It covers all steps from the submission of a support ticket to its closing. \Cref{fig:approach-italian-help-desk-log} provides an overview.
\\
\textbf{Step 1: Mining Directly-Follows Graph as an overview.} 
The Directly-Follows Graph of this log displays \num{9} activities. While activity \Task{Take in charge ticket} occurred \num{5059} times, activity \Task{Schedule intervention} only occurred \num{5} times. The main path through the DFG is \Task{Assign seriousness} $\,\to\,$ \Task{Take in charge ticket} $\,\to\,$ \Task{Resolve ticket} $\,\to\,$ \Task{Closed}. 
Other variants are evidenced though, corresponding to the the observation of anomalies (\Task{Create SW anomaly} activity), waiting (\Task{Wait} activity), or requests for an upgrade (\Task{Require upgrade} activity).
\\
\textbf{Step 2: Mining {\Declare} windows.}
This log contains \num{4579} cases that are evenly distributed over the period of four years. We set the {\WinSize} to \num{100} and the {\WinStep} to 50.  
\\
\textbf{Step 3-4: Finding Drifts and Visual Drift Overview.} 
Based on the mined {\Declare} constraints, the {\DriftMap} is generated. \Cref{fig:italianhelp} shows the overview of the drifts in the log. For the overall set of clusters, there are three major drift points detected. \Cref{fig:italianhelp:separate} shows a more fine-granular series of drift points, which can be observed within separate clusters. 
There are also many drifts that signify unregular behaviour and are probably outliers (such as drifts 9, 10 and 11 in~\cref{fig:italianhelp:separate}). In step 5 we inspect them in detail. 
\\
\textbf{Step 5: Detecting Drift Types.}
Our system correctly detects sudden drifts in the Italian help desk log,  identifying the same two drifts that were found by ProDrift~\cite{quteprints121158}, approximately in the first half and towards the end of the time span.
As illustrated by the \gls{dvd} visualization in~\cref{fig:italianhelp:alldrifts}, we additionally detect another sudden drift in the first quarter.
By analyzing the within-cluster changes (\cref{fig:italianhelp:separate}), we notice that the most erratic cluster contains an outlier, as is shown by the spikes in~\cref{fig:italianhelp:outlier}.

\begin{figure}[tbp]
	\centering
	\begin{subfigure}[t]{0.32\columnwidth}
		\centering
		\includegraphics[width=1\linewidth]{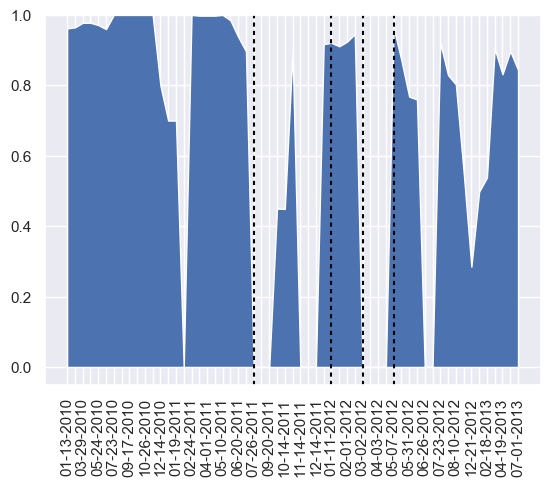}
		\caption{Cluster 9}
		\label{fig:italian-help-c9}
	\end{subfigure}%
	\begin{subfigure}[t]{0.32\columnwidth}
		\centering
		\includegraphics[width=1\linewidth]{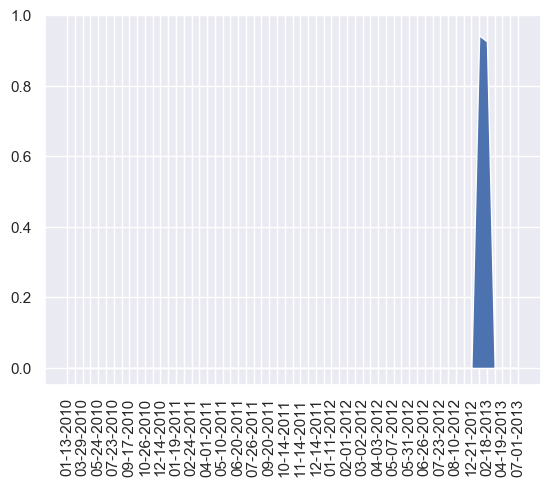}
		\caption{Cluster 12}
		\label{fig:italian-help-c12}
	\end{subfigure}
	\begin{subfigure}[t]{0.32\columnwidth}
		\centering
		\includegraphics[width=1\linewidth]{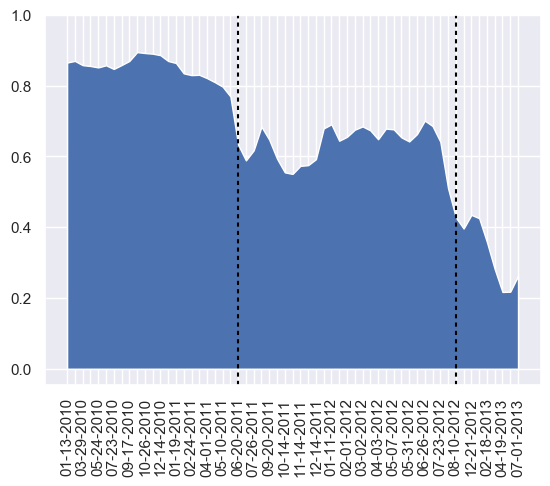}
		\caption{Cluster 15}
		\label{fig:italian-help-c15}
	\end{subfigure}
	
	
	\begin{subfigure}[t]{0.32\columnwidth}
		\centering
		\includegraphics[width=1\linewidth]{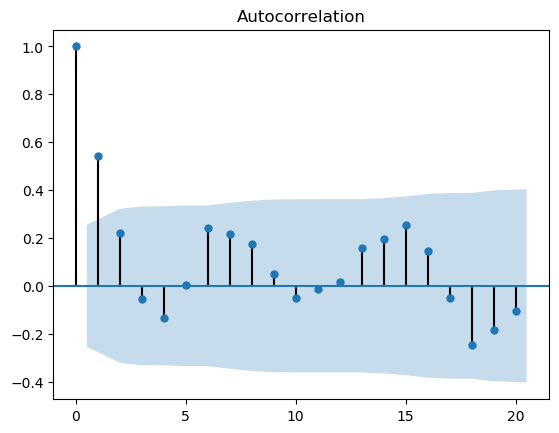}
		\caption{Autocorrelation in cluster 9}
		\label{fig:italian-help-ac9}
	\end{subfigure}%
	\begin{subfigure}[t]{0.32\columnwidth}
		\centering
		\includegraphics[width=1\linewidth]{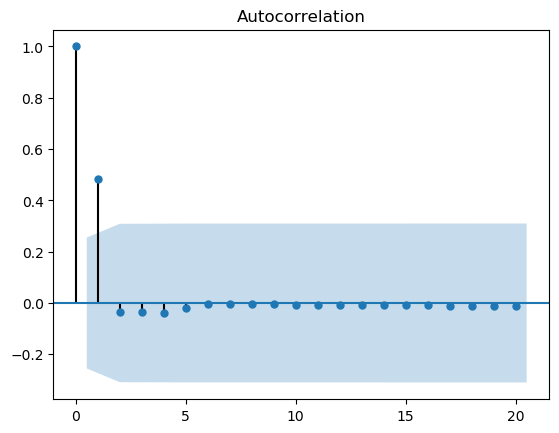}
		\caption{Autocorrelation in cluster 12}
		\label{fig:italian-help-ac12}
	\end{subfigure}
	\begin{subfigure}[t]{0.32\columnwidth}
		\centering
		\includegraphics[width=1\linewidth]{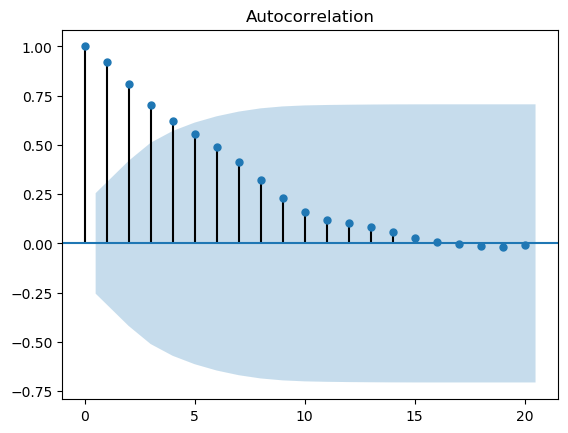}
		\caption{Autocorrelation in cluster 15}
		\label{fig:italian-help-ac15}
	\end{subfigure}
	\vspace{-1mm}
	\caption{Italian help desk autocorrelation results}
	\label{fig:aut-italian}
\end{figure}

We check for reoccurring drifts based on autocorrelation. The visualizations in~\cref{fig:aut-italian} show the autocorrelation plots of different clusters together with their {\DriftChart}s. Cluster 9 (\cref{fig:italian-help-c9}) shows seasonality, while clusters 12 and 15 (\cref{fig:italian-help-c12,fig:italian-help-c15}) do not.

Based on the Augmented Dickey-Fuller test, we discover that some of the clusters exhibit incremental drift. For example, cluster 15 has a $p$-value of \num{0.98045} indicating a unit-root, which points to non-stationarity. Indeed, we find an incremental drift with an associated decreasing trend, as shown in \cref{fig:italian-help-c15}.
The result alongside the erratic measures are shown in~\cref{table:italianticekt-err-clusters}. They highlight that cluster 9 has the most erratic drift behavior.

\begin{table}[tb]
	\caption{Italian help desk log erratic clusters.}
	\label{table:italianticekt-err-clusters}
	\resizebox{0.9\columnwidth}{!}{%
		\centering
\rowcolors{2}{white}{gray!12.5}
\begin{tabular}{r D{3} D{3}}
	\toprule
	\textbf{Drift number} & \textbf{{\Errtcsm} measure} & \textbf{A-Dickey-Fuller p-value} \\ 
	
	\midrule
    without drift & 89 & \\
    \textbf{9} & 681.4662212 & 0.000805 \\
    \textbf{11} & 578.7923308 & 0.000815 \\
    \textit{14} & 394.1381289 & 0.000168 \\
    \textit{13} & 386.3771434 & 0.130435 \\    
    \textit{7} & 287.538045 & 0.315785 \\
    \textit{10} & 256.338914 & 0.959575 \\
    \textit{16} & 174.0169479 & 0.079531 \\
    \textit{12} & 166.6380301 & 0.899872 \\
    \textbf{4} & 139.4028553 & 0.315785 \\

	\bottomrule
\end{tabular} 


	}
\end{table}

\begin{figure}[tb]
	\centering
	\begin{subfigure}{0.32\columnwidth}
		\centering
		\includegraphics[width=1\linewidth]{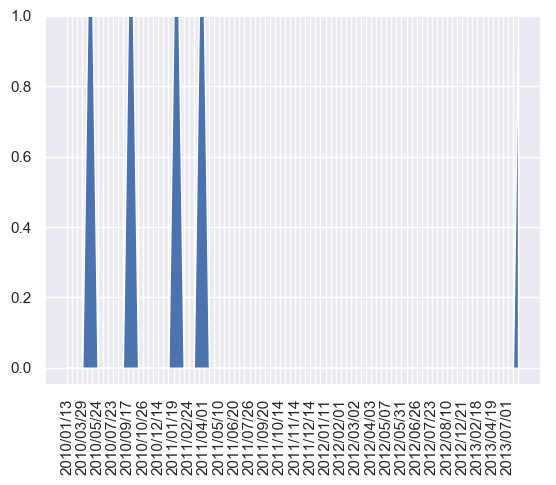}
		\caption{\Errtcsm$=$\num{578.792} for cluster 11}
		\label{fig:italianticket:11}
	\end{subfigure}
	\begin{subfigure}{0.32\columnwidth}
		\centering
		\includegraphics[width=1\linewidth]{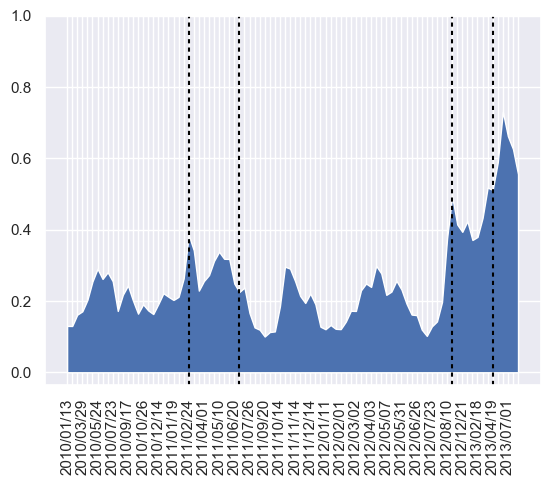}
		\caption{\Errtcsm$=$\num{139.403} for cluster 4}
		\label{fig:italianticket:4}
	\end{subfigure}
	\caption{Italian help desk log detailed clusters.}
	\label{fig:italian_ticket_detailed}
\end{figure}

\noindent\textbf{Step 6: Understanding drift behavior.} 
We further investigate the most erratic cluster. \Cref{fig:italian-help-c9} shows that its behavior of a \emph{reoccurring drift} type. 
%
During the times of peaks, the activity \Task{Create SW anomaly} always had \Task{Take in charge ticket} executed immediately beforehand. 
Also, we observe that the \Task{Assign seriousness} activity was executed before \Task{Create SW anomaly}, and no other \Task{Create SW anomaly} occurred in between.
We further analyze other clusters with 
erratic behavior as shown in~\cref{table:italianticekt-err-clusters}. \Cref{fig:italian_ticket_detailed} shows the drift for cluster 11 and cluster 4. The corresponding constraints are listed in~\cref{table:italianticekt-drifts-constrains}. 
\begin{table}[bt]
	\caption{Italian ticket log constraints; including minimum, maximum, and average confidence.}
	\label{table:italianticekt-drifts-constrains}
	\resizebox{1.0\columnwidth}{!}{
\rowcolors{2}{white}{gray!12.5}
\begin{tabular}{r r r r D{1} D{1} D{1}}
	\toprule
	\textbf{Cluster} & \textbf{Constraint} & \textbf{Activity 1}         & \textbf{Activity 2}         & \textbf{Min} & \textbf{Max} & \textbf{Mean} \\ \midrule
	\cellcolor{white} & {\ChaPrecTmp}       & \Task{Take in charge ticket} & \Task{Create SW anomaly}     & 0.0          & 100          & 42.8          \\
	\multirow{-2}{*}{1} & {\AltPrecTmp}       & \Task{Assign seriousness}    & \Task{Create SW anomaly}     & 0.0          & 100          & 49.0          \\ \midrule
	\cellcolor{gray!12.5} & {\ChaPrecTmp}       & \Task{Take in charge ticket} & \Task{Schedule intervention} & 0.0          & 100          & 9.9           \\
	\multirow{-2}{*}{\cellcolor{gray!12.5}11} & {\AltPrecTmp}       & \Task{Assign seriousness}    & \Task{Schedule intervention} & 0.0          & 100          & 9.9           \\ \midrule
	\cellcolor{white} & {\ChaRespTmp}       & \Task{Take in charge ticket} & \Task{Wait}                  & 9.4          & 69.6         & 23.2          \\
	\cellcolor{white} & {\NotSuccTmp}       & \Task{Resolve ticket}        & \Task{Wait}                  & 10           & 77.2         & 26            \\
	\cellcolor{white} & {\NotSuccTmp}       & \Task{Wait}                  & \Task{Assign seriousness}    & 10           & 78           & 26.6          \\
	\cellcolor{white} & {\NotSuccTmp}       & \Task{Wait}                  & \Task{Take in charge ticket} & 9.8          & 73.3         & 22.1          \\
	\cellcolor{white} & {\AltRespTmp}       & \Task{Assign seriousness}    & \Task{Wait}                  & 9            & 72.3         & 23.8          \\
	\cellcolor{white} & {\AltRespTmp}       & \Task{Wait}                  & \Task{Closed}                & 8.3          & 61.4         & 22.5          \\
	\cellcolor{white} & {\AltRespTmp}       & \Task{Wait}                  & \Task{Resolve ticket}        & 8.3          & 61.4         & 22.8          \\
	\multirow{-8}{*}{4} & {\UniqTmp}          & \Task{Wait}                  &                              & 9.8          & 68.6         & 25.1          \\ \bottomrule
\end{tabular} 
%
%

	}
\end{table}
\Cref{fig:italianticket:11} has four spikes, where \Task{Schedule intervention} activities occurred. Immediately before \Task{Schedule intervention}, \Task{Take in charge ticket} occurred.
Also, \Task{Assign seriousness} occurred before \Task{Schedule intervention}. 
We notice, however, that this cluster shows \emph{outlier} behavior, due to its rare changes.

%
\Cref{fig:italianticket:4} shows a \emph{gradual} drift until June 2012, and an \emph{incremental} drift afterward. 
We notice that all constraints in the cluster have \Task{Wait} either as an activation (e.g., with \AltResp{\Task{Wait}}{\Task{closed}}) or as a target (e.g., with \ChaResp{\Task{Take in charge ticket}}{\Task{Wait}}).

\begin{figure}[tbp]
	\includegraphics[width=\columnwidth]{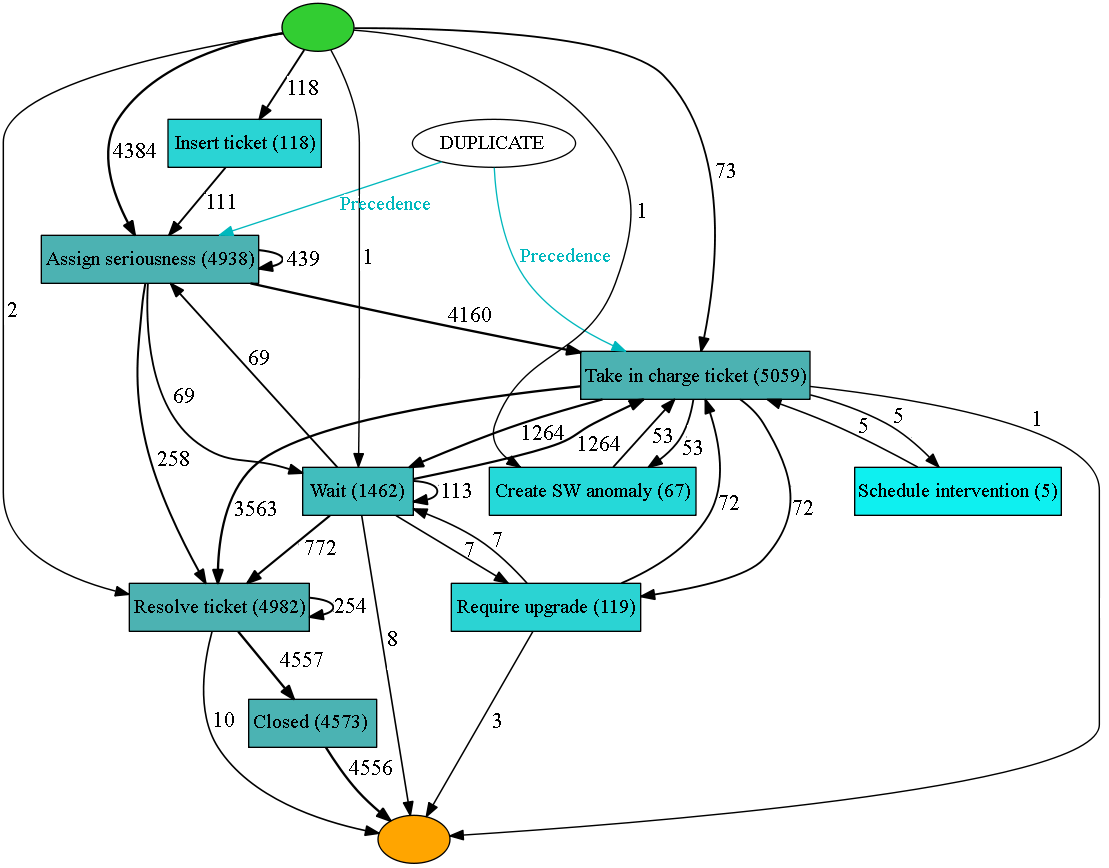}
	\vspace{-3mm}
	\caption{Extended Directly-Follows Graph of cluster 12 in the Italian help desk log.}
	\label{fig:italian-cluster12}
	\vspace{-3mm}
\end{figure}

Finally, we look at cluster 12 with its one-spike drift in \cref{fig:italian-help-c12}.
The corresponding eDFG in \cref{fig:italian-cluster12} shows that this behaviour relates to a \Task{Take in charge ticket} and \Task{Assign seriousness}.

\label{italianhelpdesk}

\subsubsection{BPI2011 event log}
\begin{figure}[t]
	\includegraphics[width=\columnwidth]{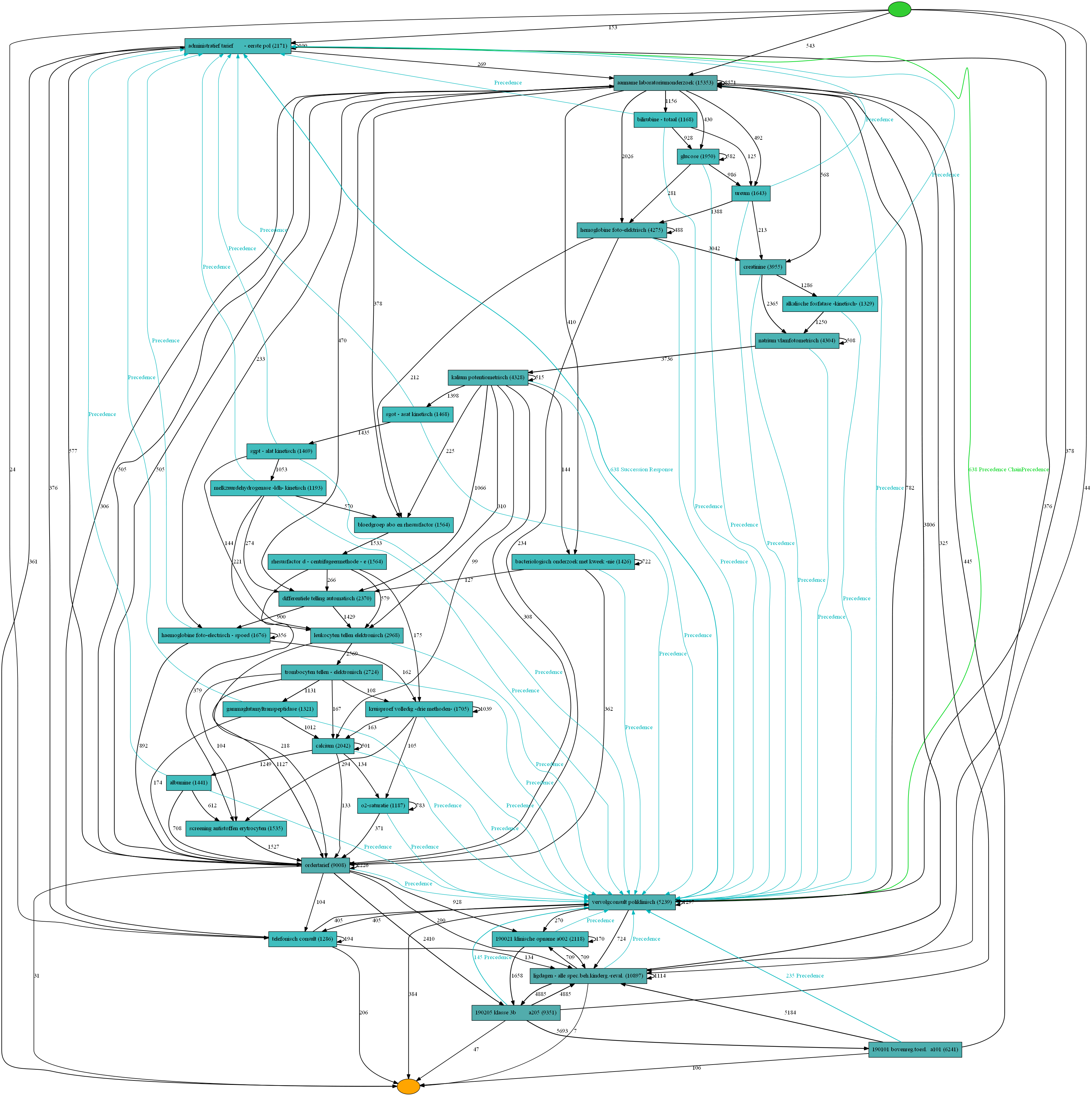}
	\vspace{-3mm}
	\caption{Cluster 16 of BPIC2011}
	\label{fig:bpic2011-dfg-cl16}
	\vspace{-3mm}
\end{figure}

%
%
BPI2011 is the log from the the Gynaecology department of a hospital in the Netherlands. 
\\
\textbf{Step 1: Mining Directly-Follows Graph as an overview.} 
The Directly-Follows Graph includes \num{34} activities. It is shown in \cref{fig:bpic2011-dfg-cl16}. The paths of the cases are largely different, such that no clear main path can be identified.
\\
\textbf{Step 2: Mining {\Declare} windows.} 
This log contains \num{1142} cases spanning over a period of three years and four month. We chose the {\WinSize} of \num{40} and the {\WinStep} of \num{20} in our analysis.
\\
\begin{figure}[tbp]
	\centering
	\begin{subfigure}[t]{0.315\columnwidth}
		\centering
		\includegraphics[width=1\linewidth]{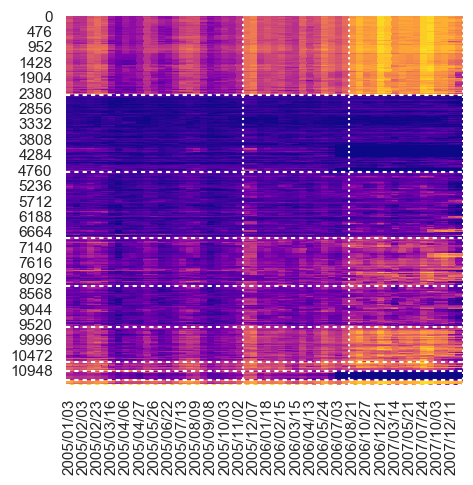}
		\caption{Overall change points}
		\label{fig:bpic2011:alldrifts}
	\end{subfigure}%
	\begin{subfigure}[t]{0.35\columnwidth}
		\centering
		\includegraphics[width=1\linewidth]{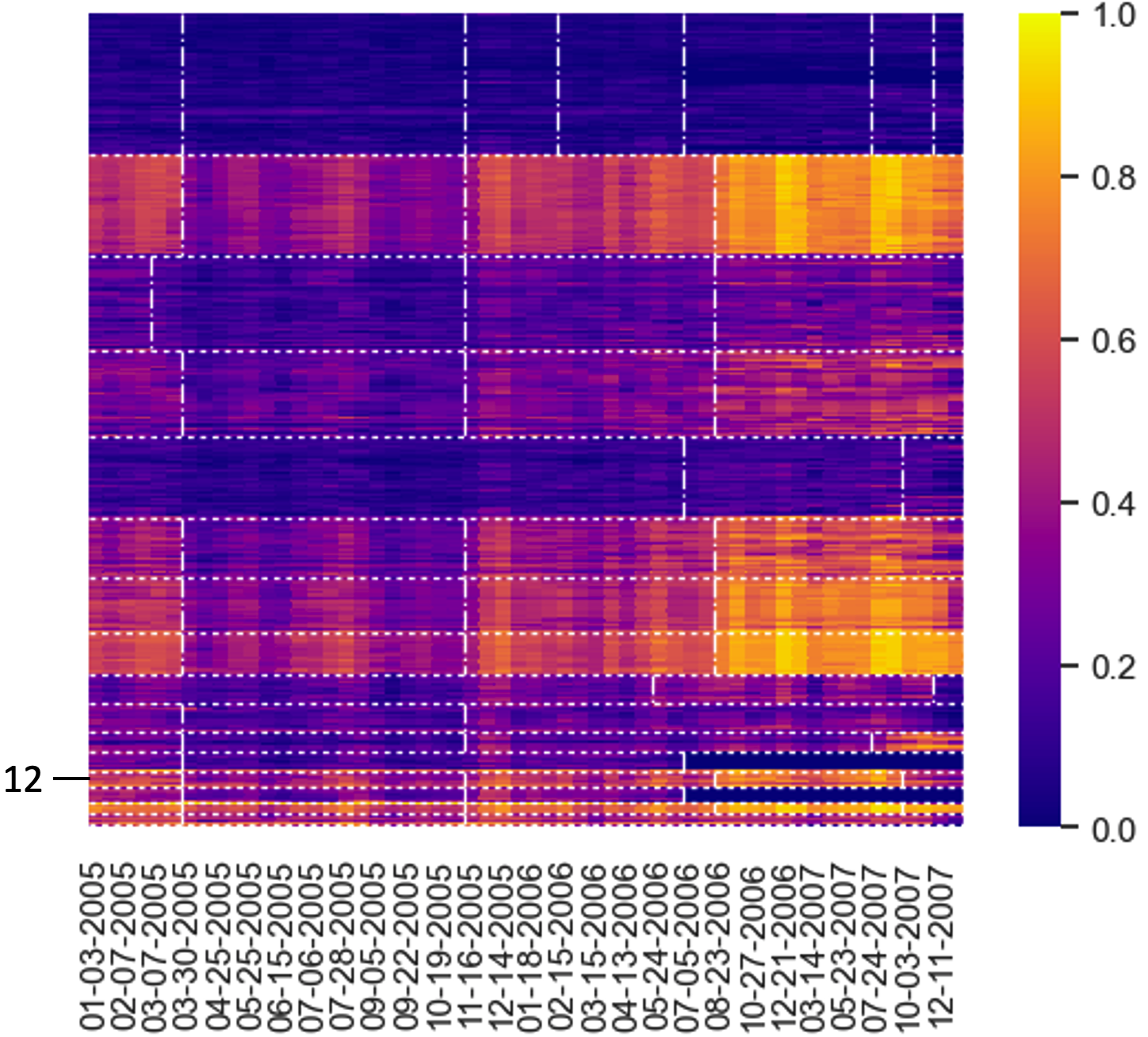}
		\caption{Drifts by cluster}
		\label{fig:bpic2011:separate}
	\end{subfigure}%
	\begin{subfigure}[t]{0.33\columnwidth}
		\centering
		\includegraphics[width=1\linewidth]{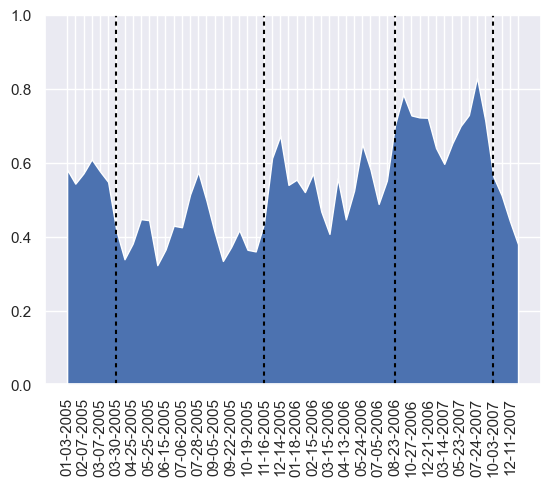}
		\caption{Most erratic cluster (cluster 12)}
		\label{fig:bpic2011:erratic}
	\end{subfigure}
	
	\caption{BPI2011 \gls{dvd} visualizations.}
	\label{fig:bpic2011}
	\vspace{-2mm}
\end{figure}
\textbf{Step 3-4: Finding Drifts and Visual Drift Overview.} 
\Cref{fig:bpic2011:alldrifts} shows the {\DriftMap} of the {\emph{BPIC2011} event log}. 
As in~\cite{DBLP:conf/er/OstovarMRHD16}, two drifts are detected towards the second half of the time span of the log.
However, in addition, our technique identifies drifting behavior at a finer granularity.
\Cref{fig:bpic2011:separate} shows the drifts pertaining to clusters of constraints.
The trend of the confidence measure for the most erratic cluster is depicted in~\Cref{fig:bpic2011:erratic}.

\begin{figure}[tbp]
	\centering
	\begin{subfigure}[t]{0.297\columnwidth}
		\centering
		\includegraphics[width=1\linewidth]{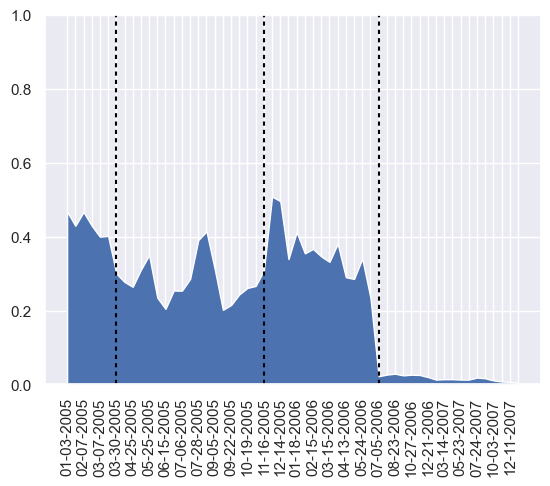}
		\caption{{\DriftChart}\\ of cluster 13}
		\label{fig:bpic11:13dc}
	\end{subfigure}%
	\hspace{1mm}
	\begin{subfigure}[t]{0.297\columnwidth}
		\centering
		\includegraphics[width=1\linewidth]{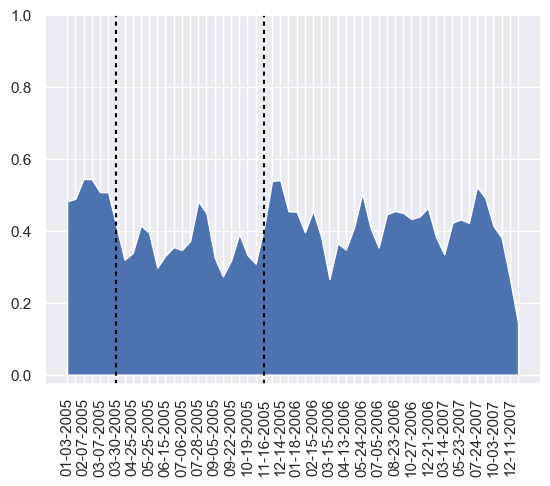}
		\caption{{\DriftChart}\\ of cluster 15}
		\label{fig:bpic11:15dc}
	\end{subfigure}%
	\hspace{1mm}
	\begin{subfigure}[t]{0.36\columnwidth}
		\centering
		\includegraphics[width=1\linewidth]{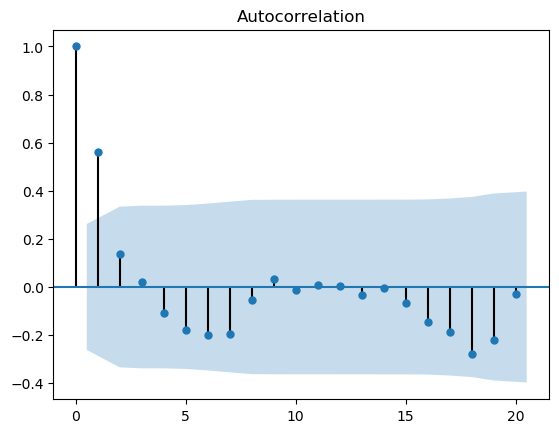}
		\caption{Autocorrelation\\ plot of cluster 15}
		\label{fig:bpic11:15aut}
	\end{subfigure}

	\caption{BPI2011 visualizations.}
	\label{fig:bpic2011-aut}
	\vspace{-2mm}
\end{figure}

While the {\DriftMap} shows that most of the drifts display increasing trends for the plots at the end of the event log timeline, \cref{fig:bpic11:13dc} highlights the opposite direction. The most erratic cluster is characterized by a confidence values that decrease from the beginning of the timeline and decreases afterwards. 
\\
\textbf{Step 5: Detecting Drift Types.} 
To better understand a particular drift, we further examine the constraints that participate in the drift. We explore statistical properties of the discovered drifts.  
We use the erratic measure to identify the strongest drifts and run sudden drift detection in order to identify the drift types.
Sudden drifts are visible in \cref{fig:bpic2011:alldrifts} that correspond to those found in~\cite{DBLP:conf/er/OstovarMRHD16}. Moreover, we are able to discover the sudden drifts for each individual cluster of behavior as shown by vertical lines in~\cref{fig:bpic2011:separate,fig:bpic2011:erratic}. 

\begin{table}[tb]
	\caption{BPI2011 erratic clusters.}
	\label{table:table-bpic2011-clusters}
	\centering
\rowcolors{2}{white}{gray!12.5}
\begin{tabular}{r r r}
	\toprule
	\textbf{Drift number} & \textbf{{\Errtcsm} measure} & \textbf{A-Dickey-Fuller p-value} \\ 
	\midrule
    without drift & 55 & \\
    \emph{12} & \num{221.7915528} & \num{0.060885} \\
    \emph{6} & \num{220.9907785} & \num{0.479091} \\
    \emph{8} & \num{215.3089575} & \num{0.546296} \\
    \emph{7} & \num{214.0067707} &\num{0.887760} \\
    \emph{1} & \num{205.8439399} & \num{0.760080} \\
	\bottomrule
\end{tabular} 

\end{table}

\begin{figure}[tb]
	\centering
		\centering
		\includegraphics[width=0.33\linewidth]{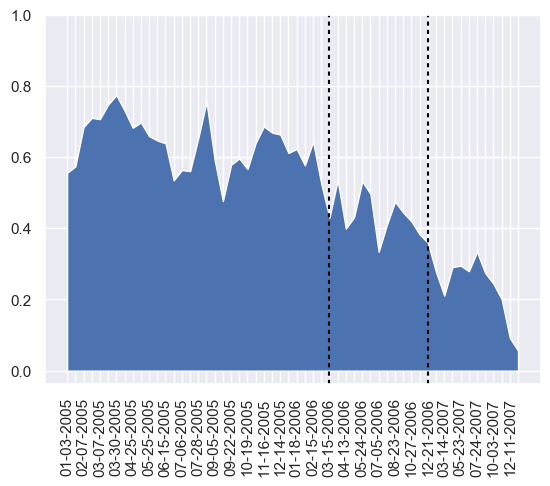}
	\caption{Cluster 16 of BPIC2011}
	\label{fig:BPIC2011:cl16}
\end{figure}

Running the autocorrelation analysis reveals that most of the drifts do not show seasonality. An exception is cluster 15. Its autocorrelation graph (\cref{fig:bpic11:15aut}) and  {\DriftChart} (\cref{fig:bpic11:15dc}) exhibit seasonality.
The Augmented Dickey-Fuller test~\Cref{table:table-bpic2011-clusters} evidences that all of the most erratic clusters are non-stationary. This means that there is a constant change in the process behavior, thus we can conclude that those drifts are incremental.
\\
\textbf{Step 6: Understanding drift behavior.} 
%
\Cref{fig:BPIC2011:cl16} illustrates the drift chart of cluster 16, which we consider for the annotation of the extended DFG 
in~\cref{fig:bpic2011-dfg-cl16}. Apparently, the majority of the drifts in this cluster refer to activity \emph{vervolgconsult poliklinisch}, which is subject to {\PrecTmp} constraints with several other activities.

\subsection{Computational Performance}\label{performance}
We have tested the computational performance of the system. 
We used a MacOS system, equipped with 2.4 GHz Dual-Core Intel Core i5 and 8 GB of RAM. \Cref{table:performance} shows the wall-clock time needed for our system to process each data set, and the basic data set characteristic. To determine the computational performance we used parameters applied in our tests from~\Cref{synthetic,real-world}. 

\begin{table}[tb]
	\caption{Characteristics of the event logs and wall-clock time performance of the system expressed in seconds. CM stands for Conditional Move, CR for Conditional Removal, CS for Conditional to Sequence, IHD for Italian Help Desk logs.}
	\label{table:performance}
	\centering
\rowcolors{2}{white}{gray!12.5}
\begin{tabular}{l r r r r r r r }
	\toprule
	\textbf{Event log} & \textbf{CM}  & \textbf{CR} & \textbf{CS} & \textbf{Loop} & \textbf{IHD} & \textbf{BPI2011} & \textbf{Sepsis} \\ \midrule
	\SeqNum & 9998  & 9999 & 2999  & 9999  & 4579  & 1142  & 1050 \\
	av.seq.l.    & 22.27  & 23.10 & 23.04 & 23.13 &  4.66  & 98.31& 14.49\\
	\ActNum & 27      & 28      & 28      & 27      & 14      & 33     & 16 \\
	\midrule
	Steps 2-4 & 34.98  & 34.16 & 27.83 & 34.77   & 34.21 & 48.77 & 17.39   \\
	Steps 1, 6 & 20.98 & 18.88  & 8.94 & 18.16   & 5.48 & 194.19 & 4.22    \\
	Step 5 & 11.24  & 10.84  & 11.86   & 10.93  & 10.76 & 15.04 & 8.29    \\ \midrule
	\textbf{Total}    & \textbf{67.24}  & \textbf{63.87} & \textbf{48.63} & \textbf{63.82} & \textbf{50.45} & \textbf{258.02} & \textbf{29.90} \\
	\bottomrule
\end{tabular} 
\end{table}

We have measured the computation time of the different steps of the algorithm. First, we measured the time needed to extract time series from the data, cluster, perform change point detection, visualize {\DriftMap} and {\DriftChart}s (Steps 2-4 of our algorithm from~\cref{sec:approach}). Second, we measured the time to build extended DFGs for each cluster (Steps 1 and 6). Third, we measured the time employed by the system to generate autocorrelation plots, finding erratic and spread of constraints measures, and performing the stationarity tests (Step 5). 

The tests show that our system if mostly affected by the number of activities, \ActNum, and the average length of the sequences in the DFG. This parameter is a key factor for the complexity of the extended DFG, as the rendering of the graph appears to be the most costly operation due to the number of {\Declare} constraints that need to be visualized for some of the clusters. Indeed, the \emph{BPI2011} event log required the highest amount of time for all steps. The Italian help desk log needed the lowest time to complete all calculations, as {\ActNum} and average sequence length is the lowest of other datasets.

\subsection{User Evaluation}\label{user}
The previous part of the evaluation highlights the accuracy of our drift detection and visualization.
Our system is designed to meet the requirements of business process analysts. The objective of our user evaluation is to collect evidence in order to judge to which extent the requirements have been effectively addressed. To this end, we conducted a user study with \num{12} process mining experts who are familiar with different tools and approaches for visualizing business process event logs.

The participants were introduced to the data set of the helpdesk case 
described in~\cref{italianhelpdesk} 
together with its Directly-Follows Graph. 
Then, the participants learned about the four major visualization techniques of our system (extended directly-follows graph, drift map, drift chart, and drift measures). We collected quantitative and qualitative data via a survey with a Likert scale and open questions.

Our quantitative evaluation builds on the established technology acceptance model~\cite{davis1989perceived,venkatesh2003user}. This model posits that the two major antecedents of technology adoption are perceived usefulness and ease of use. In our context, this essentially means that if process analysts perceive our visualization system to be easy to use and to provide useful insights, they would likely want to use it in their daily work. The user perceptions of ease of use and usefulness were assessed using the established psychometric measurement instrument with 5 and 6 question items per construct, respectively~\cite{davis1989perceived}.

\begin{figure}[htb]
	\includegraphics[width=\columnwidth]{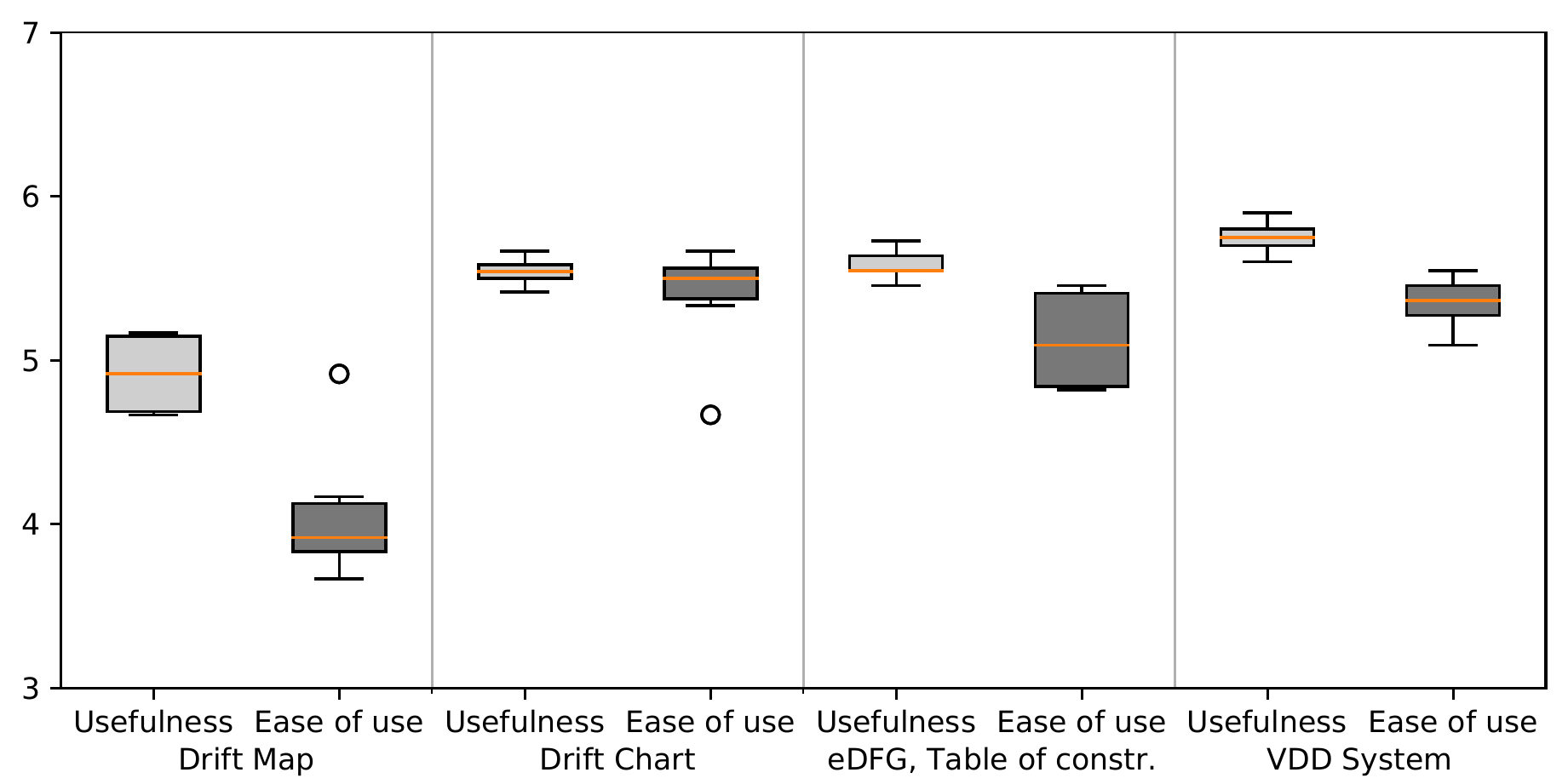}
	\caption{Boxplots of perceived ease of use and perceived usefulness according to the user study}
	\label{fig:TAM}
\end{figure}

The results of the technology acceptance assessment are presented in Fig.~\ref{fig:TAM}. We observe that both ease of use and usefulness are perceived to be close to each other, with usefulness being consistently stronger. Both measurement scales have a high average between 5 and 6, indicating that the users agree that the system is easy to use and useful. The exception is the ease of use of the drift map, which was partially judged to be difficult.

Beyond the quantitative assessment, we also collected qualitative feedback on the different visualizations of our system. Regarding the \emph{drift map}, participant P7 states that it ``\emph{visualizes in one picture a great amount of detailed information in a useful way. It allows perceiving the changes of all the behavior without query for each of them.}'' Participant P2 suggests that it ``\emph{would be nice to add the meaning of clusters.}'' To address this point, we added the feature to filter the drift map for constraints that relate to a specific activity. Regarding the \emph{drift chart}, Participant P6 notes that it ``\emph{is very easy to understand. It clearly shows the compliance of the cases with certain constraints and how it evolves over time.}'' Participant P5 suggests some indication ``\emph{if less/more traces comply with rules.}''
To address this point, we added absolute numbers showing how many cases relate to this chart. Regarding the \emph{extended DFG}, Participant P8 emphasizes that ``\emph{I like that they provide details of specific constraints. I like to visually see the process. I like the enhanced process model.}'' Participant P5 highlights that ``\emph{I see a risk of information overload.}'' We address this point by offering a functionality to filter the eDFG. Regarding the overall system, different participants expressed their perceptions on usefulness by emphasizing that the system ``\emph{provides very powerful means to explore the process change}'' (P6). Participant P8 states that ``\emph{I like to see the three visualizations together.}'' Participant P5 concludes that the information provided by the system ``\emph{certainly improves the accuracy of decisions.}''


%
%
\subsection{Discussion}
\label{subsec:discussion}

Our method addresses all the five requirements for process drift detection presented in~\cref{sec:analyst} as follows:

\smallskip
\begin{compactitem}
	\item[\ref{req:identify}]
	We evaluated our method with the synthetic logs showing its ability to identify drifts precisely;
	\item[\ref{req:categorize}]
	We developed a visualization approach based on {\DriftMap}s and {\DriftChart}s for the classification of process drifts and have shown its effectiveness for real-world logs. Our enhanced approach based on change point detection has yielded an effective way o automatically discover the exact points at which \textit{sudden} and \textit{reoccurring concept} drifts occur.
	The indicative approximation of long-running progress of \textit{incremental} and \textit{gradual} drifts was also found.
	\emph{Outliers} were detected via time series clustering;
	\item[\ref{req:drill}] 
	Using clustering, {\DriftMap}, and {\DriftChart}s, the method enables the drilling down into (rolling up out) sections with a specific behavior (general vs. cluster-specific groups of constraints);
	\item[\ref{req:quanti}]
	We introduced, and incorporated into our technique, a drift measure called {\Errtcsm} that quantifies the extent of the drift change;
	\item[\ref{req:quali}]
	To further qualitatively analyze the detected drifts, \gls{dvd} shows how the process specification looks before and after the drift (as a list of {\Declare} constraints, refer to~\cref{table:italianticekt-drifts-constrains}).
\end{compactitem}
\noindent

\subsection{Limitations}
\label{subsec:limitations}
In this section, we outline the future work directions defined by the limitations of our system.

We noticed that irregularly sampled data could affect the analysis. 
Our approach splits a log into windows of a fixed number of traces. 
The irregular data could produce graphs that have unevenly spaced timeticks. 
Taking into account the time ranges instead of the number of traces will affect our analysis. 
Different strategies for splitting the log should be investigated in future work. 


When interacting with the~\gls{dvd} system, an analyst manually identifies seasonal drifts based on the autocorrelation graphs and explores incremental drifts based on {\DriftChart}s. 
Future work will aim at automating both these tasks.

As demonstrated in \cref{performance}, the performance of the system allows for the handling of industrial datasets. However, this performance is achieved for the offline setting, when the necessary information is precomputed, and does not extend to the online setting, as new input data will trigger an overall recalculation. Extending the system to online settings is another avenue for future work.

For datasets with a large number of possible activities and a significant number of drifts, the performance of the system could be further improved by prioritizing {\Declare} constraints that get rendered as \glspl{dfg}.

Finally, the choices of algorithms for clustering and change-point detection could be informed by the input data. In the case of a large dataset, faster clustering algorithms could be selected. The analysis of such choices on the system's performance is future work.
%
%
\section{Conclusions}
\label{sec:conclusion}
%
%
In this paper, we presented a visual system for the detection and analysis of process drifts from event logs of executed business processes. Our contributions are techniques for fine-granular process drift detection and visualization.
The different visualizations of our system integrate extended Directly-Follows Graphs, {\Declare} constraints, the {\DriftMap}s and {\DriftChart}s plus several metrics and statistics for determining types of drift.

We evaluated our system both on synthetic and real-world data. 
On synthetic logs, we achieved an average F-score of \num{0.96} and outperformed all the state-of-the-art methods. 
On real-world logs, the technique describes all types of process drifts in a comprehensive manner. 
Also, the evaluation reported that our technique can identify outliers of process behavior.
Furthermore, we conducted a user study, which highlights that our visualizations are easy to interact with and useful, as perceived by process mining experts.
\ifCLASSOPTIONcompsoc
  \section*{Acknowledgments}
\else
  \section*{Acknowledgment}
\fi

This work is partially funded by the EU H2020 program under MSCA-RISE agreement 645751 ({RISE\_BPM}).
A.~Polyvyanyy was in part supported by the Australian Research Council Discovery Project DP180102839.
C.~Di Ciccio was partly supported by the MUR under grant ``Dipartimenti di eccellenza 2018-2022'' of the Department of Computer Science at Sapienza University of Rome.

\ifCLASSOPTIONcaptionsoff
  \newpage
\fi

\begin{IEEEbiography}[{\includegraphics[width=1in,height=1.25in,clip,keepaspectratio]{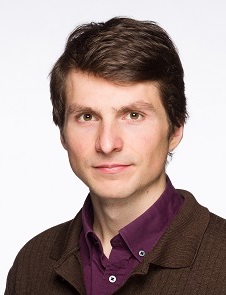}}]{Anton Yeshchenko}
is a Ph.D. student and a teaching and research associate at the Institute for Information Business at Wirtschaftsuniversit\"{a}t Wien, Austria. His research focuses on visualization research, data science, and machine learning for temporal event sequence data. He has a theoretical background in applied mathematics and computer science from Taras Shevchenko University (Ukraine) and Tartu University (Estonia). He has worked at research institute FBK in Trento developing new methods in predictive monitoring of business processes. He has published in international A-ranked conferences including BPM, ER, and CoopIS.
\end{IEEEbiography}

\begin{IEEEbiography}[{\includegraphics[width=1in,height=1.25in,clip,keepaspectratio]{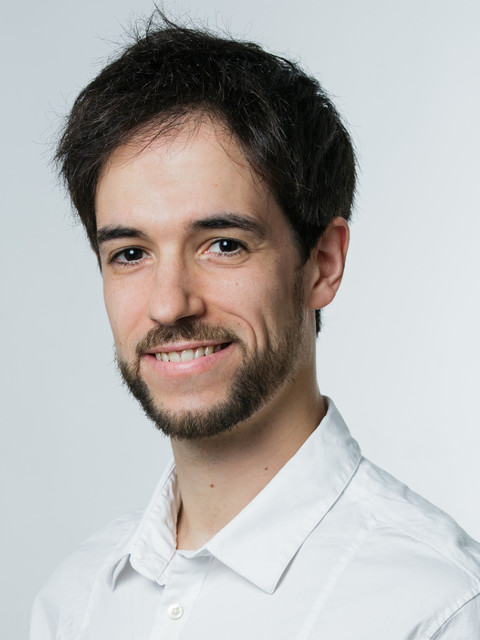}}]{Claudio Di Ciccio}
	is an Assistant Professor with the Department of Computer Science at Sapienza University of Rome. He received a PhD in Computer Science and Engineering in 2013 at Sapienza. His research interests include Process Mining, Blockchains, and Declarative Modelling. 
	His work in these fields has been published in renowned international conferences such as BPM, CAiSE, and ICSOC, 	and top journals, including Decision Support Systems, Information Systems and ACM Trans. on Management Information Systems. He received the best paper award at BPM 2015 
	and the best user paper award at ECIR 2019. 
	He is a member of the Steering Committee of the IEEE Task Force on Process Mining. 
\end{IEEEbiography}

\begin{IEEEbiography}[{\includegraphics[width=1in,height=1.25in,clip,keepaspectratio]{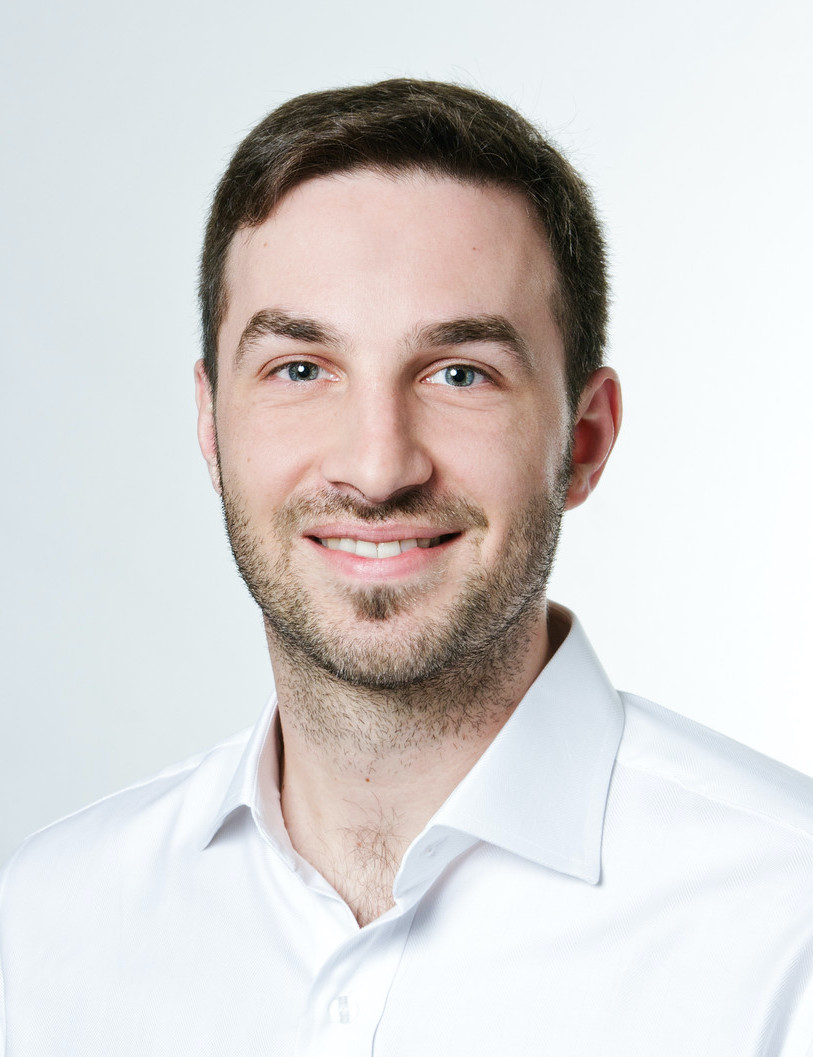}}]{Jan Mendling}
is a Full Professor with the Institute for Information Business at Wirtschaftsuniversit\"{a}t Wien, Austria. His research interests include business process management and information systems. He has published more than 400 research papers and articles, among others in the Journal of the Association of Information Systems, ACM Transactions on Software Engineering and Methodology, IEEE Transaction on Software Engineering, Information Systems, European Journal of Information Systems, and Decision Support Systems. He is a board member of the Austrian Society for Process Management, one of the founders of the Berliner BPM-Offensive, and member of the IEEE Task Force on Process Mining. He is a co-author of the textbooks Fundamentals of Business Process Management and Wirtschaftsinformatik. 
\end{IEEEbiography}
\begin{IEEEbiography}[{\includegraphics[width=1in,height=1.25in,clip,keepaspectratio]{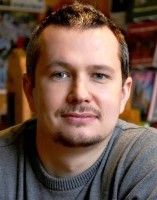}}]{Dr. Artem Polyvyanyy}
is a senior lecturer at the School of Computing and Information Systems, Melbourne School of Engineering, at the University of Melbourne (Australia). 
In March 2012, he received a Ph.D. degree 
in Computer Science from the University of Potsdam (Germany). His research and teaching interests include Information Systems, Process Modeling and Analysis, Process Mining, Process Querying, Data Science, and Algorithms. Artem Polyvyanyy has published over 80 research papers and articles. He has actively contributed to several open-source initiatives that led to a significant impact on research and practice, including jBPT, Oryx, and Apromore.
\end{IEEEbiography}






\begin{thebibliography}{10}
\providecommand{\url}[1]{#1}
\csname url@samestyle\endcsname
\providecommand{\newblock}{\relax}
\providecommand{\bibinfo}[2]{#2}
\providecommand{\BIBentrySTDinterwordspacing}{\spaceskip=0pt\relax}
\providecommand{\BIBentryALTinterwordstretchfactor}{4}
\providecommand{\BIBentryALTinterwordspacing}{\spaceskip=\fontdimen2\font plus
\BIBentryALTinterwordstretchfactor\fontdimen3\font minus
  \fontdimen4\font\relax}
\providecommand{\BIBforeignlanguage}[2]{{%
\expandafter\ifx\csname l@#1\endcsname\relax
\typeout{** WARNING: IEEEtran.bst: No hyphenation pattern has been}%
\typeout{** loaded for the language `#1'. Using the pattern for}%
\typeout{** the default language instead.}%
\else
\language=\csname l@#1\endcsname
\fi
#2}}
\providecommand{\BIBdecl}{\relax}
\BIBdecl

\bibitem{DBLP:conf/emisa/MannhardtB17}
F.~Mannhardt and D.~Blinde, ``Analyzing the trajectories of patients with
  sepsis using process mining,'' in \emph{RADAR+EMISA@CAiSE}, ser. {CEUR}
  Workshop Proceedings, vol. 1859.\hskip 1em plus 0.5em minus 0.4em\relax
  CEUR-WS.org, 2017, pp. 72--80.

\bibitem{aigner2007visual}
W.~Aigner, S.~Miksch, W.~M{\"u}ller, H.~Schumann, and C.~Tominski, ``Visual
  methods for analyzing time-oriented data,'' \emph{IEEE transactions on
  visualization and computer graphics}, vol.~14, no.~1, pp. 47--60, 2007.

\bibitem{aigner2011visualization}
W.~Aigner, S.~Miksch, H.~Schumann, and C.~Tominski, \emph{Visualization of
  time-oriented data}.\hskip 1em plus 0.5em minus 0.4em\relax Springer Science
  \& Business Media, 2011.

\bibitem{cousins1991visual}
S.~B. Cousins and M.~G. Kahn, ``The visual display of temporal information,''
  \emph{Artificial Intelligence in Medicine}, vol.~3, no.~6, pp. 341--357,
  1991.

\bibitem{viegas2004studying}
F.~B. Vi{\'e}gas, M.~Wattenberg, and K.~Dave, ``Studying cooperation and
  conflict between authors with history flow visualizations,'' in
  \emph{Proceedings of the SIGCHI conference on Human factors in computing
  systems}, 2004, pp. 575--582.

\bibitem{havre2002themeriver}
S.~Havre, E.~Hetzler, P.~Whitney, and L.~Nowell, ``Themeriver: Visualizing
  thematic changes in large document collections,'' \emph{IEEE transactions on
  visualization and computer graphics}, vol.~8, no.~1, pp. 9--20, 2002.

\bibitem{kim2010tracing}
N.~W. Kim, S.~K. Card, and J.~Heer, ``Tracing genealogical data with
  timenets,'' in \emph{Proceedings of the International Conference on Advanced
  Visual Interfaces}, 2010, pp. 241--248.

\bibitem{mendling2008metrics}
J.~Mendling, \emph{Metrics for process models: empirical foundations of
  verification, error prediction, and guidelines for correctness}.\hskip 1em
  plus 0.5em minus 0.4em\relax Springer Science \& Business Media, 2008,
  vol.~6.

\bibitem{DBLP:books/sp/Aalst16}
W.~M.~P. van~der Aalst, \emph{Process Mining - Data Science in Action}.\hskip
  1em plus 0.5em minus 0.4em\relax Springer, 2016.

\bibitem{DBLP:books/sp/DumasRMR18}
M.~Dumas, M.~{La Rosa}, J.~Mendling, and H.~A. Reijers, \emph{Fundamentals of
  Business Process Management, Second Edition}.\hskip 1em plus 0.5em minus
  0.4em\relax Springer, 2018.

\bibitem{Denisov/BPM2018:MiningConceptDriftinPerformanceSpectraofProcesses}
V.~Denisov, E.~Belkina, and D.~Fahland, ``{BPIC}'2018: {M}ining concept drift
  in performance spectra of processes,'' 2018.

\bibitem{DBLP:conf/simpda/HompesBADB15}
B.~Hompes, J.~C. A.~M. Buijs, W.~M.~P. van~der Aalst, P.~Dixit, and H.~Buurman,
  ``Detecting change in processes using comparative trace clustering,'' in
  \emph{SIMPDA 2015}, 2015, pp. 95--108.

\bibitem{DBLP:conf/s-bpm-one/SeeligerNM17}
A.~Seeliger, T.~Nolle, and M.~M{\"{u}}hlh{\"{a}}user, ``Detecting concept drift
  in processes using graph metrics on process graphs,'' in \emph{S-BPM}, 2017,
  p.~6.

\bibitem{DBLP:conf/otm/ZhengW017}
C.~Zheng, L.~Wen, and J.~Wang, ``Detecting process concept drifts from event
  logs,'' in \emph{{OTM} CoopIS}, 2017, pp. 524--542.

\bibitem{DBLP:conf/er/OstovarMRHD16}
A.~Ostovar, A.~Maaradji, M.~{La Rosa}, A.~H.~M. ter Hofstede, and B.~F. van
  Dongen, ``Detecting drift from event streams of unpredictable business
  processes,'' in \emph{ER}, 2016, pp. 330--346.

\bibitem{DBLP:journals/tkde/MaaradjiDRO17}
A.~Maaradji, M.~Dumas, M.~{La Rosa}, and A.~Ostovar, ``Detecting sudden and
  gradual drifts in business processes from execution traces,'' \emph{{IEEE}
  TKDE}, vol.~29, no.~10, pp. 2140--2154, 2017.

\bibitem{sedlmair2012design}
M.~Sedlmair, M.~D. Meyer, and T.~Munzner, ``Design study methodology:
  Reflections from the trenches and the stacks,'' \emph{{IEEE} Trans. Vis.
  Comput. Graph.}, vol.~18, no.~12, pp. 2431--2440, 2012.

\bibitem{DBLP:journals/corr/abs-2007-15272}
X.~Wang, W.~Chen, J.~Xia, Z.~Chen, D.~Xu, X.~Wu, M.~Xu, and T.~Schreck,
  ``Concept{E}xplorer: {V}isual analysis of concept driftsin multi-source
  time-series data,'' \emph{CoRR}, vol. abs/2007.15272, 2020.

\bibitem{ware2012information}
C.~Ware, \emph{Information visualization: perception for design}.\hskip 1em
  plus 0.5em minus 0.4em\relax Elsevier, 2012.

\bibitem{Aalst.etal/CSRD09:DeclarativeWFsBalancing}
W.~M.~P. van~der Aalst, M.~Pesic, and H.~Schonenberg, ``Declarative workflows:
  Balancing between flexibility and support,'' \emph{CS - R{\&}D}, vol.~23,
  no.~2, pp. 99--113, 2009.

\bibitem{DBLP:journals/tmis/CiccioM15}
C.~Di~Ciccio and M.~Mecella, ``On the discovery of declarative control flows
  for artful processes,'' \emph{{ACM} TMIS}, vol.~5, no.~4, pp. 24:1--24:37,
  2015.

\bibitem{CharlesTruonga/SParxiv:SelectiveReviewOfOfflineChangePointDetectionMethods}
C.~Truong, L.~Oudre, and N.~Vayatis, ``Selective review of offline change point
  detection methods,'' \emph{Signal Process.}, vol. 167, 2020.

\bibitem{dos2019process}
C.~dos Santos~Garcia, A.~Meincheim, E.~R.~F. Junior, M.~R. Dallagassa, D.~M.~V.
  Sato, D.~R. Carvalho, E.~A.~P. Santos, and E.~E. Scalabrin, ``Process mining
  techniques and applications-a systematic mapping study,'' \emph{Expert
  Systems with Applications}, 2019.

\bibitem{DBLP:journals/csur/GamaZBPB14}
J.~Gama, I.~Zliobaite, A.~Bifet, M.~Pechenizkiy, and A.~Bouchachia, ``A survey
  on concept drift adaptation,'' \emph{{ACM} Comput. Surv.}, vol.~46, no.~4,
  pp. 44:1--44:37, 2014.

\bibitem{TsymbalAlexey/:TheProblemOfConceptDrift:DefinitionsAndRelatedWork}
A.~Tsymbal, ``The problem of concept drift: definitions and related work,''
  \emph{Computer Science Department, Trinity College Dublin}, vol. 106, no.~2,
  p.~58, 2004.

\bibitem{DBLP:conf/caise/BauerSGGW18}
M.~Bauer, A.~Senderovich, A.~Gal, L.~Grunske, and M.~Weidlich, ``How much event
  data is enough? {A} statistical framework for process discovery,'' in
  \emph{CAISE}, 2018, pp. 239--256.

\bibitem{DBLP:journals/jcss/DeutchM12}
D.~Deutch and T.~Milo, ``A structural/temporal query language for business
  processes,'' \emph{J. Comput. Syst. Sci.}, vol.~78, no.~2, pp. 583--609,
  2012.

\bibitem{DBLP:conf/apn/PolyvyanyyWCRH14}
A.~Polyvyanyy, M.~Weidlich, R.~Conforti, M.~{La Rosa}, and A.~H.~M. ter
  Hofstede, ``The {4C} spectrum of fundamental behavioral relations for
  concurrent systems,'' in \emph{{Petri nets}}.\hskip 1em plus 0.5em minus
  0.4em\relax Springer, 2014, pp. 210--232.

\bibitem{DBLP:conf/simpda/PrescherCM14}
J.~Prescher, C.~D. Ciccio, and J.~Mendling, ``From declarative processes to
  imperative models,'' in \emph{{SIMPDA}}, ser. {CEUR} Workshop Proceedings,
  vol. 1293.\hskip 1em plus 0.5em minus 0.4em\relax CEUR-WS.org, 2014, pp.
  162--173.

\bibitem{DBLP:conf/er/YeshchenkoCMP19a}
A.~Yeshchenko, C.~D. Ciccio, J.~Mendling, and A.~Polyvyanyy, ``Comprehensive
  process drift detection with visual analytics,'' in \emph{{ER}}, ser. Lecture
  Notes in Computer Science, vol. 11788.\hskip 1em plus 0.5em minus 0.4em\relax
  Springer, 2019, pp. 119--135.

\bibitem{quteprints121158}
A.~Ostovar, S.~J.~J. Leemans, and M.~L. Rosa, ``Robust drift characterization
  from event streams of business processes,'' \emph{{ACM} Trans. Knowl. Discov.
  Data}, vol.~14, no.~3, pp. 30:1--30:57, 2020.

\bibitem{DBLP:conf/vizsec/CappersMEW18}
B.~C.~M. Cappers, P.~N. Meessen, S.~Etalle, and J.~J. van Wijk, ``Eventpad:
  Rapid malware analysis and reverse engineering using visual analytics,'' in
  \emph{VizSEC}.\hskip 1em plus 0.5em minus 0.4em\relax {IEEE}, 2018, pp. 1--8.

\bibitem{DBLP:journals/tvcg/XuMR017}
P.~Xu, H.~Mei, L.~Ren, and W.~Chen, ``Vidx: Visual diagnostics of assembly line
  performance in smart factories,'' \emph{{IEEE} Trans. Vis. Comput. Graph.},
  vol.~23, no.~1, pp. 291--300, 2017.

\bibitem{DBLP:conf/bigdataconf/GuoJCGZC19}
S.~Guo, Z.~Jin, Q.~Chen, D.~Gotz, H.~Zha, and N.~Cao, ``Visual anomaly
  detection in event sequence data,'' in \emph{BigData}.\hskip 1em plus 0.5em
  minus 0.4em\relax {IEEE}, 2019, pp. 1125--1130.

\bibitem{1316839}
W.~{van der Aalst}, T.~{Weijters}, and L.~{Maruster}, ``Workflow mining:
  discovering process models from event logs,'' \emph{TKDE}, vol.~16, no.~9,
  pp. 1128--1142, 2004.

\bibitem{DBLP:series/hci/AignerMST11}
W.~Aigner, S.~Miksch, H.~Schumann, and C.~Tominski, \emph{Visualization of
  Time-Oriented Data}, ser. Human-Computer Interaction Series.\hskip 1em plus
  0.5em minus 0.4em\relax Springer, 2011.

\bibitem{DBLP:journals/corr/abs-2006-14291}
Y.~Guo, S.~Guo, Z.~Jin, S.~Kaul, D.~Gotz, and N.~Cao, ``Survey on visual
  analysis of event sequence data,'' \emph{CoRR}, vol. abs/2006.14291, 2020.

\bibitem{DBLP:journals/tvcg/ChenXR18}
Y.~Chen, P.~Xu, and L.~Ren, ``Sequence synopsis: Optimize visual summary of
  temporal event data,'' \emph{{IEEE} Trans. Vis. Comput. Graph.}, vol.~24,
  no.~1, pp. 45--55, 2018.

\bibitem{DBLP:journals/tvcg/GuoXZGZC18}
S.~Guo, K.~Xu, R.~Zhao, D.~Gotz, H.~Zha, and N.~Cao, ``Eventthread: Visual
  summarization and stage analysis of event sequence data,'' \emph{{IEEE}
  Trans. Vis. Comput. Graph.}, vol.~24, no.~1, pp. 56--65, 2018.

\bibitem{DBLP:journals/tvcg/ZhangCD19}
Y.~Zhang, K.~Chanana, and C.~Dunne, ``Idmvis: Temporal event sequence
  visualization for type 1 diabetes treatment decision support,'' \emph{{IEEE}
  Trans. Vis. Comput. Graph.}, vol.~25, no.~1, pp. 512--522, 2019.

\bibitem{DBLP:conf/chi/WongsuphasawatGPWTS11}
K.~Wongsuphasawat, J.~A.~G. G{\'{o}}mez, C.~Plaisant, T.~D. Wang,
  M.~Taieb{-}Maimon, and B.~Shneiderman, ``Lifeflow: visualizing an overview of
  event sequences,'' in \emph{{CHI}}.\hskip 1em plus 0.5em minus 0.4em\relax
  {ACM}, 2011, pp. 1747--1756.

\bibitem{DBLP:journals/tvcg/MonroeLLPS13}
M.~Monroe, R.~Lan, H.~Lee, C.~Plaisant, and B.~Shneiderman, ``Temporal event
  sequence simplification,'' \emph{{IEEE} Trans. Vis. Comput. Graph.}, vol.~19,
  no.~12, pp. 2227--2236, 2013.

\bibitem{DBLP:journals/tvcg/LawLMB19}
P.~Law, Z.~Liu, S.~Malik, and R.~C. Basole, ``{MAQUI:} interweaving queries and
  pattern mining for recursive event sequence exploration,'' \emph{{IEEE}
  Trans. Vis. Comput. Graph.}, vol.~25, no.~1, pp. 396--406, 2019.

\bibitem{DBLP:journals/tvcg/WongsuphasawatG12}
K.~Wongsuphasawat and D.~Gotz, ``Exploring flow, factors, and outcomes of
  temporal event sequences with the outflow visualization,'' \emph{{IEEE}
  Trans. Vis. Comput. Graph.}, vol.~18, no.~12, pp. 2659--2668, 2012.

\bibitem{DBLP:journals/tvcg/TanahashiM12}
Y.~Tanahashi and K.~Ma, ``Design considerations for optimizing storyline
  visualizations,'' \emph{{IEEE} Trans. Vis. Comput. Graph.}, vol.~18, no.~12,
  pp. 2679--2688, 2012.

\bibitem{DBLP:journals/tvcg/AlbersDG11}
D.~Albers, C.~N. Dewey, and M.~Gleicher, ``Sequence surveyor: Leveraging
  overview for scalable genomic alignment visualization,'' \emph{{IEEE} Trans.
  Vis. Comput. Graph.}, vol.~17, no.~12, pp. 2392--2401, 2011.

\bibitem{DBLP:journals/tvcg/CappersW18}
B.~C.~M. Cappers and J.~J. van Wijk, ``Exploring multivariate event sequences
  using rules, aggregations, and selections,'' \emph{{IEEE} Trans. Vis. Comput.
  Graph.}, vol.~24, no.~1, pp. 532--541, 2018.

\bibitem{DBLP:conf/iui/MalikDMOPS15}
S.~Malik, F.~Du, M.~Monroe, E.~Onukwugha, C.~Plaisant, and B.~Shneiderman,
  ``Cohort comparison of event sequences with balanced integration of visual
  analytics and statistics,'' in \emph{{IUI}}.\hskip 1em plus 0.5em minus
  0.4em\relax {ACM}, 2015, pp. 38--49.

\bibitem{DBLP:conf/chi/ZhaoLDHW15}
J.~Zhao, Z.~Liu, M.~Dontcheva, A.~Hertzmann, and A.~Wilson, ``Matrixwave:
  Visual comparison of event sequence data,'' in \emph{{CHI}}.\hskip 1em plus
  0.5em minus 0.4em\relax {ACM}, 2015, pp. 259--268.

\bibitem{DBLP:conf/icpm/LeemansPW19}
S.~J.~J. Leemans, E.~Poppe, and M.~T. Wynn, ``Directly follows-based process
  mining: Exploration {\&} a case study,'' in \emph{{ICPM}}.\hskip 1em plus
  0.5em minus 0.4em\relax {IEEE}, 2019, pp. 25--32.

\bibitem{DBLP:conf/caise/LeemansFA15}
S.~J.~J. Leemans, D.~Fahland, and W.~M.~P. van~der Aalst, ``Scalable process
  discovery with guarantees,'' in \emph{{BMMDS/EMMSAD}}, ser. Lecture Notes in
  Business Information Processing, vol. 214.\hskip 1em plus 0.5em minus
  0.4em\relax Springer, 2015, pp. 85--101.

\bibitem{DBLP:journals/corr/abs-1905-06169}
A.~Berti, S.~J. van Zelst, and W.~M.~P. van~der Aalst, ``Process mining for
  python (pm4py): Bridging the gap between process- and data science,''
  \emph{CoRR}, vol. abs/1905.06169, 2019.

\bibitem{DBLP:journals/tkde/AalstWM04}
W.~M.~P. van~der Aalst, T.~Weijters, and L.~Maruster, ``Workflow mining:
  Discovering process models from event logs,'' \emph{{IEEE} Trans. Knowl. Data
  Eng.}, vol.~16, no.~9, pp. 1128--1142, 2004.

\bibitem{weijters2006process}
A.~Weijters, W.~M. van Der~Aalst, and A.~A. De~Medeiros, ``Process mining with
  the heuristics miner-algorithm,'' \emph{Technische Universiteit Eindhoven,
  Tech. Rep. WP}, vol. 166, pp. 1--34, 2006.

\bibitem{Polyvyanyy.etal/FAOC2016:ExpressivePowerBehavioralProfiles}
A.~Polyvyanyy, A.~Armas{-}Cervantes, M.~Dumas, and
  L.~Garc{\'{\i}}a{-}Ba{\~{n}}uelos, ``On the expressive power of behavioral
  profiles,'' \emph{Formal Asp. Comput.}, vol.~28, no.~4, pp. 597--613, 2016.

\bibitem{DiCiccio.etal/IS2017:ResolvingInconsistenciesRedundanciesDeclare}
C.~Di~Ciccio, F.~M. Maggi, M.~Montali, and J.~Mendling, ``Resolving
  inconsistencies and redundancies in declarative process models,'' \emph{IS},
  vol.~64, pp. 425--446, Mar. 2017.

\bibitem{DBLP:conf/cidm/MaggiMA11}
F.~M. Maggi, A.~J. Mooij, and W.~M.~P. van~der Aalst, ``User-guided discovery
  of declarative process models,'' in \emph{{CIDM}}.\hskip 1em plus 0.5em minus
  0.4em\relax {IEEE}, 2011, pp. 192--199.

\bibitem{DBLP:conf/icpm/CecconiGCMM20}
A.~Cecconi, G.~D. Giacomo, C.~D. Ciccio, F.~M. Maggi, and J.~Mendling, ``A
  temporal logic-based measurement framework for process mining,'' in
  \emph{{ICPM}}.\hskip 1em plus 0.5em minus 0.4em\relax {IEEE}, 2020, pp.
  113--120.

\bibitem{DBLP:journals/is/MaggiCFK18}
F.~M. Maggi, C.~Di~Ciccio, C.~Di~Francescomarino, and T.~Kala, ``Parallel
  algorithms for the automated discovery of declarative process models,''
  \emph{Inf. Syst.}, vol.~74, no. Part 2, pp. 136--152, 2018.

\bibitem{DBLP:journals/is/SmedtWSV18}
J.~D. Smedt, J.~D. Weerdt, E.~Serral, and J.~Vanthienen, ``Discovering hidden
  dependencies in constraint-based declarative process models for improving
  understandability,'' \emph{Inf. Syst.}, vol.~74, no. Part 1, pp. 40--52,
  2018.

\bibitem{Aghabozorgi:2015:TCD:2799194.2799230}
S.~R. Aghabozorgi, A.~S. Shirkhorshidi, and Y.~W. Teh, ``Time-series clustering
  - {A} decade review,'' \emph{Inf. Syst.}, vol.~53, pp. 16--38, 2015.

\bibitem{killick2012optimal}
R.~Killick, P.~Fearnhead, and I.~A. Eckley, ``Optimal detection of changepoints
  with a linear computational cost,'' \emph{Journal of the American Statistical
  Association}, vol. 107, no. 500, pp. 1590--1598, 2012.

\bibitem{box2015time}
G.~E. Box, G.~M. Jenkins, G.~C. Reinsel, and G.~M. Ljung, \emph{Time series
  analysis: forecasting and control}.\hskip 1em plus 0.5em minus 0.4em\relax
  John Wiley \& Sons, 2015.

\bibitem{cheung1995lag}
Y.-W. Cheung and K.~S. Lai, ``Lag order and critical values of the augmented
  dickey--fuller test,'' \emph{Journal of Business \& Economic Statistics},
  vol.~13, no.~3, pp. 277--280, 1995.

\bibitem{meyer2015nested}
M.~Meyer, M.~Sedlmair, P.~S. Quinan, and T.~Munzner, ``The nested blocks and
  guidelines model,'' \emph{Information Visualization}, vol.~14, no.~3, pp.
  234--249, 2015.

\bibitem{DBLP:conf/icpm/YeshchenkoMCP20}
A.~Yeshchenko, J.~Mendling, C.~D. Ciccio, and A.~Polyvyanyy, ``{VDD:} {A}
  visual drift detection system for process mining,'' in \emph{{ICPM} Doctoral
  Consortium / Tools}, ser. {CEUR} Workshop Proceedings, vol. 2703.\hskip 1em
  plus 0.5em minus 0.4em\relax CEUR-WS.org, 2020, pp. 31--34.

\bibitem{davis1989perceived}
F.~D. Davis, ``Perceived usefulness, perceived ease of use, and user acceptance
  of information technology,'' \emph{MIS quarterly}, pp. 319--340, 1989.

\bibitem{venkatesh2003user}
V.~Venkatesh, M.~G. Morris, G.~B. Davis, and F.~D. Davis, ``User acceptance of
  information technology: Toward a unified view,'' \emph{MIS quarterly}, pp.
  425--478, 2003.

\end{thebibliography}
\end{document}